\documentclass{aa} 

\usepackage{graphicx}
\usepackage{txfonts}
\usepackage{xfrac}
\usepackage[colorlinks=true,urlcolor=blue,citecolor=blue,
            linkcolor=blue]{hyperref}
\usepackage{mhchem}
\usepackage{array}

\def\aap{A\&A}

\def\apjl{ApJL}
\def\apjs{ApJS}

\def\la{\mathrel{\mathchoice {\vcenter{\offinterlineskip\halign{\hfil
$\displaystyle##$\hfil\cr<\cr\sim\cr}}}
{\vcenter{\offinterlineskip\halign{\hfil$\textstyle##$\hfil\cr
<\cr\sim\cr}}}
{\vcenter{\offinterlineskip\halign{\hfil$\scriptstyle##$\hfil\cr
<\cr\sim\cr}}}
{\vcenter{\offinterlineskip\halign{\hfil$\scriptscriptstyle##$\hfil\cr
<\cr\sim\cr}}}}}
\def\ga{\mathrel{\mathchoice {\vcenter{\offinterlineskip\halign{\hfil
$\displaystyle##$\hfil\cr>\cr\sim\cr}}}
{\vcenter{\offinterlineskip\halign{\hfil$\textstyle##$\hfil\cr
>\cr\sim\cr}}}
{\vcenter{\offinterlineskip\halign{\hfil$\scriptstyle##$\hfil\cr
>\cr\sim\cr}}}
{\vcenter{\offinterlineskip\halign{\hfil$\scriptscriptstyle##$\hfil\cr
>\cr\sim\cr}}}}}

\def\Vl{V_{\!\ell}}
\def\vth{v_{\rm th}}

\def\vdreq{{v^{\hspace{-0.8ex}^{\circ}}_{\rm dr}}}

\def\tstop{\tau_{\rm stop}}

\def\half{\hspace*{0.2pt}\sfrac{1\hspace*{-0.4pt}}{\hspace*{0.2pt}2}}

\def\rhom{\rho_{\rm m}}
\def\nH{n_{\langle\rm H\rangle}}
\def\rhoL{\rho L}
\def\ek{\epsilon_k}
\def\Vs{V_{\!s}}
\newcommand{\Edes}{E_{\rm des}}
\newcommand{\nsurf}{n_{\rm surf}}
\newcommand{\msp}{m_{\rm sp}}
\newcommand{\Pvap}{P_{\rm vap}}
\newcommand{\freq}{\nu_0}

\newcommand{\centerimage}[2][]{$\vcenter{\hbox{\includegraphics[#1]{#2}}}$}

\begin{document} 

\title{The oxygen cold trap in protoplanetary discs}

\author{P.~Woitke\inst{1}
}

\institute{Space Research Institute, Austrian Academy of Sciences,
           Schmiedlstr.~6, A-8042, Graz, Austria
}

\date{Received July 24$^{\rm th}$ 2026; accepted November 24$^{\rm th}$ 2026}

\abstract{The diffusive transport of oxygen in protoplanetary discs
  followed by sequestration into ice in the midplane -- a process
  known as the cold trap -- can substantially alter the chemical
  environment where planets form.  This paper presents a new
  time-dependent modelling framework that allows us to consistently
  simulate kinetic gas-phase chemistry, ice growth and desorption,
  turbulent diffusive mixing, and gravitational settling of icy
  particles with a changing particle size distribution over
  million-year timescales in one-dimensional vertical
  columns. Applying our model to a T\,Tauri disc across radii from 1
  to 100\,au reveals that vertical cold trapping becomes an effective
  mechanism once the diffusion-to-settling timescale ratio exceeds
  unity, simultaneously driving the gas-phase C/O up toward unity, and
  the gas-phase C/H down, by up to two orders of magnitude at large
  distances. All models feature the development of an upward
  counter-flux due to the turbulent up-whirling of icy particles
  followed by ice sublimation above the horizontal icelines, which
  brings the cold trapping to a halt and eventually establishes a
  relaxed vertical structure after about $10^4$ to $10^6$\,yrs,
  depending on the radial distance.  The ice formation leads to a
  strongly peaked size distribution around 1\,$\mu$m, which contains
  the majority of the condensed mass in the disc after ice formation,
  and changes the dynamical behaviour of the grains substantially.  We
  conclude that the vertical cold trap can provide an important part
  of the explanation why evolved discs observed with ALMA often show
  high C/O ratios as inferred from strong emissions from hydrocarbon
  molecules in the outer disc, such as \ce{C2H} and \ce{c-C3H2}.
  However, in none of our models does the C/O ratio exceed unity,
  which is a consequence of CO-blocking, a chemical process where water
  molecules naturally stop forming once C/O approaches unity.}


\maketitle


\section{Introduction}
\label{sec:intro}
The chemical composition of the gas, and the element composition of
the dust and ice in protoplanetary discs are the building blocks for
nascent planets, setting the stage for their subsequent evolution and
ultimate habitability.  Among the most crucial volatile species is
water (\ce{H2O}), which can condense beyond the snowline around
150\,K, and can be transported both radially and vertically with the
drifting and settling icy grains.  The transport of water ice is key
to establishing the chemical gradients that define the planet-forming
environment.

Protoplanetary discs are not vertically homogeneous. The upper layers
are exposed to intense stellar UV radiation, creating a warm
photodissociation region (PDR) where molecules are broken down into
atoms.  Outside of just a few au, gaseous water is only found in trace
concentrations in discs as shown by both thermo-chemical models
\cite{Aikawa2002, Woitke2009, Walsh2010} and observations; for a
review see \cite{vanDishoeck2021}.  {Herschel/HIFI observations,
  for example by \cite{Bergin2010}, \cite{Hogerheijde2011} and
  \cite{Du2017} yielded one detection (TW\,Hya) but mostly upper
  limits far below model predictions, which was interpreted as water
  ice locked in grown, settled grains}. Instead, oxygen is
predominantly present in form of CO and oxygen atoms.  Water ice is
not stable in the upper layers as either the small grains residing
there are too warm, or UV photo-dissociation destroys the water ice.
In stark contrast, the dense and cold midplane is shielded from this
radiation, providing a quiescent environment where volatiles can
freeze onto the surfaces of dust grains. {Water ice has been
  detected in the outer disc of HD\,142527 \citep{Min2016} and in
  absorption in edge-on discs with JWST \citep{Sturm2023}.  In these
  deeper regions,} gaseous water molecules would be stable against
photo-dissociation, but the cold temperatures cause them to freeze out
almost instantaneously. {Even the younger Class~I discs show
  little water vapour \citep{Harsono2020}.}  However, these layers have
been shown to only begin relatively deep in the disc surface, at
vertical optical extinctions of about $A_{\rm V}^{\rm ver}\ga$\,5-10,
forming a 'horizontal snowline'. Such deep layers are notoriously
difficult to observe.

This vertical dichotomy presents a fundamental puzzle: since the
direct gas-phase formation of gaseous \ce{H2O} lacks an effective
neutral-neutral pathway at temperatures $T\!<\!200\,$K
\citep{Woitke2009b}, oxygen must be transported from the atomic
reservoir in the upper layers to the midplane to form water ice. Many
dynamical models for water ice formation and transport, however, have
ignored this crucial chemical step.  Previous models have explored
various aspects of this problem.  For instance, detailed chemical
networks have successfully predicted the vertical stratification of
key species in static disc models
\citep[e.g.][]{Woitke2009,Fogel2011,Bruderer2012}. \cite{Oberg2011}
developed a clear model based on the simple but powerful idea of
sequential freeze-out along the midplane, resulting in gaseous $\rm
C/O\!\approx\!1$ between the \ce{CO2} and CO icelines, a concept that
is challenged, however, by the vertical dichotomy. Separately, the
vertical transport through turbulent diffusion
and gravitational settling has been extensively studied by
e.g. \cite{Dullemond2004} and \cite{Riols2018}, and it has been shown
that icy grains can be constantly mixed over the horizontal snowline
\citep{Woitke2022} releasing water molecules locally.  However, a
self-consistent, time-dependent treatment that simultaneously couples
the gas-phase chemistry, gas-grain interactions (including ice
formation and desorption), and the vertical dynamics of both gas and
dust has remained a challenge.  Without such a model, the long-term
efficiency of sequestering atmospheric oxygen into midplane water ice
-- a process we term the {\em cold trap} -- cannot be robustly
quantified.

High-resolution observations with the Atacama Large Millimeter Array
(ALMA) have revealed a surprisingly rich hydrocarbon chemistry in the
upper layers at 100\,au radii of many protoplanetary discs, with
bright emission from molecules such as \ce{C2H}, \ce{c-C3H2} and CS
\citep[e.g.][]{Miotello2017, LeGal2021}.  The high abundance of these
species is challenging to explain with standard chemical models
assuming a solar C/O ratio ($\sim\!0.5$), as most carbon should be
locked in CO.  Consequently, these observations have led multiple
studies to conclude that the gas in planet-forming discs is
significantly oxygen-depleted in the outer regions, with a gas-phase
C/O ratio equal to or slightly greater than one \citep{Bergin2015,
  Kama2016}. {Gas-phase CO may be depleted by 1-2 orders of
  magnitude in many discs \citep{Bergin2013, Zhang2020b, Bergner2020}.}
A recent study of HD\,143006 by \cite{Terwisga2026} has
used a large data set of 14 ALMA lines to derive $\rm
C/O\!=\!1.05_{-0.03}^{+0.1}$ by extensive thermo-chemical disc
modelling.  A leading hypothesis to explain this oxygen deficiency is
indeed the cold trap: the efficient sequestration of oxygen into water
ice followed by gravitational settling toward the shielded midplane,
effectively removing it from the observable gas.

In this paper, we examine the effectiveness of the vertical cold trap
for oxygen in protoplanetary discs. We have developed a new
time-dependent, one-dimensional (1D, vertical) model that simulates
the evolution of gas, dust, and ice over timescales of millions of
years. Our model combines a targeted chemical rate network with
gas-phase chemistry, kinetic ice formation, surface hydrogenation,
thermal and UV photo-desorption, turbulent diffusive mixing of gas and
icy grains, and gravitational settling of grains with changing size
distribution. Previous models, such as \cite{Krijt2016a,Krijt2016b},
have shown that the transport and sequestration of volatile-rich
solids can strongly affect the observable gas-phase abundances in
discs. {In particular, \cite{Krijt2020} coupled simplified
  time-dependent chemistry for the major C- and O-bearing ices H$_2$O,
  CO, CO$_2$, CH$_4$, and CH$_3$OH to dust growth and pebble drift in
  2D models, and demonstrated that chemical processing and the radial
  transport of icy pebbles together can alter the C/O ratio in the
  outer disc and the warm molecular layers. Their work emphasised the
  importance of radial pebble drift and volatile redistribution across
  the CO snowline, but necessarily employed a reduced chemical
  description and did not address the detailed vertical balance
  between settling, turbulent re-mixing, ice growth, and desorption of
  small icy grains.}

{More recently, \cite{Robinson2026} studied the CO iceline and
  surface in a thermodynamically coupled, vertically resolved 2D
  disc model, including dust evolution, ice--vapour exchange, ice
  opacities, and radiative transfer. They showed that vertical
  structure is essential for modelling volatile evolution: vapour is
  deposited over a wide radial range rather than only at the midplane
  iceline radius, and vertical diffusion from a vapour reservoir
  above the ice surface can stabilise the CO iceline instability
  found in earlier 1D models. However, their study focused on CO phase
  exchange only, with a simplified chemical model
  for the phase transition between CO vapour and ice.}

{Recent work by \cite{vanClepper2022} followed a different
  approach more similar to what is assumed in this paper.  The authors
  developed a 1D vertical model coupling a large gas-grain chemical
  network to pebble growth. In their models, gas and small dust
  diffuse vertically, while pebbles form near the midplane and
  instantly remove the icy material from further chemical exchange
  with the gas. They found that pebble growth can deplete CO in the
  surface layers by one to two orders of magnitude and can raise the
  gas-phase C/O ratio only to unity, but not beyond. This work
  demonstrated the importance of coupling vertical transport, kinetic
  chemistry, and dust evolution.}

Several other studies have explored the chemical evolution of volatile
ices on dynamically evolving dust, but they have primarily focused on
the radial transport of planet-forming material in the disc midplane
in the frame of 1D hydrodynamical disc models.
For instance, \cite{Eistrup2018} demonstrated that local grain growth
drastically alters the grain surface area available for ice chemistry,
shifting the dominant carriers of carbon and oxygen.
\cite{Molyarova2021,Molyarova2026} modelled the volatile transport
and chemical metamorphoses of ices on radially drifting pebbles,
showing how cosmic-ray driven processing and radial dust trapping
combine to shape the inner disc C/O ratio and metallicity. {These
global radial models provide important insight into the redistribution
of volatiles by pebble drift, but they often simplify the vertical
structure or assume that the midplane ice reservoir is already
established. Conversely, vertical models with detailed chemistry and
pebble growth have not yet quantified whether the cold trap remains a
one-way sink once the icy particles themselves are allowed to diffuse,
settle, grow, sublimate, and change their size distribution
self-consistently.}

The key question addressed here is therefore whether vertical cold
trapping leads to permanent sequestration of oxygen in the shielded
midplane, or whether turbulent up-whirling of the icy grains creates
an upward counter-flux that limits or even nullifies the effect.

\section{Model}
\label{sec:method}

Our model is based on the DiffuDrift code developed by
\cite{Woitke2020,Woitke2025}, which simulates the formation of cloud
particles consisting of several solid materials in a 1D vertical
planetary atmosphere. Here we consider a mixture of a 'passive'
material, which represents the refractory dust that is not affected by
chemistry, and several ice species, including water ice.  Before we
explain how to combine particle growth, settling and diffusion
with kinetic chemistry in Sect.\,\ref{sec:kin_chem}, we first
summarise the basics of the DiffuDrift model.

\subsection{Moment method, net growth, settling, and diffusion}

DiffuDrift simulates the particles in terms of their size moments
\begin{equation}
  \rhoL_j = \int_{\Vl}^\infty\!\!f(V)\,V^{j/3}\,dV
  \quad\quad\mbox{$(j\!=\!0,1,2,3)$}\ ,
\end{equation}
where $V\,\rm[cm^3]$ is the volume of a particle and
$f(V)\,\rm[cm^{-6}]$ is the particle size distribution
function. $\rho\rm\,[g/cm^3]$ is the gas mass density, and $\Vl$ is
the lower integration boundary for the particles.  The unit of $L_j$
is $[\rm cm^j/g]$, for example $L_0$ is the number of particles per
gram of gas, and $L_3$ is the total volume of the particles per gram
of gas.

To keep track of the changing material composition of the particles
{with volumes $V\!=\!\sum_s \Vs$, where $s$ is an index for the
  different solid materials, here $s\!=\!\rm\{passive, ice\#1, ice\#2,
  ...\}$, we introduce the following additional variables}
\begin{equation}
  \rhoL_3^s = \int_{\Vl}^\infty f(V)\,\Vs\,dV 
  \quad\quad\mbox{$(s\!=\!1,\,...\,,S)$}\ ,
\end{equation}
such that $L_3\!=\!\sum_s L_3^s$. {The key assumption is that the
  material composition of the particles}
\begin{equation}
  b_{\rm  mix}^s = \frac{\Vs}{V} = \frac{L_3^s}{L_3}
\end{equation}
{is independent of their size, while it is allowed to change with
  position and time, so there is no history of layers.}

A DiffuDrift simulation advances these size moments in time in a
given atmospheric 1D vertical structure as
\begin{eqnarray}
  \frac{d}{dt}\big(\rhoL_j\big)
  &=&
      \underbrace{\Vl^{j/3} J_\star}_{\rm nucleation}
   ~+~ \underbrace{\int_{V_\ell}^{\infty}\!\frac{\partial V}{\partial t}
                  f(V)\,V^{j/3}\,dV}_{\rm growth\;\&\;sublimation}\,
   ~+~ \underbrace{\frac{\partial(\rhoL_j)}{\partial t}\Bigg|_{\rm coag} 
                 }_{\rm coagulation}
  \nonumber\\
  &+& 
  \underbrace{\frac{\partial}{\partial z}
    \int_{V_\ell}^{\infty}\!\vdreq\,f(V)\,V^{j/3}\,dV}_{\rm settling}  
  ~+~\underbrace{\frac{\partial}{\partial z}\,
      \bigg(D\,\rho\,\frac{\partial L_j}{\partial z}\bigg)}_{\rm diffusion} 
  \ .\label{eq:mom}
\end{eqnarray}
In the application of DiffuDrift presented in this paper, we neglect
coagulation and nucleation. Considering particles with
radii between sub-microns and millimetres in protoplanetary discs, we
are safely in the Epstein regime of large Knudsen numbers
\begin{equation}
  Kn = \frac{\bar{\ell}}{2a} \gg 1 \ ,
\end{equation}
where $\bar{\ell}$ is the mean free path length of the gas particles
and $a$ is the particle radius, see Eq.\,(11) in \cite{Woitke2003}. In
this case we can write the net growth rates as
\begin{equation}
  \frac{\partial \Vs}{\partial t}
  ~=~  4\pi a^2\,n^s_{\rm key}\,v^s_{\rm rel}\,\alpha_s\,V_0^s\;\gamma_s(T) \ ,
  \label{eq:growth}
\end{equation}
which is the change of the volume of a particle due to the
deposition/sublimation of material $s$ on/from its surface. The total
volume change caused by such chemical surface processes is $\partial
V/\partial t = \sum_s \partial \Vs/\partial t$.  $a$ is the particle
radius ($V\!=\!4\pi/3\,a^3$), and $\alpha_s\!\approx\!1$ is the
sticking probability. In this paper, we only consider basic growth
reactions where a supersaturated vapour condenses directly into ice of
the same kind, for example for water we have $s\!=\!\ce{H2O\#}$.  In
this case, the key molecule that limits the growth is \ce{H2O} itself
with a molecular particle density of $n^s_{\rm key}\!=\!n_{\ce{H2O}}$,
and the key molecule's relative thermal velocity is $v^s_{\rm
  rel}\!=\!\sqrt{kT/(2\pi\,m^s_{\rm key})}$, where $m^s_{\rm key}$ is
the mass of the key molecule for the growth of solid $s$.  $V_0^s$ is
the monomer volume of the condensed species $s$. For water, we find
$V_0\!=\!3.22\times 10^{-23}\rm\,cm^3$ assuming an ice density of
$\rho_{\ce{H2O\#}}\!=\!\rm 0.93\,g/cm^3$.

The saturation factor $\gamma(T)$ accounts for thermal desorption as
inverse process of the adsorption
\begin{equation}
  \gamma_s(T) = \left(1-\frac{1}{S(T)}\right)\;\times\;
  \left\{\begin{array}{cl}
  1 &                \mbox{\!\!if $S(T)\geq 1$}\\
  \!\! b_{\rm mix}^s & \mbox{\!\!if $S(T)< 1$} \ ,
  \end{array}\right.
  \label{eq:desorb}
\end{equation}
where the supersaturation ratio, for example for water, is
$S(T)\!=\!p_{\ce{H2O}}/p_{\rm vap}^{\ce{H2O}}(T)$ and $p_{\rm
  vap}^{\ce{H2O}}(T)$ is the water vapour saturation
pressure. Equation~(\ref{eq:desorb}) was introduced by
\cite{Woitke2020}. It assumes that the net growth
(adsorption\,$-$\,thermal desorption) rate always changes sign where
$S\!=\!1$ and that the desorption rate is proportional to the
probability to find condensate $s$ on the surface. The relations
bewteen the vapour pressures and the desorption energies are explained
in App.\,\ref{app:pvap} and material properties are listed in
Table~\ref{tab:icedata}.

In the large Knudsen number limiting case, the vertical settling
velocity (also called equilibrium drift velocity or final fall speed)
is given by
\begin{equation}
  \vdreq(a) = -g_z\,\frac{\rhom\,a}{\rho\,\vth} \ ,
  \label{eq:vdr}
\end{equation}
where $g_z\!=\!G M_\star\,z/r^3$ is the vertical gravitational
acceleration, $z$ the height above the midplane, $r=\sqrt{x^2+z^2}$
the radius and $x$ is the distance to the rotation axis.  The varying
material density of the particles is $\rhom\!=\!\sum_s b_{\rm mix}^s
\rho_{s}$, where $\rho_s$ is the material mass density of the pure solid
$s$. The mean gas thermal velocity is
$\vth\!=\!\sqrt{8\,kT/(\pi\,\mu)}$ where $\mu$ the mean gas particle
mass.

{Since $\partial\Vs/\partial t\!\propto\!a^2$ and
  $\vdreq\!\propto\!a$ are powerlaws according to
  Eqs.\,(\ref{eq:growth}) and (\ref{eq:vdr})}, we can analytically
solve the integrals in Eq.\,(\ref{eq:mom}) to obtain expressions
$\propto\!j\,L_{j-1}$ for the net growth and $\propto\!L_{j+1}$ for
the settling terms, see \cite{Woitke2003} and \cite{Helling2006}.
{An analog procedure was introduced by \cite{Woitke2025}, which
  uses a representation of the size moments by two particles}
\begin{equation}
  \rhoL_j = \sum\limits_{p=1}^2 n_p V_{\!p}^{j/3}\quad\quad\mbox{$(j=0,1,2,3)$} \ ,
  \label{eq:double-delta}
\end{equation}
{where the two particle densities and volumes
  $\{n_1,n_2,V_1,V_2\}$ can be computed from the knowns
  $\{L_0,L_1,L_2,L_3\}$ at any time. These two particles "represent"
  the whole size distribution, they have the correct total number
  density, size, area and volume as given by the size moments, so
  knowing them is equivalent to knowing the moments.  We have shown in
  \cite{Woitke2025} that the two-particle representation leads to the
  same solution as the above described analytic solution of the
  integrals for growth and settling, when all particles are in the
  same hydrodynamical regime, either in the large Knudsen number
  limiting case (Epstein regime), as considered in this paper, or in
  the viscous small Knudsen number limiting case (Stokes regime), see
  Apps.~A and B in \cite{Woitke2025}. Equation~(\ref{eq:double-delta})
  is then applied to the settling term with a power-correction factor}
\begin{equation}
  \int_{V_\ell}^{\infty}\!\vdreq\,f(V)\,V^{j/3}\,dV
  = P_{\rm c}(j\!+\!1) \sum\limits_{p=1}^2 n_p V_p^{j/3} \vdreq(V_p)
  \label{eq:moment_drift}
\end{equation}
that can be considered as the closure condition for our moment system.
The power-correction factor is unity, $P_{\rm c}(j)\!=\!1$, for
$j=\{0,1,2,3\}$, but has a value slighly larger than one for
$j\!=\!4$. {The power-correction scheme uses our knowledge about
  how the settling velocity scales with radius $a$. In the large
  Knudsen number limiting case $\vdreq\!\propto\!a$, so we apply
  $P_{\rm c}(3+1)$ to correct. If the dust size distribution function
  is very narrow, a power-correction is not required, $P_{\rm
    c}(4)\!\to\!1$, but if it is broad there will be a small
  correction $P_{\rm c}(4)>1$ because the $4^{\rm th}$ moment, as
  required in Eq.\,(\ref{eq:moment_drift}) for $j\!=\!3$, is slightly
  underpredicted by the 2 representative particles. We characterise
  the width of the size distribution function by $L_1L_3/(L_2)^2$ and
  use a plausible functional shape of the size distribution function
  to work out the difference between the $4^{\rm th}$ moment and the
  $4^{\rm th}$ moment predicted by the two representative particles as
  function of width. In App.~\ref{app:power_correct}, we detail which
  functional shape of the size distribution was chosen for this
  paper.}

{Another verification test for our moment closure is described in
  App.~\ref{app:networks}, where we model the ice formation at
  selected points without settling and diffusion. The initial
  condition is a powerlaw size distribution in this case, and 
  all particles grow by the same $\Delta a$. We work out the moments
  analytically and compare these values with the results found in the
  model after long simulation times}. The full set of equations solved
in this paper, including all transport terms and the kinetic chemistry
described in Sect.\,\ref{sec:kin_chem} is listed in
App.~\ref{app:fulleqs}.

{$D\rm\,[cm^2/s]$ is the eddy-diffusion constant caused by turbulent
mixing. We use a constant value for our simulations in this paper,
depending on radius $r$ and viscosity parameter $\alpha$, following
the $\alpha$-disc model by \cite{Shakura1973}}
\begin{equation}
  D = \alpha\,c_{\rm S}\,H_p \ ,
  \label{eq:Diff}
\end{equation}
{where $c_{\rm S}\!=\!\sqrt{kT/\mu}$ is the isothermal sound velocity
in the midplane and $H_p\!=\!c_{\rm S}/\Omega$ the gas scale height,
$\Omega$ the Keplerian circular frequency, for more details see
Sect.~2.9 in \cite{Woitke2024}.}

The modelling is then carried out by explicit time steps $\Delta t$ by
means of an operator splitting method, {see Fig.~2 in
  \cite{Woitke2025}}, doing $\half\,\Delta t$ diffusion,
$\half\,\Delta t$ settling, $\Delta t$ chemistry and particle growth,
$\half\,\Delta t$ settling and $\half\,\Delta t$ diffusion, and so on.
For more details. All gas and dust components
of the solution vector $\vec{y}$ (see Sect.\,\ref{sec:kin_chem}) are
diffused with the same diffusion coefficient, but only the size
moments $L_j$ and $L_3^s$ are settled.

\subsection{Kinetic Chemistry}
\label{sec:kin_chem}
In this paper, we have replaced DiffuDrift's chemistry, growth \&
sublimation module based on thermo-chemical equilibrium by a new
module that involves kinetic chemistry {instead}.  In the previous
DiffuDrift version, the solution vector was $\vec{y} =
\{\rhoL_0,...\,,\rhoL_3; \rhoL_3^1,...\,,\rhoL_3^S;
\rho\epsilon_1,...\,,\rho\epsilon_K\}$, where $\ek$ are the abundances
of the elements in the gas phase.  The time derivatives $d\ek/dt$
account for the consumption of the elements due to the formation and
growth of the various solid materials. The time derivative
$d\vec{y}/dt$ was calculated by assuming that all molecular abundances
adjust immediately to $n_i=n_i(T,p,\ek)$, as is true in instantaneous
thermo-chemical equilibrium. The new solution vector is
\begin{equation}
  \vec{\tilde{y}} = \{\rhoL_0,...\,,\rhoL_3;
             ~\rhoL_3^1,...\,,\rhoL_3^S;
             ~\rho\epsilon_1,...\,,\rho\epsilon_K;
             ~n_1,...\,,n_I\}
  \label{eq:solvec}
\end{equation}
{where $i=\{1,...\,,I\}$ is a new index for the molecules explicitly
taken into account, $k=\{1,...\,,K\}$ the index for the elements
characterising the background gas, and $s=\{1,...\,,S\}$ the index for
the solid and ice materials. All three components (particles, element
abundances and explicit molecules) are diffused and transported.}  The
kinetic chemical rate equations will be used to advance $n_i(t)$ and
$\ek(t)$, including the consumption by ice formation. Conversely, the
$\epsilon_k$ now describe the abundances of all other ``background''
molecules not explicitely taken into account.

\subsection{Involving an external kinetic chemistry code}
\label{sec:external}

The following kinetic treatment of the molecular concentrations $n_i$
requires to have completed an external kinetic chemistry simulation
for the atmospheric column under investigation, which computes the
reaction kinetics for a set of molecules ${\cal M}$, typically
consisting of a few hundred chemical species, {and select a subset
  ${\cal S} \subset {\cal M}$ of molecules for DiffuDrift}, which must
include at least those from which DiffuDrift calculates the
supersaturation ratios, the growth rates, and the nucleation rates
\begin{equation}
  n_i \in {\cal S} \subset {\cal M} \ .
\end{equation}
In the external chemical model, the particle density of every species
is evolved in time as
\begin{equation}
  \frac{\partial n_i}{\partial t} = \Gamma_i - \Lambda_i \ ,
\end{equation}
{where $\Gamma_i$ and $\Lambda_i$ are the total formation and
  destruction rates of species $i$ computed from all chemical
  reactions. For the interface to DiffuDrift, we need to lump these
  individual reaction rates into effective rates and subdivide them
  into the following different classes of chemical processes}
\begin{eqnarray}
  \Gamma_i &\!\!=\!\!& \Gamma_i^\# + \Gamma_i^{\rm gas}  
       ~=~ \Gamma_i^\# + \Gamma^{(0)}_i
        + \sum\limits_{j\in{\cal S}} \Gamma^{(1)}_{j\to\,i}\,n_j
        + \sum\limits_{j,k\in{\cal S}} \Gamma^{(2)}_{j,k\to\,i}\,n_j n_k \ ,
  \nonumber\\
  \Lambda_i &\!\!=\!\!& \Lambda_i^\# + \Lambda_i^{\rm gas}
        = \Lambda_i^\# + \Lambda^{(0)}_i + \Lambda^{(1)}_{i\to\rm any}\,n_i
        + \sum\limits_{j\in{\cal S}} \Lambda^{(2)}_{i,j\to\rm any}\,n_i n_j 
  \nonumber\ .
\end{eqnarray}
The ice rates include the release of molecules by thermal and
photo-desorption $\Gamma_i^\#$ and the consumption of molecules by
surface growth reactions (adsorption) $\Lambda_i^\#$.  These rates (if
present in the chemistry code) will be replaced by the
growth \& sublimation rates calculated in DiffuDrift.  All other
kinetic rates in the gas phase (gas-kinetic, photo- and
Xray-reactions, ...) are sub-divided into constant (``background'')
rates, $\Gamma^{(0)}_i$ and $\Lambda^{(0)}_i$ (these do not involve
any of the selected molecules as reactants), rates that scale linearly
with the concentration of one of the selected molecules,
$\Gamma^{(1)}_{j\to\,i}$ and $\Lambda^{(1)}_{i\to\rm any}$, and rates
that scale with the product of the concentrations of two selected
molecules $\Gamma^{(2)}_{j,k\to\,i}$ and $\Lambda^{(2)}_{i,j\to\rm
  any}$.  Here, the designation $j\to i$ means a reaction of selected
$j$ with an unselected molecule (or with a photon) that produces one
selected $i$ plus additional products. $j,k\to i$ means a reaction of
two selected molecules $j$ and $k$ producing an $i$ plus additional
products.  $i\to\rm any$ means a reaction of one selected molecule
$i$ with an unselected molecule (or with a photon) that results in any
products, and $i,j\to\rm any$ means a reaction of two selected
molecules $i$ and $j$ that results in any products.

The coefficients $\{\Gamma^{(0)}_i, \Gamma^{(1)}_{j\to i},
\Gamma^{(2)}_{j,k\to i}, \Lambda^{(0)}_i, \Lambda^{(1)}_{i\to\rm any},
\Lambda^{(2)}_{i,j\to\rm any}\}$ are saved as function of height $z$
to a file which is then read by DiffuDrift and interpolated onto the
vertical grid points used by DiffuDrift.  {Thus, in DiffuDrift, there
  is no list of reactions -- only these lumped reaction coefficients,
  from which the right hand sides of the solution vector can be
  calculated as explained in the next section.  The external
  kinetic chemistry code only needs to be run once, resulting in an
  exchange file that contains these effective rate coefficients as
  function of height, from which the DiffuDrift models are
  initialised.}

\subsection{Time derivatives for chemistry and particle growth}
\label{sec:chemeq}

The time derivatives of the solution vector $\vec{\tilde{y}}$ during a
chemistry, growth \& sublimation time step are given by
\begin{eqnarray}
  \frac{\partial n_i}{\partial t} &\!\!=\!\!&
          \Gamma_i^{\rm gas} - \Lambda_i^{\rm gas} 
      ~+~ \Gamma_i^\# - \Lambda_i^\#
      ~-~ N_\ell^s\!\sum\limits_s \delta_{s,i} J_\star^s
  \label{eq:dndt} \\[-0.5mm]
  \rho\frac{\partial \ek}{\partial t} &\!\!=\!\!&
      - \sum\limits_{i\,\in\cal{S}} \nu_{i,k}\,
      \big(\Gamma_i^{\rm gas} - \Lambda_i^{\rm gas}\big) 
  \label{eq:dekdt} \\ 
  \rho\frac{\partial L_3^s}{\partial t} &\!\!=\!\!&
          V_0^s\,\big(\Gamma_s^\# - \Lambda_s^\#\big)
      ~+~ \Vl\,J_\star^s
  \label{eq:dL3s} \\[1.5mm]
  \rho\frac{\partial L_j}{\partial t} &\!\!=\!\!&
          \Vl^{j/3} J_\star
      ~+~ \frac{j}{3} \sum\limits_{p=1}^2 n_p V_p^{j/3-1}\,
         \frac{\partial V_p}{\partial t}
  \label{eq:dLj}
\end{eqnarray}
where $\nu_{i,k}$ is the stoichiometric factor of element $k$ in
molecule $i$ and $\delta_{s,i}\!=\!1$ when molecule $i$ is the
volatile of ice $s$, otherwise $\delta_{s,i}\!=\!0$. {We do not
  consider diffision here, because DifuDrift computes the diffusion of
  molecules and particles in a separate routine using the operator
  splitting method. The negative sign in Eq.\,(\ref{eq:dekdt})
  accounts for the consumption of elements in the background gas when
  forming a selected molecule $i$.} Using the two-particle representation
(Eq.\,\ref{eq:double-delta}), the total net gain of condensed volume
per time and $\rm cm^3$ is
\begin{equation}
  \sum_{p=1}^2 n_p \frac{\partial V_{\!p}}{\partial t}
  ~=~ \sum_s V_0^s\,\big(\Gamma_s^\# - \Lambda_s^\#\big)
  \label{eq:ice_rates}
\end{equation}
such that $\partial L_3/\partial t = \sum_s \partial L_3^s/\partial t$
is valid according to Eqs.\,(\ref{eq:dL3s}) and (\ref{eq:dLj}), with
$J_\star=\sum_s J_\star^s$ and $N_\ell^s\!=\!V_\ell/V_0^s$.
Equation~(\ref{eq:ice_rates}) is used to compute the ice adsorption
and desorption rates from $\partial V_p/\partial t$ for
Eq.\,(\ref{eq:dndt}) as
\begin{equation}
  \Gamma_i^\#-\Lambda_i^\# =
  - \sum\limits_s \delta_{s,i}\,\big(\Gamma_s^\#-\Lambda_s^\#\big) \ ,
\end{equation}
where $\Lambda_s^\#-\Gamma_s^\#$ is the total net growth rate [solid
  units/s/cm$^3$] of material $s$.  Equations (\ref{eq:dndt}) to
(\ref{eq:dL3s}) describe how the elements are redistributed during one
time step among the {selected} molecules $(\to n_i)$, the background gas
$(\to\ek)$ and the condensed materials $(\to L_3^s)$ as a consequence
of local gas phase and surface chemical processes. It is assumed here
that a direct conversion of background gas into ice is not possible,
the ice is assumed to form only via the included molecules. Therefore,
the $\epsilon_k$ only change when the concentrations of the included
molecules change by gas-phase chemical reactions. As the reader may
verify, by considering $({\rm Eq.}\,\ref{eq:dekdt}) +
\sum_i\nu_{i,k}\,({\rm Eq.}\,\ref{eq:dndt}) +
\sum_s\nu_{s,k}/V_0^s\,({\rm Eq.}\,\ref{eq:dL3s})$, all right hand
side terms cancel, and we find the element conservation equation
\begin{equation}
  \rho\,\ek ~+~ \sum\limits_{i\,\in\cal{S}} \nu_{i,k}\,n_i ~+~
  \sum\limits_{s=1}^S \frac{\nu_{s,k}}{V_0^s}\,\rhoL_3^s ~=~ \rm const
  \label{eq:element_conserve} \ ,
\end{equation}
where $\nu_{s,k}$ is the stoichiometric factor of element $k$ in
material $s$.  In the old implementation \citep{Woitke2025},
Eq.\,(\ref{eq:element_conserve}) was the same except for the second
term, which was missing. The constant in
Eq.\,(\ref{eq:element_conserve}) can depend on height $z$ and time
$t$, but will only change by the settling of the condensates. During
one chemistry time step $\Delta t$, const remains constant,
demonstrating that elements can only change place between included
molecules, background gas, and the particles, but they cannot be
created by any chemical processes including phase transitions. Since
all components of the solution vector are diffused with the same
diffusion constant, diffusion does not change the total gas $+$
condensate element abundances either. 
Equation~(\ref{eq:element_conserve}) also makes clear that the
definition of the ``gas element abundances'' $\ek$ has to be revised
here.  Now, $\ek$ is the {\em number of nuclei of element $k$ per gas
  mass in the background gas}, which consists of all species not
contained in the set $\cal{S}$ of explicitly treated molecules.

An interesting limiting case is reached when we include all molecules
${\cal S}\!\equiv\!{\cal M}$. In that case, we have no background
rates ($\Gamma^{(0)}_i\!=\!\Lambda^{(0)}_i\!=\!0$), still some linear
rates ($\Gamma^{(1)}_{j\to\,i}, \Lambda^{(1)}_{i\to\rm any}$) because
of the UV-photo and X-ray processes, but mostly the quadratic rates
$\Gamma^{(2)}_{j,k\to\,i}$ and $\Lambda^{(2)}_{i,j\to\rm any}$.  And
there is no background gas: $\ek\!=\!0$. Since all molecules are
included, Eq.\,(\ref{eq:dekdt}) reads $\partial\,0/\partial t = 0$,
because the net effect of all gas-phase chemical reactions on the
total gas element abundances vanishes due to element conservation.

\subsection{Ice photo-desorption}
\label{sec:photodesorption}

An additional important process in discs is the ice UV-photo
desorption.  We take this process into account by passing the UV
photon fluxes $\chi_{\rm UV}\,\rm [photons/cm^2/s]$ in the Draine band
$91.2-205$\,nm from the external chemistry code to DiffuDrift along
with all the other chemical rate constants described in
Sect.~\ref{sec:kin_chem}. We then compute the photo-desorption rates
in DiffuDrift with the same approach and assumptions as in the
external chemistry code.  In case of ProDiMo, this is
\begin{equation}
  \frac{\partial \Vs}{\partial t}\bigg|_{\rm pd} =
  - \pi a^2\,\chi_{\rm UV}\,Y_{\rm pd}^s(T)\,b_{\rm mix}^s V_0^s \ ,
  \label{eq:UVdesorb}
\end{equation}
where the photo-desorption yield for water ice is taken from
\cite{Oberg2009}:
\begin{equation}
  Y_{\rm pd}^{\ce{H2O}\#}(T) = 1.3\times10^{-3} + \frac{3.2\times10^{-5}}{T}
  \ ,
\end{equation}
{where $T$ is the temperature in Kelvin.} In order to calculate
these rates in DiffuDrift, the UV photon flux $\chi_{\rm UV}$ is also
passed from the chemical reference model.

\subsection{Ice hydrogenation}
\label{sec:hydrogenation}

Another special process is the growth of an ice by surface
hydrogenation reactions, for example
\begin{align}
  \begin{array}{c}
    \displaystyle\rm O \\
    \scriptstyle R_1\displaystyle\!\downarrow\uparrow\scriptstyle\ R_2 \\[1mm]
    \displaystyle\rm O\#\\[1mm]
    \ \\[1mm]
    \ 
  \end{array}  
  \!\!\!\!\stackrel{\it R_3}{\longrightarrow}\!\!
  \begin{array}{c}
    \displaystyle\rm OH \\
    \scriptstyle R_4\displaystyle\!\downarrow\uparrow\scriptstyle\ R_5 \\[1mm]
    \displaystyle\rm OH\#\\[1mm]
    \ \\[1mm]
    \ 
  \end{array}  
  \!\!\stackrel{\it R_6}{\longrightarrow}\rm~H_2O\#
  \label{eq:H2Ohg}\\[-15mm]\nonumber   
\end{align}
which we cannot directly model in DiffuDrift because the in-between
ice species, such as O\#, may not be thermally stable. The net rate of
this reaction chain is 
\begin{equation}
  \frac{\partial \Vs}{\partial t}\bigg|_{\rm hyd} \!=
  4\pi a^2\left(n_{\rm O}R_1
  \frac{n_{\rm H}R_3}{R_2+n_{\rm H}R_3} + n_{\rm OH}R_4\right)
  \frac{n_{\rm H}R_6}{R_5+n_{\rm H}R_6}\,V_0^s \ ,
  \label{eq:hydrogenation}
\end{equation}
where $n_{\rm O}$ and $n_{\rm H}$ are the particle densities of atomic
oxygen and hydrogen in the gas phase, $R_1$ and $R_4$ are the adsorption
rates of O and OH per surface, $R_2$ and $R_5$ are the sums of thermal
and photo desorption rates of O\#, and OH\#, respectively, $R_3$ is
the hydrogenation rate of O\# per $n_{\rm H}$, and $R_6$ is the
hydrogenation rate of OH\# per $n_{\rm H}$. These rate constants
$\{R_1,\,...\,,R_6\}$ are passed from the chemical reference model and
applied in the DiffuDrift as an additional growth reaction.

The two fractions in Eq.\,(\ref{eq:hydrogenation}) are branching
ratios -- they express the probability to advance to the next
hydrogenation step. For high temperatures, these branching ratios
become tiny because of the short residence times of O\# and OH\# on
the surface.  In case of \ce{CH4}\#, there are two additional
hydrogenation steps, two additional branching ratios, and 12 rate
constants need to be passed to DiffuDrift. If any of the species
involved in reaction\,(\ref{eq:H2Ohg}) are not among the selected species,
their densities are taken from the reference chemical model after
scaling as explained in Sect.\,\ref{sec:scaling}.

The volume shrink rates due to UV-photo desorption
$\partial\Vs/\partial t|_{\rm pd}$ and the growth rates due to
hydrogenation $\partial\Vs/\partial t|_{\rm hyd}$ are to be
added to $\partial \Vs/\partial t$ as given for adsorption and
thermal desorption by Eq.\,(\ref{eq:growth}). The respective
production and consumption rates of the affected molecules are
added to Eqs.\,(\ref{eq:dndt}) and (\ref{eq:dekdt}).

\subsection{Scaling the background production rates}
\label{sec:scaling}

The $\Gamma^{(0)}$ and $\Gamma^{(1)}$ rates describe the production of
the selected species from the background gas, and from a combination
of the background gas and the selected species, respectively.  Since
the background gas can be consumed (or transported away) in the
DiffuDrift simulations, $\epsilon_k$ will generally start to deviate
from the value $\epsilon_k^{\rm ref}$ found in the chemical reference
model that calculated the coefficients $\Gamma$, which is a wanted
design feature.  Therefore, the background rates need to be scaled
with $\epsilon_k/\epsilon_k^{\rm ref}$ to make sure that the
background rates $\Gamma\!\to\!0$ when the background gas gets
consumed, otherwise $\epsilon_k\!<\!0$ would be possible.  We assume
that these rates scale with the element abundances contained in the
background gas $\epsilon_k$ in the following way, {inspired by the law
  of mass action}
\begin{eqnarray}
  f_k &\!\!=\!\!& \epsilon_k/\epsilon_k^{\rm ref} \\*[-2mm]
  f_i &\!\!=\!\!& \left(\prod\limits_k \Big(f_k\Big)^{\nu_{i,k}}
                  \right)^{\frac{2}{\sum \nu_{i,k}}} \\*[-1mm]
  \Gamma^{(0)}_i
      &\!\!=\!\!& f_i\;\Gamma^{(0),\rm ref}_i \\
  \Gamma^{(1)}_{j\to\,i}
      &\!\!=\!\!& \sqrt{f_i}\;\,\Gamma^{(1),\rm ref}_{j\to\,i} 
\end{eqnarray}  
where $\epsilon_k^{\rm ref}$ are the element abundances in the
background gas as found in the chemical reference model, and
$\Gamma^{(0),\rm ref}_i$ and $\Gamma^{(1),\rm ref}_i$ are the constant
and linear production rates as computed by the external chemical
code and passed to the DiffuDrift model.

\subsection{The external kinetic chemistry model}
\label{sec:chemistry}

\begin{figure*}
  \begin{tabular}{ccc}
    \hspace*{-4mm}
    \includegraphics[width=66.0mm,trim=40 15 65 330,clip]
                    {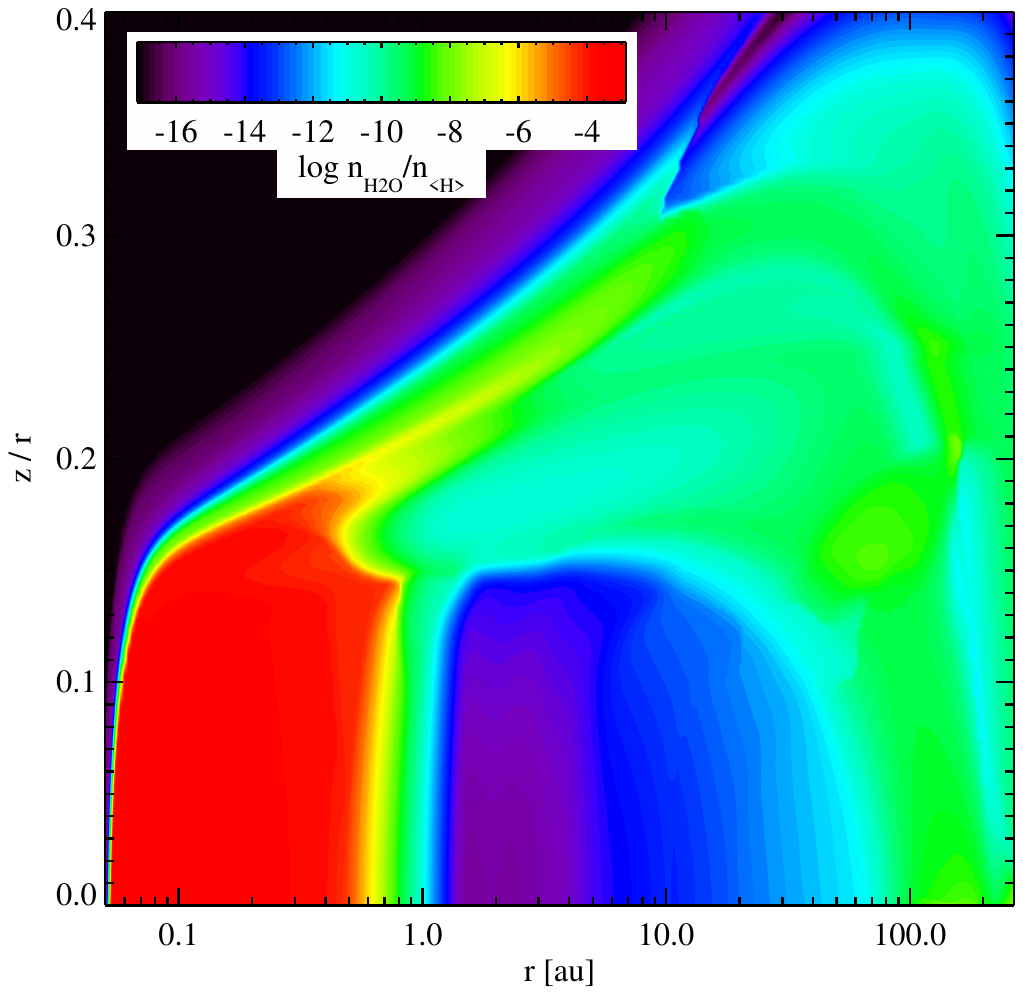} &
    \hspace*{-6mm}
    \includegraphics[width=60.7mm,trim=80 15 65 330,clip] 
                    {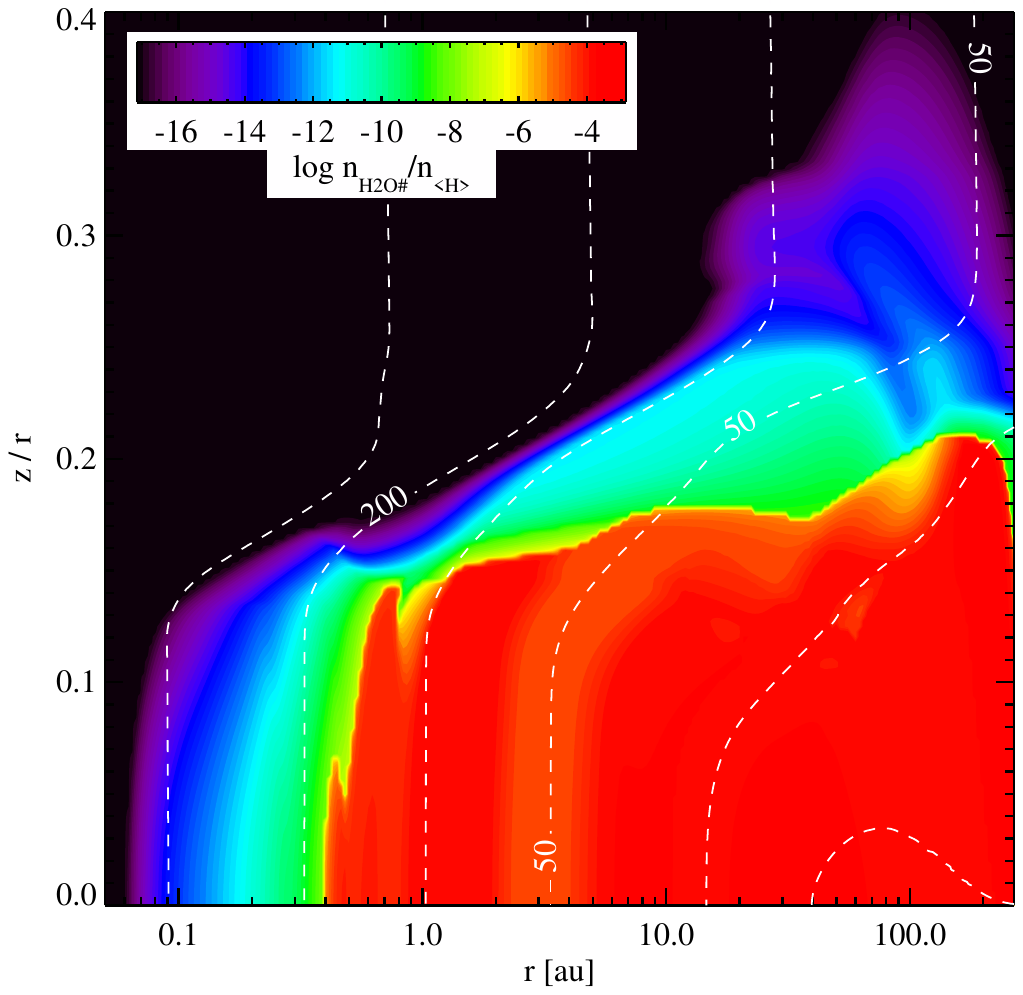} &
    \hspace*{-6mm}
    \includegraphics[width=59.5mm,trim=42 -15 0 0,clip] 
                    {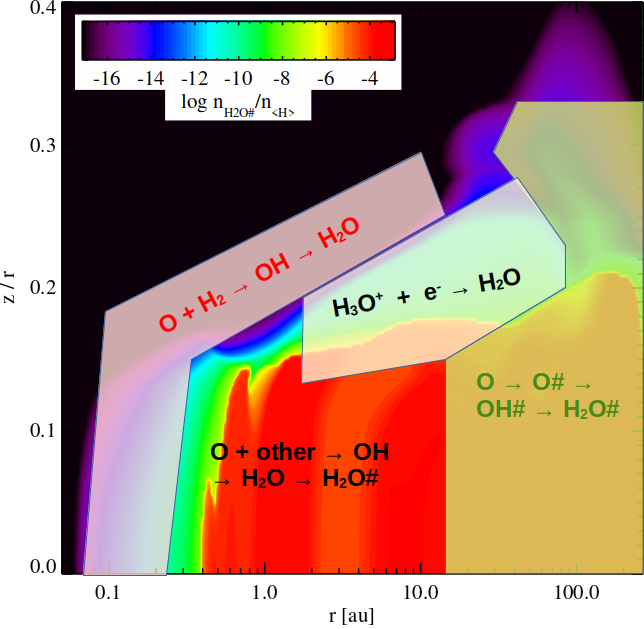} \\*[-4mm]
  \end{tabular}
  \caption{Water and water ice concentrations, and the main chemical
    pathways to form water ice in the ProDiMo chemical reference model
    for an example T\,Tauri disc.}
  \label{fig:disc_water}
\end{figure*}

The chemical reference model used in this paper is ProDiMo
\citep{Woitke2009, Woitke2016} based on the DIANA chemical standard
network of \cite{Kamp2017}, but with some modifications. In this
paper, the main purpose of the network is to correctly predict the
timescale of the ice formation from the available chemicals, in the
midplane and elsewhere, for a few important ices with a simple network
suitable for the subsequent dynamical model.

From the original DIANA chemical network with 13 elements, 171 gas and
64 ice species, see Table~4 in \cite{Kamp2017}, we have therefore
eliminated all Si, Fe, Ne, Ar and PAH species, and all species with
more than 2 carbon atoms. Concerning the ices, we concentrate on 4 ice
species in this paper: \ce{H2O}\#, \ce{CO}\#, \ce{CO2}\# and
\ce{CH4}\#.  However, in order to have the hydrogenation pathways in
the network, we also included \ce{O}\#, \ce{OH}\#, \ce{C}\#, \ce{CH}\#
\ce{CH2}\# and \ce{CH3}\#.  The latter ice species are often not
stable in a thermodynamical sense, they may have supersaturation
ratios $S\!<\!1$ and can hence not be included in DiffuDrift. These
species only stay for a short while on the surface of the dust
particles, before they thermally desorb, but during that short
residence time, they can drive important reactions on the surface that
are not possible in the gas phase. In order to reduce that complexity,
these six auxiliary ice species are not included in the DiffuDrift
model.  Instead, the hydrogenation reactions are treated as additional
growth rates by means of a stationary mini-network, the flux through
which is fully determined by the abundance of gas species and branching
ratios on the surface, see Sect.~\ref{sec:hydrogenation}.

Furthermore, we have eliminated \ce{O2}, \ce{O2+} and \ce{O2H+} from
the network used in this paper. It is a well-known issue in
astrochemical modelling that the early formation of \ce{O2} via the
low-barrier $\rm O + OH \to O_2 + H$ reaction is fast and unavoidable,
but then the destruction of \ce{O2} is extremely slow because of the
extraordinary stability of the \ce{O2} double-bond. In shielded
conditions at low temperatures, {\ce{O2} can freeze out and
  undergo a sequence of surface hydrogenation reactions through
  \ce{O2H}\# and \ce{H2O2}\# to eventually convert \ce{O2} into water
  ice \citep[e.g.][]{Miyauchi2008,Cuppen2010}.} We have verified this
pathway with ProDiMo using a more extended and sophisticated network
that includes surface chemistry.  If these surface reactions are
lacking, however, like in our simple network used here, \ce{O2} forms
a dead-end, and a large fraction of oxygen stays locked in \ce{O2} for
very long times. Since this artefact would dominate the gas phase
oxygen element abundance after ice formation that we want to quantify
with our model, we have decided to eliminate \ce{O2}.

The resulting chemical network has 8 elements, 119 gas phase and 10
ice species. A complete list of all included chemical species is
provided in Table~\ref{tab:networks}.  There are 1840 gas-phase
chemical reactions in this network, imported from from UMIST\,2022
\citep{Millar2024}, 151 UV-photo reactions from the Leiden database
\citep{Heays2017}, and 330 additional reactions from a ProDiMo
specific collection that include the ice adsorption and desorption
reactions \citep{Woitke2009}, excited molecular hydrogen chemistry
\citep{Kamp2017}, a few selected collider and 3-body termolecular
reactions \citep{Kanwar2025}, and 78 X-ray reactions
\citep{Aresu2011}.

\subsection{Vapour pressures}

\begin{table}
  \caption{Thermodynamical ice data in ProDiMo and DiffuDrift
    \label{tab:icedata}}
  \vspace*{-3mm}\hspace*{-1.5mm}
  \resizebox{91mm}{!}{
  \begin{tabular}{c|c|c}
  \hline
  &&\\[-2ex]  
  & ProDiMo\!\!\!\!\!\! & DiffuDrift \\
  & $E_{\rm des}\rm\,[K]$ & $\ln(p_{\rm vap})\rm\,[dyn/cm^2]$ \\
  &&\\[-2ex]
  \hline
  &&\\[-2ex]  
  \!\!\ce{H2O}\# & \!\!4800 \small\citep{Brown2007}\!\!
    & $21.40 - 4800/T + 0.5\ln T$\!\! \\[0.5mm] 
  \!\!\ce{CO2}\# & \!\!2990 \small\citep{Garrod2006}\!\!
    & $21.17 - 2990/T + 0.5\ln T$\!\! \\[0.5mm]
  \ce{CO}\#  & 1150 \small\citep{Collings2004}
    & $20.69 - 1150/T + 0.5\ln T$\!\! \\[0.5mm]
  \ce{CH4}\# & 1090 \small\citep{Herrero2010} 
    & $20.66 - 1090/T + 0.5\ln T$\!\! \\
  \hline
  \end{tabular}}
\end{table}

Table~\ref{tab:icedata} lists the parameters for the vapour pressures
of the four ices considered in this paper.  In order to treat the
thermal ice stability in DiffuDrift as similar as possible to
ProDiMo, we have generated new fits of the vapour pressures using the
same desorption energies and other assumptions as in ProDiMo, for
details see App.\,\ref{app:pvap}.  It is
noteworthy that the water desorption energy by \cite{Brown2007}
considers the stability of water ice on a silicate surface, whereas
for \ce{CO2} and \ce{CO}, a mixture with water ice is considered by
\cite{Garrod2006} and \cite{Collings2004}, respectively, but the
desorption energy of \ce{CH4} by \cite{Herrero2010} is for the pure
substrate.

\subsection{Chemical pathways to form water ice}

Figure \ref{fig:disc_water} summarises the predictions of the ProDiMo
chemical reference model for an example T\,Tauri disc concerning the
formation of water and water ice. We see that the gaseous water is
mostly predicted to be present only inside of the snowline which is
located at about 0.5\,au in this passive disc model.  The ice is
restricted to a height $z/r\la 0.18$, forming a horizontally and
vertically restricted ``ice-block''. This is partly because the small
grains become too warm above the shadow casted by the inner disc, and
partly because of ice photo-desorption.  The right plot shows the main
chemical pathways to form water molecules and water ice as
described in the following paragraphs.

\paragraph{(1) The neutral-neutral path via OH:}

The most basic chemical pathway to form water from the abundant oxygen
atoms in discs is the neutral-neutral reaction chain via OH
\begin{eqnarray}
  & \rm O ~\stackrel{H_2}{\longrightarrow}~ OH
          ~\stackrel{H_2}{\longrightarrow}~ H_2O
          ~\longrightarrow~ H_2O\# \ .
  \label{reac:OH}
\end{eqnarray}
The first reaction step has an activation energy of about 1300\,K,
which means that this path in practise becomes very inefficient below
about 200\,K \citep{Woitke2009b}. In contrast, the second reaction is
barrier-free. Therefore, since the freeze-out (the last reaction step
in Eq.\,\ref{reac:OH}) requires temperatures $\la\!150\,$K, this
pathway cannot explain the water ice formation locally; it would
require to form the water molecules in a warm place, and transport
them to a cooler place where they can freeze out.

However, OH can also be formed by other neutral-neutral reactions with
no or very low activation energies in the gas phase, and then continue
to form water ice according to the second and third reactions in
Eq.\,(\ref{reac:OH}). The following reactions seem to provide the most
efficient pathway to form water ice in the dense disc midplane out to about
10\,au ($T\!\ga\!30\,$K):
\begin{eqnarray}
  \quad
  \rm O ~+~ H ~+~ M  &{\longrightarrow}& \rm OH ~+~ M \\
  \rm O ~+~ H        &{\longrightarrow}& \rm OH ~+~ h\nu \ .
  \label{reac:OHform}
\end{eqnarray}
The termolecular (3-body) reaction and the radiative association
reaction are both barrier-free. Once OH is available, another
efficient way to form water molecules at high densities is radiative
association with atomic hydrogen $\rm OH + H \to H_2O ~+~ h\nu$.

\paragraph{(2) The ion-neutral path via \ce{H3O+}:}

This reaction pathway starts with an ionisation of \ce{H2}
by cosmic rays or X-rays, followed by fast reactions
with atomic oxygen and molecular hydrogen to form \ce{H3O+}
\begin{eqnarray}
  & \rm H_2 ~\stackrel{CR}{\longrightarrow}~ H_2^+
          ~\stackrel{H_2}{\longrightarrow}~ H_3^+ \ ,\\
  & \hspace*{-3mm}\rm O \stackrel{H_3^+}{\longrightarrow}~ OH^+
          \stackrel{H_2}{\longrightarrow}~ H_2O^+
          \stackrel{H_2}{\longrightarrow}~ H_3O^+
          \stackrel{e^-}{\longrightarrow}~ H_2O
          \longrightarrow~ H_2O\#   \ ,
\end{eqnarray}
which eventually recombines with a free electron to form \ce{H2O}
which can freeze out. Since this reaction path is barrier-free, it can
operate at any temperature, also at low temperatures.  However, since
it is cosmic-ray driven, it operates on long timescales. Furthermore,
it might be rather inefficient in the midplane where the
electron concentration can be very low.

\paragraph{(3) The cold surface hydrogenation path:}
Oxygen atoms can stick to grains directly where they can be stepwise
hydrogenated to form water ice on the surface without ever forming water
molecules in the gas phase
\begin{eqnarray}
  & \rm O ~\longrightarrow~ O\#
          ~\stackrel{H}{\longrightarrow}~ OH\#
          ~\stackrel{H}{\longrightarrow}~ H_2O\# \ .
\end{eqnarray}
This reaction path requires cold conditions to allow the oxygen atoms
to stay on the surface for long enough to catch a H-atom before they
desorb thermally. The surface residence timescale is a matter of
temperature and desorption energy of O\# (about 800\,K), which
effectively limits this formation path to regions in the disc where
the temperature is lower than about 30\,K. Since the desorption energy
of OH\# is about 3.5 times higher, the third reaction step is likely
to follow the second one eventually. The formulation of the
hydrogenation rates follows Eq.\,(7) in \cite{Visser2011}.

\subsection{Ice formation timescales}

Figure~\ref{fig:ChemModel} shows the initial chemical process of ice
formation during the first 100\,000\,yrs from initially mostly atomic
concentrations taken from a ProDiMo molecular cloud model (see
Table~\ref{tab:initconc}) at 3 selected sites in the disc. We see that
the ice formation proceeds with a pace that is similar to the
\ce{H2}-formation on grain surfaces, which mostly depends on density.
While at 1\,au, it takes only about $10^{-3}$\,yrs (one day) to form
the ice, the same process takes about 1000\,yrs at 100\,au.  {These
  timescales are consistent with water already being missing at the
  Class~I stage \citep{Harsono2020}.}

\subsection{Summary of model approach}

In summary, the new DiffuDrift code can simulate chemical
reaction-diffusion systems with condensation and gravitational
settling in 1D vertical geometry. The code keeps track of the changing
material composition of the particles and the kinetically changing
chemical composition of the gas phase, and takes into account the size
change of the particles when deposition or sublimation occurs. Gas
and particles are equally diffused, but only the particles are
settled.  The code is also equipped with modules for nucleation and
coagulation, but these are not used in this paper. The code uses
moments of the size distribution function to deal with the changing sizes
of the particles.

One key assumption is that all materials grow simultaneously on the
surfaces of all particles, which are treated as uniformly mixed,
i.e.\ there are no layers of different materials in the model. This
means that the water ice considered in this paper can be found as
small inclusions with equal probability at any depth in the particles,
and only a fraction thereof is present at the chemically active
surface concerning thermal and UV-photo desorption. This is expressed
by the factor $b_{\rm mix}^s$ in Eqs.\,(\ref{eq:desorb}) and
(\ref{eq:UVdesorb}). {Some support for this hypothesis comes from
  \cite{Sturm2023} who observed the CO ice being trapped in the
  \ce{H2O}/\ce{CO2} ice mixture.}

The kinetic gas phase chemistry is modelled by adopting the rates from
a reference chemical model, where the number of explicitly included
molecules can vary between one and the full number of considered
species in the reference chemical model. All non-included molecules in
the reference chemical model are lumped together into a background-gas
which is part of the modelling and is also diffused. All modules
(chemistry \& growth, diffusion, settling) are mass and element
conservative, see details about the remaining numerical deviations
after long simulations in Sects.\,\ref{sec:settled} and
\ref{sec:mainresults}. {All models in this paper have been run
  with DiffuDrift git version {\tt e420474} from October 12$^{\rm th}$
  2026, and ProDiMo git version {\tt 0b8d7f80} from August 2$^{\rm
    nd}$ 2026.}

\begin{figure*}
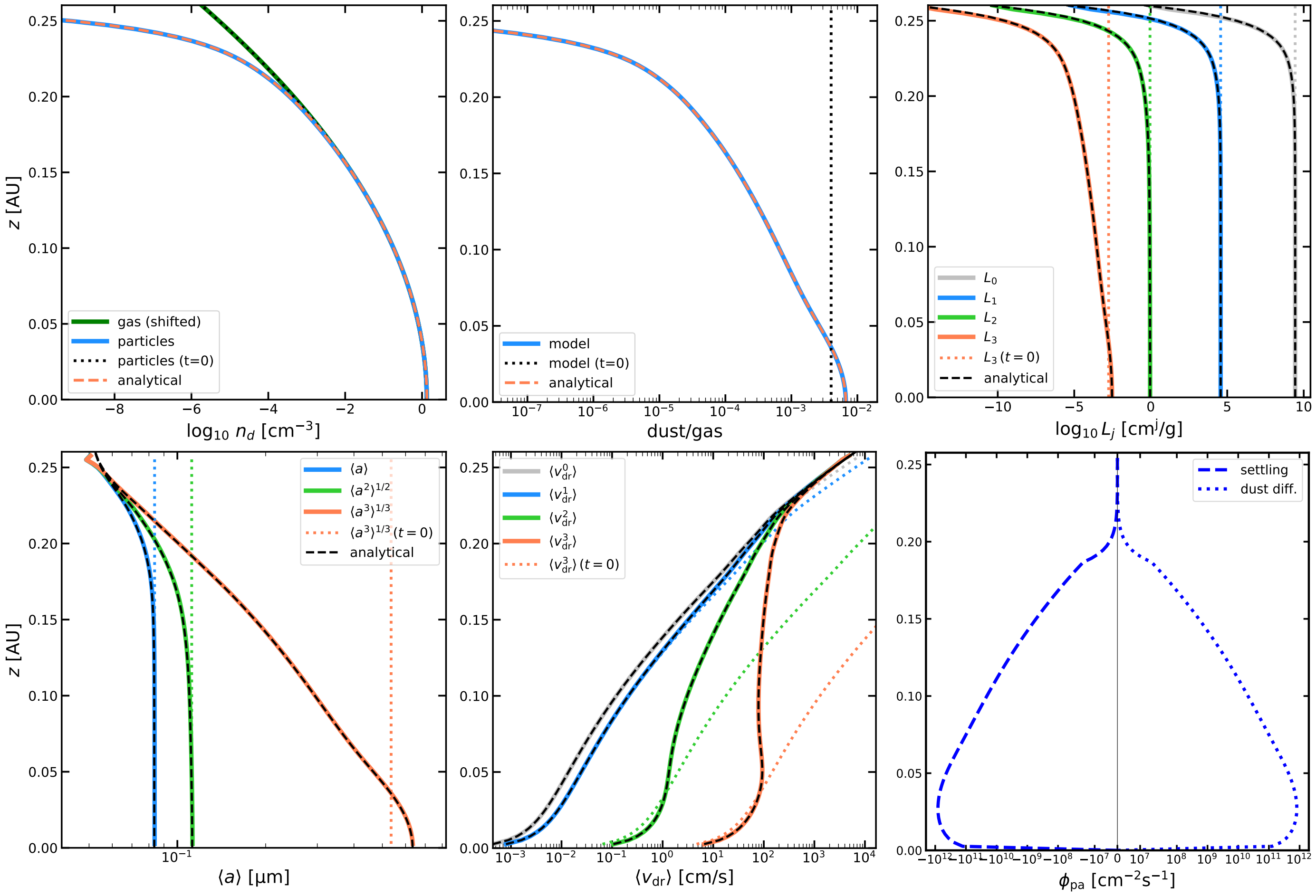

  \begin{tabular}{ccc}
    \hspace*{-2mm}
    \includegraphics[page=7 ,height=62mm,trim=15 15 15 15,clip]
                    {testcase.pdf} &
    \hspace*{-4mm}
    \includegraphics[page=8 ,height=62mm,trim=35 15 15 15,clip] 
                    {testcase.pdf} &
    \hspace*{-4mm}
    \includegraphics[page=10,height=62mm,trim=35 15 15 15,clip] 
                    {testcase.pdf} \\[-2mm]
    \hspace*{-2mm}
    \includegraphics[page=11,height=62mm,trim=15 15 15 15,clip]
                    {testcase.pdf} &
    \hspace*{-4mm}
    \includegraphics[page=12,height=62mm,trim=35 15 15 15,clip]
                    {testcase.pdf} &
    \hspace*{-3mm}
    \includegraphics[page=2,height=63mm,width=59mm,trim=30 14 5 1,clip]
                    {testfluxes.pdf} 
  \end{tabular}
  \caption{Relaxation of chemically inert 'passive' dust particles in
    a vertical column at $r\!=\!1\,$au, with an initial powerlaw size
    distribution function of $f(a)\propto\!a^{-3.5}$ and dust/gas mass
    ratio of 0.004, towards an equilibrium structure, where downward
    settling is balanced by upward diffusion.  The resulting particle
    properties are plotted after a simulation time of
    10\,000\,yrs. The initial state is plotted with dotted lines,
    the analytic solution is shown with dashed lines. The figure in
    the lower right shows the vertical fluxes
    $\phi\rm\,[cm^{-2}s^{-1}]$ (negative $=$ downward, positive $=$
    upward) of passive, condensed \ce{MgSiO3}-units as a proxy of the
    refractory dust.}
  \label{fig:test}
\end{figure*}

\section{Results}

\subsection{Gravitational settling of passive particles}
\label{sec:settled}

To test our code, we first consider the basic case of passive dust
particles, which are chemically inert and also do not coagulate nor
fragment, they are only affected by diffusion and gravitational
settling.  This is the classical case of dust settling as discussed,
for example, by \cite{Dubrulle1995}, \cite{Riols2018}, and
\cite{Woitke2024}.  We consider a vertical column of gas in a
protoplanetary disc at a distance of $r\!=\!1\,$au with a Gaussian
density distribution
\begin{equation}
  \rho(z) \propto \exp\bigg(-\frac{z^2}{2H^2}\bigg)
\end{equation}
characterised by a gas scale height $H\!=\!0.05\,$au.  The
temperature/density structure is imported from a ProDiMo model for a
standard passive T\,Tauri star, see further details in
Table~\ref{tab:columns}.  A viscosity parameter $\alpha\!=\!10^{-3}$
is used to derive the diffusion coefficient. The model domain extends
from $z\!=\!0$ to a maximum height of about $z_{\rm
  max}\!=\!0.27\,{\rm au}\!=\!5.5\,H$. The characteristic diffusion
and settling timescales are
\begin{equation}
  \hspace*{3mm}
  \tau_{\rm diff} = \bigg(\int_0^{z_{\rm max}}\!\!\!\frac{dz}{\sqrt{D(z)}}\bigg)^2
  ~\quad,\quad~
  \tau_{\rm settle}(a) =\!\!\int_{H}^{z_{\rm max}}\!\!\frac{dz}{\vdreq(a,z)} \ .
  \label{eq:timescales}
\end{equation}
When we assume the temperature to be constant as a function of height,
which is generally not recommended, we can solve these integrals
analytically. Introducing the surface density $\Sigma=\int
\rho(z)\,dz\,$ the results are
\begin{eqnarray}
  \tau_{\rm diff} &\!\!=\!\!& \frac{z_{\rm max}^2}{\alpha\,c_{\rm S}\,H}
  \label{eq:tau_diff} \ ,\\
  \tau_{\rm settle}(a) &\!\!=\!\!& 0.112\,\frac{\Sigma\,\vth\,r^3}{G
      M_\star\,H\,\rhom\,a} \ ,
  \label{eq:tau_sett}
\end{eqnarray}
where we have assumed $z_{\rm max}\!\gg\!H$.
The characteristic diffusion timescale results to be $\tau_{\rm
  diff}\!\approx\!10^4$\,yrs in our model at 1\,au. The size-dependent
characteristic settling timescale $ \tau_{\rm settle}(a)$, i.e.\ the
time required for a particle to settle from the top down to one scale
height, ranges from 60\,yrs for a 1\,mm grain to $6\times10^5\,$yrs
for a $0.1\,\mu$m grain. More information about the properties of the
considered disc column at 1\,au can be found in Table~\ref{tab:columns}.

We consider an initial dust size distribution
$f(a,t\!=\!0)\!\propto\!a^{-3.5}$ from 0.05\,$\mu$m to 3\,mm at every
height with a local dust/gas mass ratio of 0.004 as expected for
refractory dust condensation from a solar composition gas
\citep{Woitke2018}.  We apply zero-flux boundary conditions to all
variables $\{\,\rhoL_0,...\,,\rhoL_3\}$ for both the upper and the
lower boundaries at $z\!=\!0$ and $z\!=\!z_{\rm max}$, respectively.

\begin{table}
  \caption{Column properties imported from the ProDiMo disc model
    \label{tab:columns}}
  \vspace*{-3mm}\hspace*{-1.5mm}
  \resizebox{91mm}{!}{
  \begin{tabular}{l|ccc}
  \hline
  &&&\\[-2.2ex]  
  &                                1\,au & 10\,au & 100\,au \\
  &&&\\[-2.5ex]  
  \hline
  &&&\\[-2.2ex]
  gas col.\,density $\rm[g/cm^2]^{(1)}$  & 906  & 59   & 0.35 \\
  H nuclei density $(z\!=\!0)\,\rm[cm^{-3}]^{(2)}$\!\!\! &
    $2\times10^{14}$ & $9\times10^{11}$ &  $4\times10^{8}$ \\
  dust midplane temp. $\rm[K]$     & 104  & 25   & 5.1 \\
  dust top temp. $\rm[K]$          & 424  & 142  & 62.5 \\
  scale height $H\rm\,[au]^{(3)}$   & 0.05 & 0.71 & 10 \\
  vertical extension $z_{\rm max}/H$   & 5.2  & 4.9  & 3.7 \\ 
  diffusion constant $\rm[cm^2/s]^{(4)}$ & $4.6\times10^{13}$ &
                 $3.5\times10^{14}$ & $2.0\times10^{15}$ \\
  diffusion timescale $\rm[yr]^{(5)}$  & $1.1\times10^{4}$ &
                 $3.0\times10^{5}$  & $4.8\times10^{6}$ \\
  settling.\,time $(a\!=\!1\rm\mu m)~[yr]^{(5)}$ & $6.6\times10^{4}$ &
                 $1.9\times10^{5}$ & $5.3\times10^{4}$ \\
  \hline
  \end{tabular}}\\[1mm]
  \footnotesize
  $^{(1)}$: The column density structure is set as $\Sigma(r)\!\propto\!r^{-1}
  \exp{\big(-r/{\rm 30\,au}\big)}$ with a total disc mass of
  $0.02\,M_\odot$ between 0.05\,au and 250\,au.\\
  $^{(2)}$: A Gaussian vertical density profile
  $\propto\exp\big(-z^2/(2\,H^2)\big)$ is assumed.\\
  $^{(3)}$: The scale height is set as
  $H(r) = {\rm 10\,au}\;\big(r/{\rm 100\,au}\big)^{1.15}$.\\
  $^{(4)}$: Derived from viscosity parameter $\alpha\!=\!10^{-3}$, see
  Eq.\,(\ref{eq:Diff}).\\
  $^{(5)}$: characteristic timescales according to
  Eq.\,(\ref{eq:timescales}).\\*[-7mm]
\end{table}
%
%

As we have shown in \cite{Woitke2024}, the stationary equilibrium
between diffusion and gravitational settling has an analytic solution
for every particle size given by
\begin{equation}
  \frac{d}{dz}\bigg(\ln\frac{f(a,z)}{\rho(z)}\bigg)
  ~=~ \frac{\vdreq(a,z)}{D(z)} \ .
  \label{eq:analytic}
\end{equation}
Equation (\ref{eq:analytic}) is used in the following way to obtain a
quasi-analytical size distribution function $f(a,z)$ after settling:
\begin{enumerate}
\setlength{\itemsep}{1.4pt} 
\item Setup the unsettled size distribution function as
  $f_0(a,z)\!=\!\rho(z)\,a^{-3.5}$, i.e.\ with an arbitrary
  proportionality constant.
\item For every $a$, start with $f(a,0)=f_0(a,0)$ and numerically
  integrate Eq.\,(\ref{eq:analytic}) upwards to obtain $f(a,z)$.
\item For every $a$, normalise $f(a,z)$ such that the total column
  density is the same as before settling $\int\!f(a,z)\,dz =
  \int\!f_0(a,z)\,dz$.
\item Compute the total gas and dust mass column densities,\linebreak
  $\Sigma\!=\!\int\!\rho(z)\,dz\,$ and $\Sigma_{\rm
  d}\!=\!\iint\!f(a,z)\,m(a)\,da\,dz$, where $m(a)$ is the mass of one
  dust particle of size $a$, and normalise the entire $f(a,z)$ to
  obtain the requested dust/gas ratio $\Sigma_{\rm d}/\Sigma$.
\end{enumerate}
This way, we can compute analytical solutions for several derived
mean particle properties, such as
\begin{eqnarray}
  \mbox{particle density:} &
     n_{\rm d}(z) = \int\!f(a,z)\,da \ ,\\
  \mbox{size moments:} &
     \rhoL_j(z) = \int\,f(a,z)\,V^{j/3}\,da \ ,\\
  \mbox{mean particle sizes:} &
     \langle a^j\rangle(z) = \displaystyle
     \frac{\int\!f(a,z)\,a^j\,da}
          {\int\!f(a,z)\,da}\ ,\\
  \mbox{mean drift velocities:} & \!\!
     \langle v_{\rm dr}^j\rangle(z) = \displaystyle
     \frac{\int\!f(a,z)\,\vdreq(a,z)\,V^{j/3}\,da}
          {\int\!f(a,z)\,\,V^{j/3}\,da} \ ,\label{eq:meanvdr}
\end{eqnarray}
which can all be expressed in terms of the size moments $L_j$ and are hence
predicted by the model, whereas our model does not directly predict
the size distribution function. The mean drift velocities are of
particular interest, because they are used in the settling module
to transport the size moments downwards, and when $\langle v_{\rm
  dr}^j\rangle$ increases with $j$, which it normally does, it means
that the settling leads to an accumulation of larger particles at the
bottom. 

Figure~\ref{fig:test} shows the results of this benchmark test, after
letting the code run for 10\,000\,yrs.  The model uses 100 equidistant
$z$-points and executes about 67\,000 time steps, which takes
altogether about 55\,sec on a Dell Laptop. In the very beginning, the
code needs to make tiny time steps of order $\sim\!100\,$sec to settle
the big grains from the upper layers, where they do not belong. The
total mass of the passive particles still equals the initial total
mass with a relative error $<\!2\times10^{-8}$ at the end of the
simulation.

As the results show, the model relaxes towards the analytical solution
with high precision. Only at the very top of the column, where the
gradients become very steep and the gas becomes almost dust-free,
there are some minor deviations. The dust/gas mass ratio has about
doubled in the midplane, but decreases with height
by about 3 orders of magnitude up to $\sim\!4\,H$, where it
starts to drop very quickly as even the smallest grains in the model
(0.05\,$\mu$m) can no longer be sustained by turbulent up-whirling.
The bottom right figure shows the perfect balance in the final state
between upward mixing and downward settling fluxes, spanning 5 orders
of magnitude.

\subsection{Full simulations of icy particles}
\label{sec:mainresults}

The setup of our full model calculations for icy particles with
kinetic chemistry is similar to the test case for passive particles
described in Sect.\,\ref{sec:settled}: We import the
density/temperature structure of a selected vertical column from
ProDiMo and set the constant diffusion coefficient according to
Eq.\,(\ref{eq:Diff}) with $\alpha\!=\!10^{-3}$. Some characteristics
of the three selected columns at 1\,au, 10\,au and 100\,au are listed
in Table~\ref{tab:columns}. From the characteristic timescales for
diffusion and settling, it is clear that we have to simulate the
column at 1\,au for about $10^4$\,yrs, the column at 10\,au for about
$10^5$\,yrs, and the column at 100\,au for about $10^6$\,yrs to reach
a steady state.

As initial condition for the particles $L_j(z,t\!=\!0)$, we use a
powerlaw size distribution of passive dust particles from $0.05\,\mu$m
to $3\,$mm with $\rm dust/gas\!=\!0.004$, and set the initial ice
concentrations to zero $L_3^s(z,t\!=\!0)\!=\!0$ for all $s$ except for
the passive component.  We then select a chemical network to be
imported from ProDiMo from the available choices ``tiny'', ``medium''
and ``full'', see Table~\ref{tab:networks}.  In
Appendix~\ref{app:networks}, we discuss the pros and cons of these
networks, and decided to use the full network for the main results
presented in this paper, {which implies that there is no
  background gas $\epsilon_k$ in these models. The full set of
  equations solved in this case is listed in App.~\ref{app:fulleqs}.
  All required quantities are imported from ProDiMo, i.e.\, $\rho(z)$,
  $T(z)$, the various contribtions to the rate constants
  $\Gamma_i^{\rm gas}(z)$ and $\Lambda_i^{\rm gas}(z)$, the reference
  quantities such as $\epsilon_k^{\rm ref}(z)$, and the hydrogenation
  and photo-desorption parameters$(z)$.  These quantities are
  interpolated onto the DiffuDrift grid and smoothed during
  initialisation using a box-car filter over neighboring cells.}
Next, we use constant initial concentrations of all selected molecules
as $n_i(z,t\!=\!0) = n_i^0$ to set up the molecular concentrations and
element abundances.

These initial atomic and molecular concentrations $n_i^0$ are listed
in Table~\ref{tab:initconc}.  We have computed them from a
time-dependent ProDiMo model for a molecular cloud with parameters
$\nH\!=\!10^4\rm\,cm^{-3}$, $T\!=\!10\,$K and optical extinction
$A_V\!=\!10$, see details in App.\,\ref{app:initconc}. The initial
concentrations are also visible on the left side of
Fig.\,\ref{fig:ChemModel} at $t\!=\!10^{-8}$\,yr.


As boundary conditions we use zero-flux (closed box) conditions for
all components of the solution vector $\vec{y}$ (see
Sect.\,\ref{sec:kin_chem}). We check that the total element abundances
contained in gas and ice, integrated over the whole column, do not
change significantly over time in the models (relative errors
$<\!10^{-7}$), that the total mass of the passive material component
remains constant (relative error $<\!10^{-8}$), and that the net local
charge stays close to zero everywhere (relative errors $<\!10^{-6}$)
at all times.

\begin{figure*}[!thbp]
  \begin{tabular}{ccc}
    {\sf 1\,au} & {\sf 10\,au} & {\sf 100\,au} \\[-1mm]
    \hspace*{-4mm}
    \includegraphics[page=1,height=48.5mm,trim= 0 0 0 10,clip]
                    {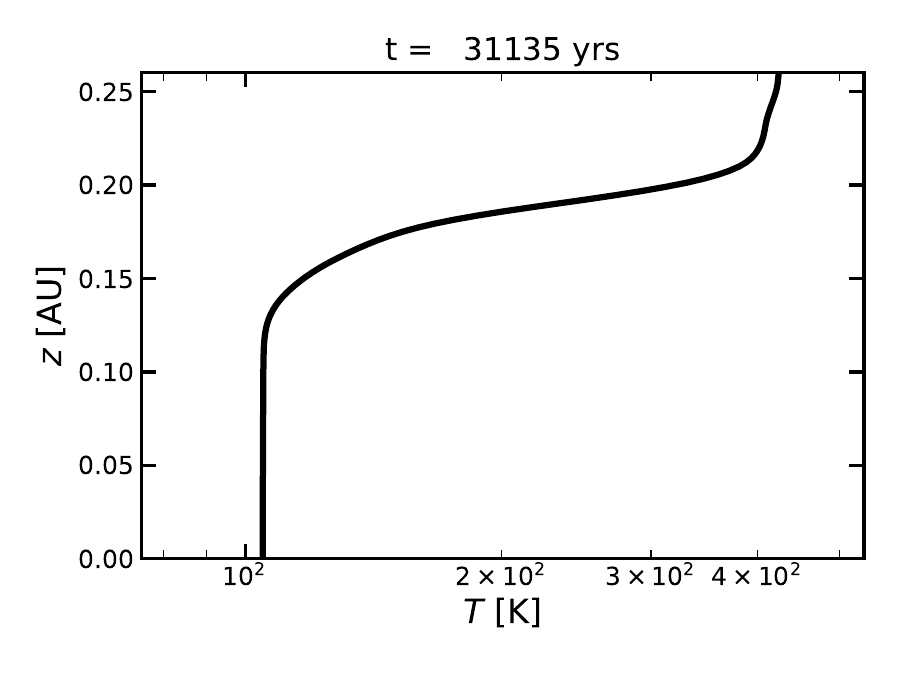} &
    \hspace*{-6mm}
    \includegraphics[page=1,height=48.5mm,trim=35 0 0 10,clip] 
                    {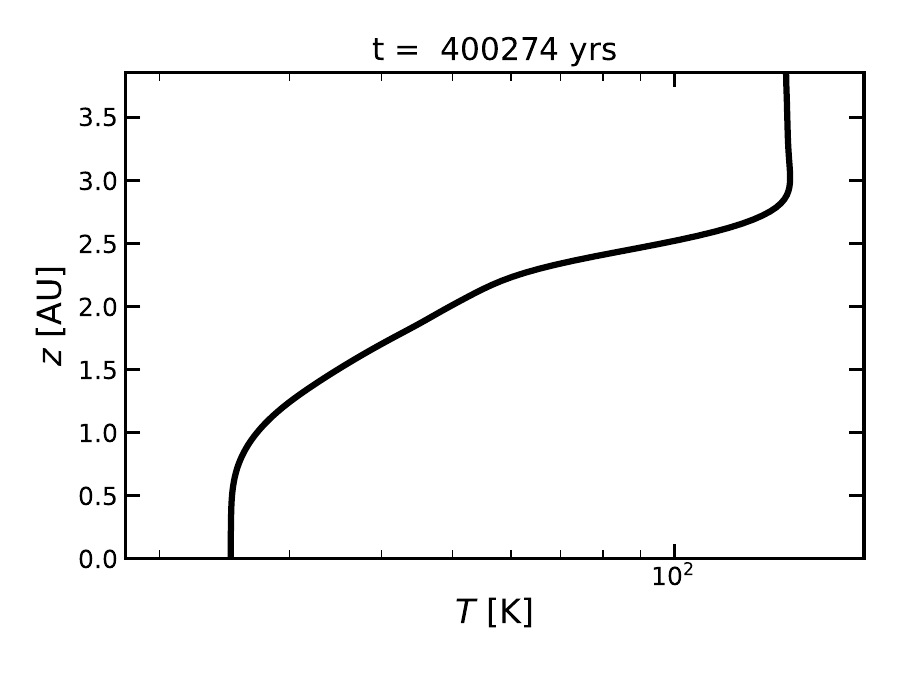} &
    \hspace*{-6mm}
    \includegraphics[page=1,height=48.5mm,trim=35 0 0 10,clip] 
                    {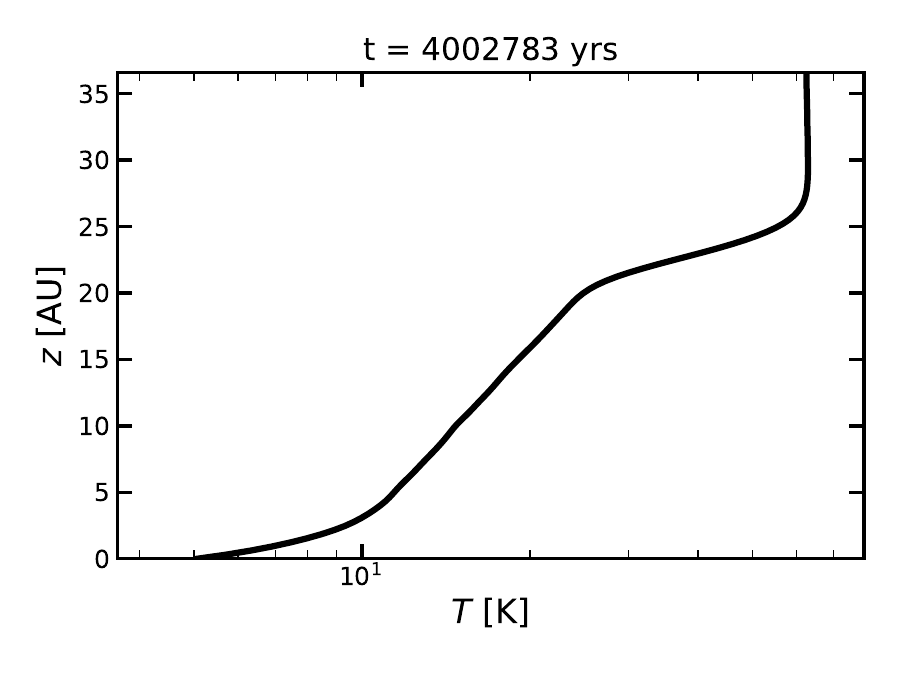} \\*[-4mm]
    \hspace*{-4mm}
    \includegraphics[page=5,height=48.5mm,trim= 0 0 0 10,clip]
                    {out_1AU.pdf} &
    \hspace*{-6mm}
    \includegraphics[page=5,height=48.5mm,trim=35 0 0 10,clip] 
                    {out_10AU.pdf} &
    \hspace*{-6mm}
    \includegraphics[page=5,height=48.5mm,trim=35 0 0 10,clip] 
                    {out_100AU.pdf} \\*[-4mm]
    \hspace*{-4mm}
    \includegraphics[page=18,height=65mm,trim= 0 0 0 10,clip]
                    {out_1AU.pdf} &
    \hspace*{-6mm}
    \includegraphics[page=18,height=65mm,trim=35 0 0 10,clip] 
                    {out_10AU.pdf} &
    \hspace*{-6mm}
    \includegraphics[page=18,height=65mm,trim=35 0 0 10,clip] 
                    {out_100AU.pdf} \\*[-4mm]
    \hspace*{-4mm}
    \includegraphics[page=17,height=65mm,trim= 0 0 0 10,clip]
                    {out_1AU.pdf} &
    \hspace*{-6mm}
    \includegraphics[page=17,height=65mm,trim=35 0 0 10,clip] 
                    {out_10AU.pdf} &
    \hspace*{-6mm}
    \includegraphics[page=17,height=65mm,trim=35 0 0 10,clip] 
                    {out_100AU.pdf} \\*[-4mm]
  \end{tabular}
  \caption{Relaxed vertical structures at 1\,au, 10\,au and 100\,au,
    after modelling chemistry, ice formation, diffusion and settling
    with DiffuDrift for about 30000\,yrs, 400\,000\,yrs, and $4\times
    10^6$\,yrs, respectively. The top panel shows the temperature
    structure iported from the ProDiMo model, the second panel shows
    the gas element abundances, and the lower two rows show the
    chemical composition of the gas phase for oxygen and carbon. Here,
    each atom and molecule has a unique colour. The overplotted thin
    dashed lines show the concentrations of the same molecules in the
    original ProDiMo model.}
  \label{fig:fullmodels}
  \vspace*{5mm}
\end{figure*}
\begin{figure*}
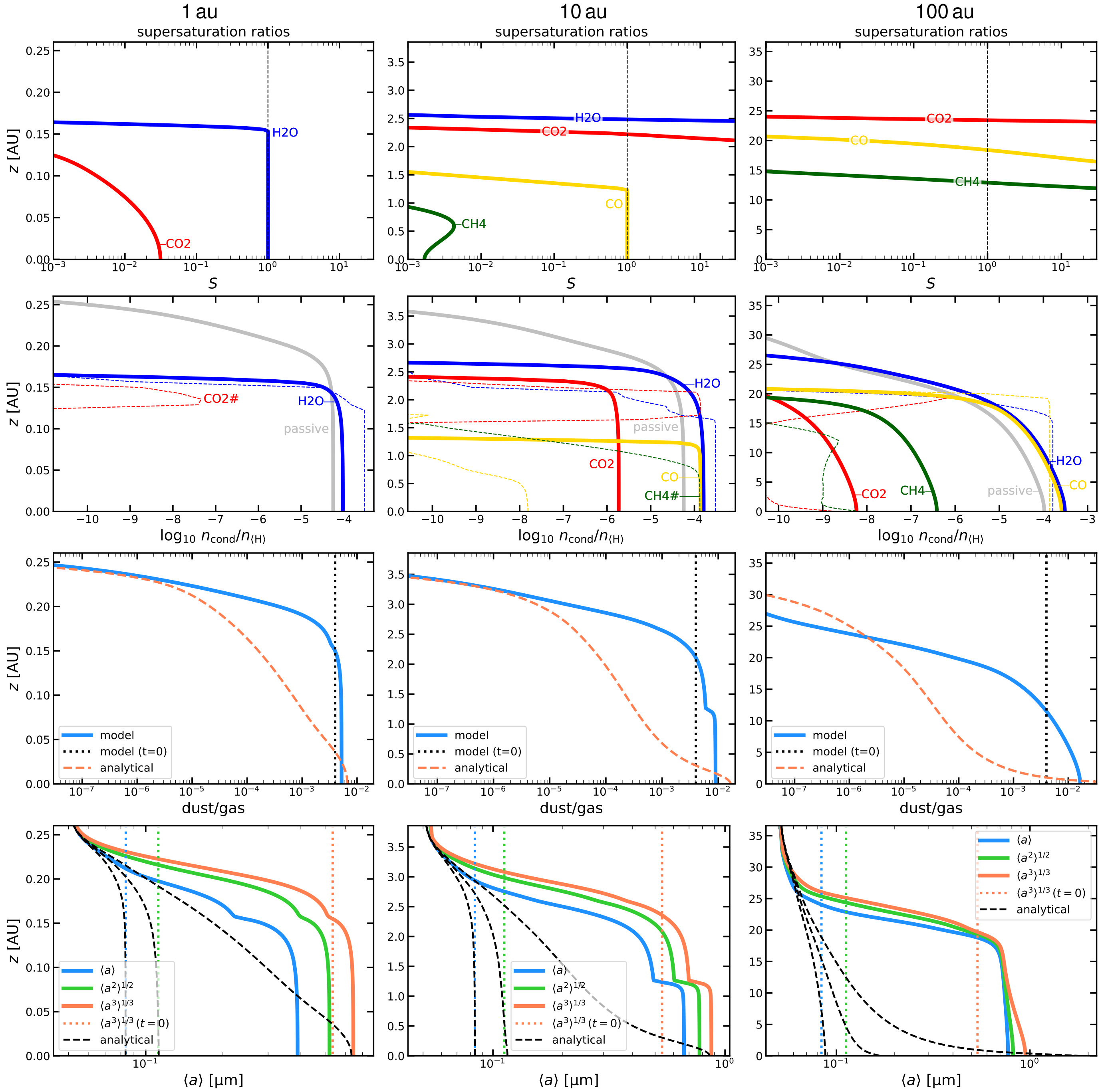

  \vspace*{-2mm}
  \begin{tabular}{ccc}
    {\sf 1\,au} & {\sf 10\,au} & {\sf 100\,au} \\[-1mm]
    \hspace*{-4mm}
    \includegraphics[page=7,height=48mm,trim= 0 0 0 10,clip]
                    {out_1AU.pdf} &
    \hspace*{-6mm}
    \includegraphics[page=7,height=48mm,trim=35 0 0 10,clip] 
                    {out_10AU.pdf} &
    \hspace*{-6mm}
    \includegraphics[page=7,height=48mm,trim=35 0 0 10,clip] 
                    {out_100AU.pdf} \\*[-4mm]
    \hspace*{-4mm}
    \includegraphics[page=14,height=45mm,trim= 0 0 0 30,clip]
                    {out_1AU.pdf} &
    \hspace*{-6mm}
    \includegraphics[page=14,height=45mm,trim=35 0 0 30,clip] 
                    {out_10AU.pdf} &
    \hspace*{-6mm}
    \includegraphics[page=14,height=45mm,trim=35 0 0 30,clip] 
                    {out_100AU.pdf} \\*[-4mm]
    \hspace*{-4mm}
    \includegraphics[page=9,height=48mm,trim= 0 0 0 10,clip]
                    {out_1AU.pdf} &
    \hspace*{-6mm}
    \includegraphics[page=9,height=48mm,trim=35 0 0 10,clip] 
                    {out_10AU.pdf} &
    \hspace*{-6mm}
    \includegraphics[page=9,height=48mm,trim=35 0 0 10,clip] 
                    {out_100AU.pdf} \\*[-4mm]
    \hspace*{-4mm}
    \includegraphics[page=12,height=48mm,trim= 0 0 0 10,clip]
                    {out_1AU.pdf} &
    \hspace*{-6mm}
    \includegraphics[page=12,height=48mm,trim=35 0 0 10,clip] 
                    {out_10AU.pdf} &
    \hspace*{-6mm}
    \includegraphics[page=12,height=48mm,trim=35 0 0 10,clip] 
                    {out_100AU.pdf} \\*[-4mm]
  \end{tabular}
  \caption{continued from Fig.\,\ref{fig:fullmodels}. The top panel
    shows the supersaturation ratios $S_i\!=\!p_i/p_{\rm vap}(T)$. The
    second panel shows the solid concentrations in terms of condensed
    units per H-nucleus, where the thin dashed lines show the results
    of the chemical reference ProDiMo model. {Here, the symbol \#
      flags those ice species which are only found in the ProDiMo
      model, but not in the DiffuDrift model where these ices do not
      occur (only at 1\,au)}. The lower three panels show particle
    properties similar to Fig.\,\ref{fig:test}, where the dotted lines
    are the initial conditions and the dashed lines are the expected
    relaxed particle properties without ice formation {from the
    analytical solution for passive particles explained in
    Sect.\,\ref{sec:settled}. dust/gas means the local
    (dust$+$ice)-to-gas mass ratio.}}
  \label{fig:fullmodels2}
  \vspace*{-2mm}
\end{figure*}

\subsubsection{The evolution of element abundances and chemistry}

The top panel of Fig.~\ref{fig:fullmodels} shows the dust temperature
structures $T(z)$ for the three columns at 1\,au, 10\,au and 100\,au,
imported from ProDiMo. We note that, except for some thermal
velocities, the simulations do not require to know the gas temperature
structure because, concerning the chemistry, $T_{\rm gas}$ is
imprinted into the chemical rate constants which are imported from
ProDiMo as function of height. What DiffuDrift requires are only the
vapour pressures that are calculated with the dust temperature.  All
particles at the same point are assumed to have the same temperature
independent of size.

The second panel shows the gas-phase element abundances in the final
relaxed states. At $r\!=\!1\,$au, there is no long-term effect of the
mid-plane ice formation on $\epsilon_{\rm C}$, and just a weak effect
on $\epsilon_{\rm O}$ in the lower disc layers, see second panel from
the top in Fig.~\ref{fig:fullmodels}.  This behaviour changes dramatically at
10\,au where the midplane is cooler. We observe that the gas phase C/O
approaches unity in a height interval between about $z\!=\!1.2\,$au
and 2\,au. In this interval, \ce{H2O} and \ce{CO2} ices are present,
and $\rm C/O\!\approx\!1$ resembles the principal freeze-out behaviour
between the \ce{CO2} and CO icelines as established by
\cite{Oberg2011}.  Below $z\!=\!1.2\,$au, CO also freezes out, and
both $\epsilon_{\rm C}$ and $\epsilon_{\rm O}$ drop sharply by about
two orders of magnitude, {but C/O remains smaller than one}. At
the top of the $r\!=\!10\,$au model, the gas phase C/O ratio has
slightly increased to 0.65 without the existence of any ices locally.
This is the cold trap effect.

\begin{figure*}[!thbp]
  \begin{tabular}{ccc}
    {\sf 1\,au} & {\sf 10\,au} & {\sf 100\,au} \\
    \hspace*{-4mm}
    \includegraphics[page=1,height=49.5mm,trim= 0 0 34 10,clip]
                    {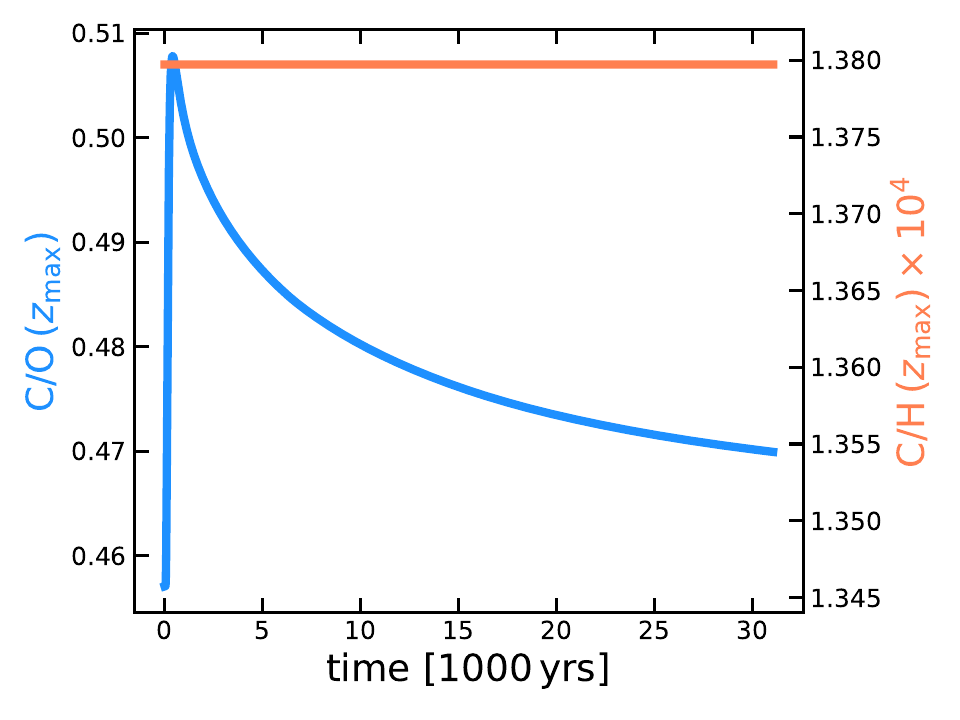} &
    \hspace*{-3mm}
    \includegraphics[page=1,height=49.5mm,trim=34 0 34 10,clip] 
                    {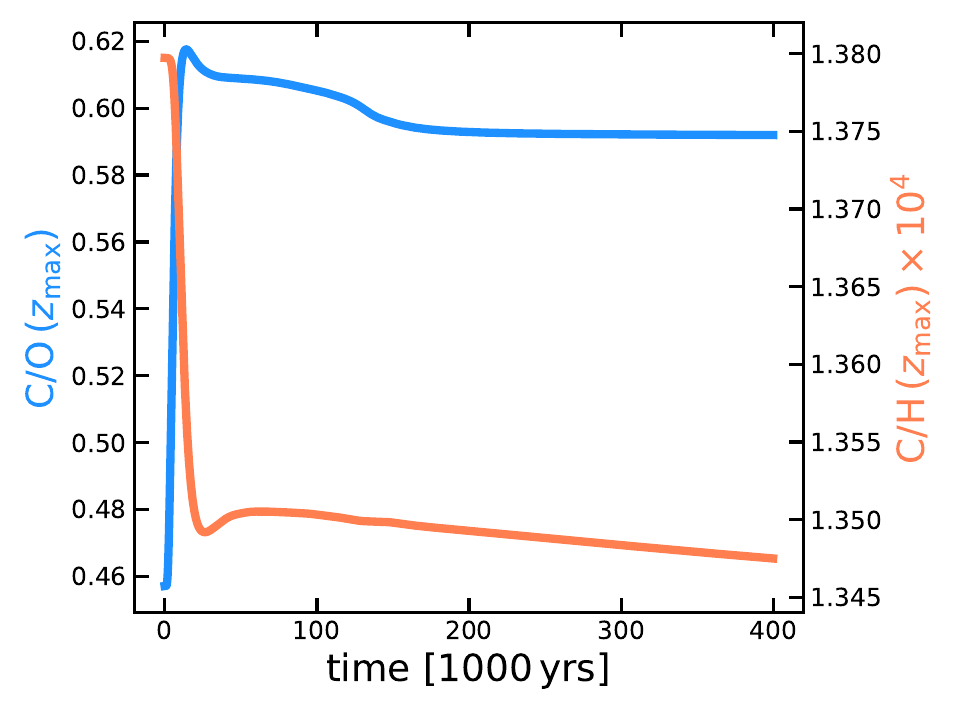} &
    \hspace*{-3mm}
    \includegraphics[page=1,height=49.5mm,trim=34 0 0 10,clip] 
                    {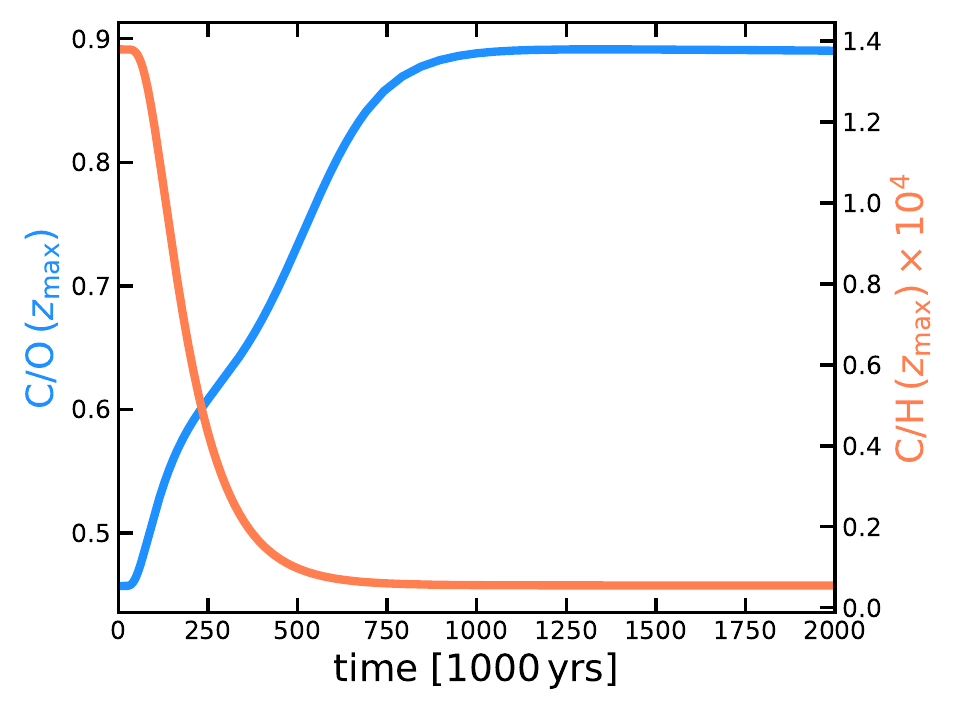} \\*[-3mm]
  \end{tabular}
  \caption{The gas phase C/O and C/H ratios at the top ($z\!=\!z_{\rm
      max}$) as function of simulation time in the three models for
    1\,au, 10\,au, and 100\,au. {The relaxation process takes
      about 30\,000\,yr at 1\,au, 200\,000\,yr at 10\,au, and 1\,Myr at
    100\,au.}}
  \label{fig:CtoO}
\end{figure*}

At $r\!=\!100\,$au the cold trap effect is very strong. We find $\rm
C/O\!\approx\!0.89$ at the top, where both $\epsilon_{\rm C}$ and
$\epsilon_{\rm O}$ are now smaller than $\epsilon_{\rm N}$ by one
order of magnitude.  $\epsilon_{\rm C}$ and $\epsilon_{\rm O}$
continue to decrease by another two orders of magnitude below
$z\!\approx\!18\,$au once all four considered ices are locally
present. However, even in the $r\!=\!100\,$au model, the C/O ratio
never crosses unity.  Similar results have been obtained by
\cite{vanClepper2022} using their ice formation and transport models
with detailed kinetic chemistry.  This is due to an effect known as
CO-blocking \citep{Burrows1999,Oberg2011}. Once \ce{H2O} is fully
condensed, the remaining oxygen in the gas phase is almost completely
tied up in CO molecules, and C/O approaches unity. To increase C/O
beyond one, oxygen would have to be extracted from the CO and
converted into a separate oxygen-bearing species such as \ce{H2O},
while the associated carbon remains in the gas in form of \ce{C},
\ce{C+}, \ce{CH4}, or hydrocarbons. Subsequent freeze-out of the newly
formed \ce{H2O} could then, in principle, lead to $\rm
C/O\!\gg\!1$. In thermochemical equilibrium, such a conversion of CO
into \ce{CH4} and \ce{H2O} indeed occurs at sufficiently low
temperatures, typically below $500-800$\,K depending on pressure and
elemental composition \citep{Woitke2018}.  However, in kinetic
chemistry, breaking the CO triple-bond is extremely difficult in the
UV-shielded midplane. X-rays can do that in the inner disc
\citep{Walsh2015,Woitke2024}, but beyond the snowline, CO destruction
is mainly driven by cosmic rays, through the production of \ce{He+}
followed by $\rm He^+ + CO \to C^+ + O + He$. The corresponding
timescale can be Myr or longer, depending on the adopted cosmic ray
ionisation rate \citep[e.g.][]{Helling2014, Schwarz2018,
  Bosman2018}. This is partly due to a mechanism that we refer to as
the {\em CO self-healing mechanism}: CO destruction does not release
oxygen alone: it also creates reactive carbon, either as \ce{C} or
\ce{C+}. This carbon efficiently recycles the O-bearing intermediates
back into CO through reactions such as $\rm C + OH \to CO + H$, see
e.g.\ \cite{Aikawa1999}, \cite{Bosman2018} and \cite{Schwarz2018}.

\begin{table}
  \caption{Composition of the planet-forming gas$^{(1)}$.}
  \label{tab:gascomp}
  \vspace{-2mm}
  \begin{tabular}{r|p{65mm}}
    \hline
    &\\[-2.2ex]  
    1\,au & 
      \ce{H2} (84\%), He (16\%), \ce{CO2} (120\,ppm),
      CO (112\,ppm), \ce{N2}$^{(2)}$ (66\,ppm), O-atoms (0.4\,ppm) \\
    &\\[-2ex]  
    10\,au &
      \ce{H2} (84\%), He (16\%), O-atoms (96\,ppm), \ce{N2} (67\,ppm), 
      CO (27\,ppm) \\
    &\\[-2ex]  
    100\,au &  
      \ce{H2} (84\%), He (16\%), \ce{N2} (67\,ppm), H-atoms (0.9\,ppm)\\[1mm]
    \hline
  \end{tabular}\\
  \footnotesize
  $^{(1)}$: found at the bottom of the DiffuDrift models.\\
  $^{(2)}$: nitrogen ices not included in this study. 
\end{table}  

The third and forth panels of Fig.~\ref{fig:fullmodels} show the
chemical results. We find substantial deviations from the results of
the ProDiMo reference model, which are caused by the vertical
diffusion and the transport of frozen molecules in the icy grains that
are both not included in ProDiMo. The principle effects of vertical
diffusion on the chemistry have been described by \cite{Woitke2022},
and we find many similarities in the results presented in this
paper. The most abundant species, such as \ce{H2} and CO, the neutral
atoms in higher layers, and the ices in the midplane, are transported
both up and down, and at the locations where these abundant chemicals
finally decompose, for example by photo processes or thermal
desorption, the release of reaction products create a more active
chemistry, with a richer mixture of ionised, atomic, molecular and ice
species. The midplane results are summarised in
Table~\ref{tab:gascomp}. In particular we find:
\begin{enumerate}
\setlength{\itemsep}{1.4pt}
\item {While water molecules freeze out in the midplane at 1\,au, and
  follow the the expected saturation curve $S\!\approx\!1$, CO and \ce{CO2}
  molecules are still around}. The cosmic-ray
  driven unblocking of the CO molecule followed by water and water ice
  formation, which is important in the ProDiMo model to create a
  carbon-rich chemistry in the midplane even at 1\,au, turns out to be
  an artefact of the assumption of steady state chemistry. The
  diffusion is simply the faster process.
\item There are over-concentrations of OH and \ce{H2O} above the
  horizontal snowline, where water ice sublimates, and water
  molecules are locally released by thermal and photo-desorption.
  This effect is best visible at $z\!\approx\!0.17\,$au in the 1\,au
  model, but also present in the other models.
\item In addition to water ice, the model at 10\,au has also \ce{CO2}
  ice and even starts to build CO ice below $z\!\la\!1.2\,$au. This
  causes a sharp drop of the concentrations of all oxygen-bearing
  molecules. {However, DiffuDrift finds oxygen atoms to be the
    most important oxygen carrier in the midplane at 10\,au which
    prevents the C/O-ratio to cross unity. The oxygen atoms are
    provided by downward diffusion and are only slowly converted and
    sequestered into the ice.  This result is in contrast to the
    original ProDiMo model in time-independent mode withut diffusion,
    which finds a carbon-rich $\rm C/O>1$ gas with abundant HCN and
    hydro-carbon molecules, such as \ce{CH4}, HCN, \ce{C2H2} and
    \ce{C2H4}.}
\item The model at 100\,au is featured by concentrations of \ce{C+},
  \ce{C}, \ce{O} and \ce{CO} that are down by about two orders of
  magnitude in the upper regions with respect to the ProDiMo reference
  model, because of the cold trap effect. The midplane stays
  oxygen-rich $\rm(C/O\!\la\!1)$ at $r\!=\!100\,$au.
\end{enumerate}

\begin{figure*}
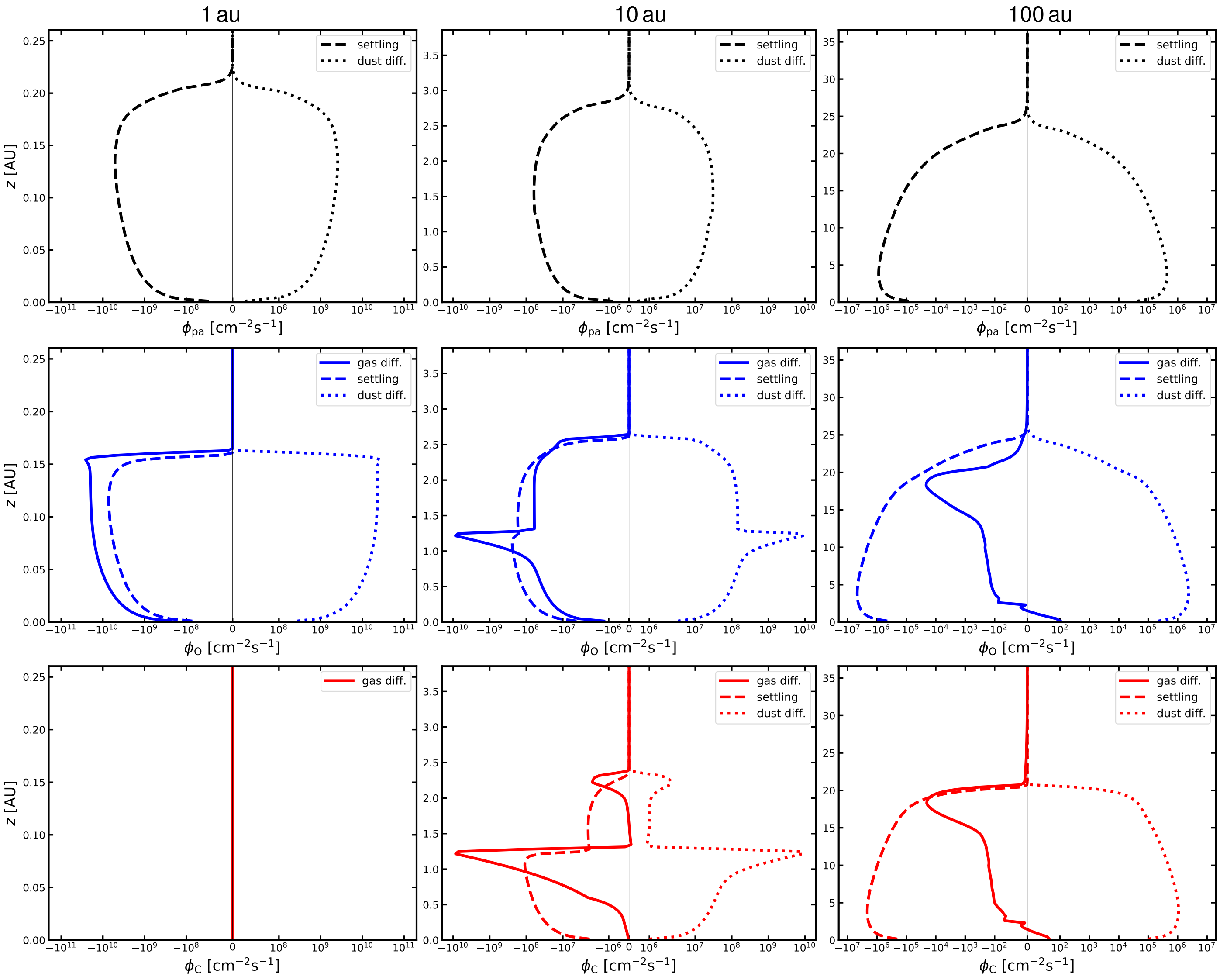

  \begin{tabular}{ccc}
    {\sf 1\,au} & {\sf 10\,au} & {\sf 100\,au} \\
    \hspace*{-4mm}
    \includegraphics[page=2,height=49mm,trim= 0 0 0 10,clip]
                    {fluxes_1AU.pdf} &
    \hspace*{-6mm}
    \includegraphics[page=2,height=49mm,trim=29 0 0 10,clip] 
                    {fluxes_10AU.pdf} &
    \hspace*{-6mm}
    \includegraphics[page=2,height=49mm,trim=29 0 0 10,clip] 
                    {fluxes_100AU.pdf} \\*[-2mm]
    \hspace*{-4mm}
    \includegraphics[page=6,height=49mm,trim= 0 0 0 10,clip]
                    {fluxes_1AU.pdf} &
    \hspace*{-6mm}
    \includegraphics[page=6,height=49mm,trim=29 0 0 10,clip] 
                    {fluxes_10AU.pdf} &
    \hspace*{-6mm}
    \includegraphics[page=6,height=49mm,trim=29 0 0 10,clip] 
                    {fluxes_100AU.pdf} \\*[-2mm]
    \hspace*{-4mm}
    \includegraphics[page=8,height=49mm,trim= 0 0 0 10,clip]
                    {fluxes_1AU.pdf} &
    \hspace*{-6mm}
    \includegraphics[page=8,height=49mm,trim=29 0 0 10,clip] 
                    {fluxes_10AU.pdf} &
    \hspace*{-6mm}
    \includegraphics[page=8,height=49mm,trim=29 0 0 10,clip] 
                    {fluxes_100AU.pdf} \\*[-2mm]
  \end{tabular}
  \caption{Vertical element fluxes in the final relaxed state of the
    modelled columns {at 1\,au, 10\,au and 100\,au. The first row
    shows the fluxes of the passive material, and the second and third
  rows show the element fluxes of oxygen and carbon, respectively.}}
  \label{fig:fluxes}
\end{figure*}

\subsubsection{The evolution of dust and ice}

Figure ~\ref{fig:fullmodels2} continues to show the results of the
same models for the dust and ice. The top panel shows the
supersaturation ratios $S$ and the second panel shows the passive
(refractory dust) and ice concentrations. The 1\,au model has only
water ice, whereas the models at 10\,au and 100\,au have all four
considered ice species. In the 1\,au and 10\,au models, the horizontal
icelines are connected to $S\!=\!1$. Below these lines, the molecular
partial pressures adjust to the vapour pressure by rapid adsorption
and thermal desorption, resulting in $S\!\ga\!1$. However, in the
cooler 10\,au model, the vapour pressures of \ce{H2O} and \ce{CO2}
become so tiny that the actual molecular partial pressures can easily
exceed them, e.g. by diffusion, and $S\!\gg\!1$ becomes possible. In
the 100\,au model, $S\!\to\!1$ entirely looses its regulating effect,
because the ices rather form by hydrogenation and are destroyed by
UV-photo desorption.  For example, there is no $S\!=\!1$ intersection
for water ice anymore -- the water in the 100\,au column remains
highly supersaturated at every height for all times.

There is a strong impact of the ice formation on the particle size
distribution as shown in the lower two panels of
Fig.\,\ref{fig:fullmodels2}.  According to the initially assumed
powerlaw size distribution $f(a)\!\propto\!a^{-3.5}$, the dust surface
is mostly provided by the small sub-micron particles. After the ice
has condensed, which covers all surfaces about equally, these tiny
particles have now grown at least to 0.4\,$\mu$m in the 1\,au model
({see details in Table~\ref{tab:Lmom}}), and to 0.8\,$\mu$m in the
100\,au model.  All particles that are even smaller than that are now
entirely missing. The total condensed mass is then mostly contained in
these micron-sized particles.  This is different from the initial
powerlaw size distribution, according to which the condensed mass is
mostly contained in the big particles.

And this has important consequences for the dynamical behaviour of the
icy particles. These micron-sized icy particles can easily be lofted
by turbulence, and since they contain most of the condensed mass,
the total (dust$\,+\,$ice)/gas mass ratio increases substantially (up
to 2 orders of magnitude) at intermediate heights by ice formation,
with consequences for the opacity structure. {We note again here
  that the DiffuDrift models presented in this paper do not include
  coagulation. Although DiffuDrift is equipped with a charging and
  coagulation module \citep{Woitke2025}, the counter process of
  fragmentation would be required here, too, and that is not yet
  implemented. For models with coagulation and fragmentation, see
  \cite{Stammler2017} and \cite{Robinson2026}. Since ice deposition is
  based on gas-grain collisions, it is intrinsically faster than
  coagulation based on dust-dust collisions. Therefore, we expect
  that some of the effects discussed here, like the removal of the
  very small grains and the overpopulation of the 1\,$\mu$m size bins
  might still be there when coagulation and fragmentation is taken
  into account, although coagulation and fragmentation will certainly
  smooth out these effects on the long run.}

The model at 1\,au features a sudden increase of $\langle
a\rangle$ around $z\!=\!0.16\,$au, and the model at 10\,au has two of
these size steps at 2.6\,au and 1.2\,au.  These steps are related to
the horizontal icelines of \ce{H2O} and \ce{CO} (the latter only for
the 10\,au model).  In contrast, the 100\,au model rather features
a constant mixture of all four incuded ices.  In all cases, the
ice-to-dust ratio approaches about 0.9 by mass, and about 3.5 by
volume in the midplane.  {This can be compared to \cite{Min2016} who
found that about 80\% of the oxygen is in ice in HD\,142527}.

\subsubsection{The cold trap effect}

A cold trap works by adding an actively cooled surface to a gas
container with partly condensable molecules. These molecules will
diffuse toward the cold surface, condense out, and will continue to be
trapped there for as long as the surface stays cold. This is routinely
used in laboratory experiments to remove unwanted water vapour. The
analogy to a vertical column in a protoplanetary disc is obvious. The
cold surface is provided by the dust particles in the cold midplane
below a much warmer disc surface that is also hostile to ice because
of the UV irradiation.  Figure~\ref{fig:CtoO} shows the evolution of
the gas phase C/O ratio at the top of the models. In all cases, we see
a rapid increase followed by a relaxation towards a final constant
value. The times required to reach these final values coincide well
with the maximum of the two characteristic timescales for settling and
diffusion in Table~\ref{tab:columns}, {which is about $10^4$ to
  $10^6$ yrs}. The moment C/O starts to increase significantly, C/H
decreases.  These two gas element ratios are tightly correlated.
{In comparion, the observed CO depletion appears within
  $\sim\!1$\,Myr \citep{Bergner2020, Zhang2020b}, and C/H has been
  observed to decrease by about one order of magnitude outside of
  70\,au in HD 163296 \citep{Zhang2020a}}.

\subsubsection{The vertical fluxes of elements}

The reason why the cold trap effect does not continue forever, which
would quickly lead to a complete depletion of oxygen (and carbon) from
the upper disc layers, is shown in Fig.\,\ref{fig:fluxes}.  For a
classical cold trap, the cold surface must be at rest.  However, the
micron-sized particles in protoplanetary discs are diffusing up and
down in the same way as the gas. Once oxygen and carbon have diffused
down into the cold and shielded midplane, and have deposited to form
the ice, the icy particles diffuse up and sublimate again at the
respective horizontal icelines of the various ice materials.

The different vertical element fluxes in the final relaxed state are
depicted in Fig.~\ref{fig:fluxes}. Once again, we observe a perfect
flux balance concerning the passive dust material between upward
diffusion and downward settling, as already shown in
Fig.\,\ref{fig:test} and discussed in Sect.\,\ref{sec:settled}.
Concerning the elements oxygen and carbon, there are now three
competing fluxes: the downward transport of the elements contained in
the ice by settling; the upward diffusion of the same elements in the
ice by particle diffusion; and the diffusive flux of the elements in
the gas phase, which can be either upward or downward depending on the
gradients in the gas phase. At the top of a new ice layer, the
elements disappear into the ice, creating a steep gas concentration
gradient that causes downward gas diffusion. This is the very reason
for the cold trap effect. In the early stages of the model, the gas
that condenses to form the ice is replenished by a diffusive flux from
above. However, in the final relaxed state, the sum of the three
fluxes balances out for each element.  In particular, the loss of gas
that forms the ice is balanced by an influx of sublimating icy
particles, bringing the cold trap effect to a halt.

\section{Summary and conclusions}
\label{sec:summary}

This paper has analysed the cold trap mechanism for oxygen and carbon
caused by vertical diffusive transport and sequestration into ice in
the cold midplane of protoplanetary discs. We have introduced an
extension of the DiffuDrift code \citep{Woitke2020,Woitke2025} to
include kinetic chemistry. The new code can consistently simulate
kinetic chemistry, cloud formation, diffusion, particle growth and
gravitational settling in time-dependent models for 1D vertical
columns.  In future applications, we intend to apply this code to
cloud formation in planetary atmospheres, but in this paper, we have
applied it to the problem of ice formation on dust particles in a
series of vertical columns of a protoplanetary disc.

The chemical rates are provided by an external chemistry code that
needs to be executed on the same physical structure of the considered
1D vertical column.  We have presented a new concept how to reduce
heavy chemical networks by means of introducing a ``background gas''
that contains all molecules not explicitly selected.  The chemical
rates are separated into three different classes, the rates between
the included chemicals, the rates between these chemicals and the
background gas, and the rates between the chemicals and the
condensates. The latter are replaced by the growth and sublimation
rates as formulated by means of supersaturation ratios in DiffuDrift.

As long as the resulting chemical structure remains similar to the
structure computed by the external chemical code, this method should
be fast and accurate. The chemical composition of the background gas
is the key here, because this composition enters into the computation
of the effective chemical background rates by the external code.  We
have tested this new approach with a tiny network (12 species), and a
medium network (24 species), in comparison to a full ProDiMo network
with 119 gas species and 10 ice species. We ran a hard test where we
evolved the chemistry from a simple initial molecular cloud
composition toward the complex molecular composition with ices in the
disc midplane. The level of agreement was found to be very
satisfactory, however, since the swapping of an oxygen-rich to a
carbon-rich composition of the disc midplane due to ice formation can
change the composition of the background gas substantially, we have
decided to use the full ProDiMo network for all main models presented
in this paper.

\begin{figure}
  \vspace*{1mm}
  \hspace*{-1mm}
  \includegraphics[page=1,width=90mm,trim= 10 10 10 10,clip]
                  {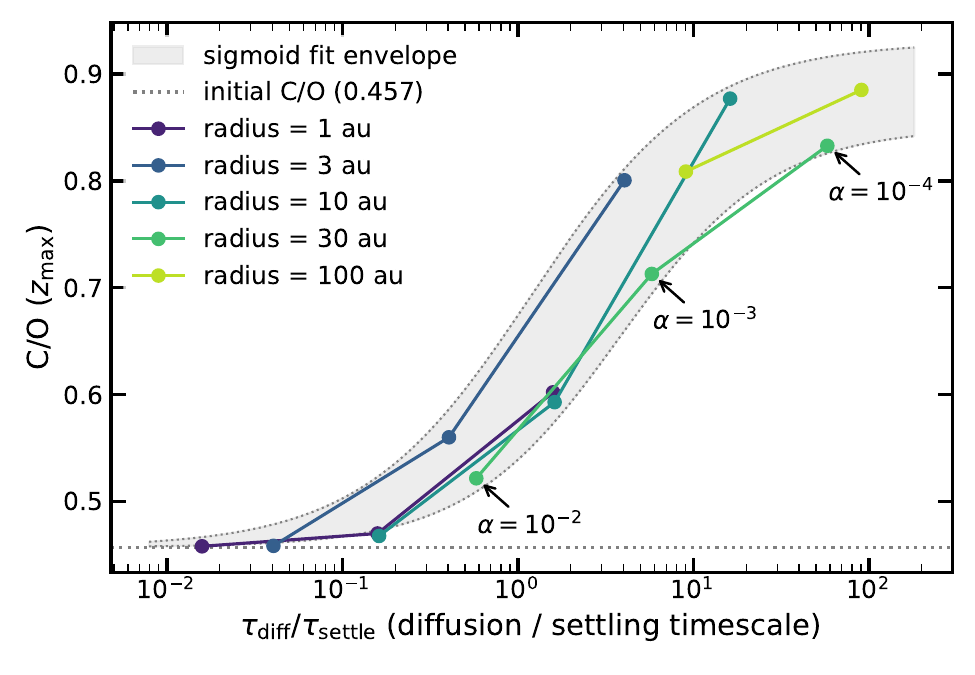} \\
  \hspace*{-1mm}
  \includegraphics[page=2,width=90mm,trim= 12 10 10 10,clip]
                  {Vs_timescales.pdf} \\[-5mm]
  \caption{The gas phase C/O and C/H ratios at the top of the model
    $z\!=\!z_{\rm max}$ as function of the ratio of
    diffusion-to-settling timescales {at 1\,$\mu$m}, see
    Eq.\,(\ref{eq:timescales}).}
  \label{fig:Vs_timescales}
  \vspace*{-3mm}
\end{figure}

In the main application of this paper we explored the effectiveness of
the cold trap, where a mostly atomic oxygen and carbon gas diffuses
downward, forms OH, \ce{H2O}, CO, \ce{CO2}, and other complex
hydro-carbon molecules, and some of these molecules can deposit on the
surface of the refractory dust particles in deeper layers, once these
layers are sufficiently cold and shielded from the UV-field. We found
the icy particles to quickly develop a counter-flux that brings the
cold trap to a halt and generates a final stationary solution after
about $10^4$ to $10^6$\,yrs, depending on the distance to the
star. The icy particles are whirled up by turbulence and the ice
sublimates at the horizontal icelines, either because the small grains
become too warm in the upper illuminated layers or the UV-irradiation
causes photo-desorption. In the final relaxed state, the counter-flux
by upward particle diffusion balances the downward transport by
settling and gas diffusion.

The resulting properties of the relaxed structures depend critically
on the ratio of the diffusion\,/\,settling timescales as defined by
Eq.\,(\ref{eq:timescales}), see Fig.\,\ref{fig:Vs_timescales}. For
this figure, we have computed additional models at 3\,au and 30\,au,
and have run each column three times, with viscosity parameter
$\alpha\!=\!10^{-2}$, $10^{-3}$, and $10^{-4}$.  However, the model at
100\,au with $\alpha\!=\!10^{-4}$ already has a diffusion timescale of
$5\times10^7$\,yrs, which exceeds the lifetime of protoplanetary
discs, so we discarded that model.  The results of all other 14 models
are shown in Fig.\,\ref{fig:Vs_timescales}.  We see that the C/O and
C/H ratios at the top of the columns $(z\!=\!z_{\rm max})$ remain
unaffected as long as $\tau_{\rm diff}\!\ll\!\tau_{\rm settle}$. For
$\tau_{\rm diff}/\tau_{\rm settle}\!\ga\!1$, however, C/O increases
toward unity and C/H drops sharply. None of the models show $\rm
C/O\!>\!1$ at the top, although this can happen locally in the disc
midplane.  This is a consequence of the implemented
kinetic chemistry, where the gas will naturally stop to produce larger
amounts of OH, \ce{H2O} and \ce{CO2} once $\rm C/O\!\to\!1$ is reached
(CO-blocking).

Using Eqs.\,(\ref{eq:tau_diff}) and (\ref{eq:tau_sett}), the timescale
ratio is given by
\begin{align}
  \frac{\tau_{\rm diff}}{\tau_{\rm settle}} \approx~~&
  2.75~\left(\frac{\alpha}{0.001}\right)^{-1}\!\beta^2\,
  \left(\frac{\Sigma}{1000\rm\,g\,cm^{-2}}\right)^{-1}
  \left(\frac{r}{1\rm\,au}\right)^{-1} \nonumber\\ \times\,
 &\left(\frac{T}{100\rm\,K}\right)^{-1} 
  \left(\frac{M_{\star}}{M_{\odot}}\right)\,
  \left(\frac{\rho_m}{2\rm\,g\,cm^{-3}}\right)\,
  \left(\frac{a}{1\,\mu\rm m}\right)
  \label{eq:tau_ratio}
\end{align}
with geometric aspect ratio $\beta\!=\!z_{\rm max}/r\!\approx\!0.3$ in our
models. Equation~(\ref{eq:tau_ratio}) shows that in many cases,
protoplanetary discs are on the verge of $\tau_{\rm
  diff}\!>\!\tau_{\rm settle}$, in which case the cold trap effect can
be expected to become relevant.  This happens in particular in the
tenuous outer disc where the surface density tapers off exponentially,
$(\Sigma/1000\,{\rm g\,cm^{-2}})\times(r/1\rm\,au)\!\ll\!1$, and
$T\!\ll\!100\,$K. In particular, protoplanetary discs with low
turbulence $\alpha$ are prone to exhibiting the cold trap effect.

\cite{Krijt2020} combined a reduced chemical network centered on five
ice and vapour species with 2D diffusion, settling, radial pebble
drift and coagulation/fragmentation.  They found large CO depletion
factors up to 100 in the warm molecular layer beyond 20\,au after
3\,Myr. {These results, which are similar to ours, are principally
  consistent with observations \citep{Du2017, Zhang2020a,
    Bergin2016}.}  However, \cite{Krijt2020} also found C/O ratios as
large as $3-10$, which is inconsistent with our results $\rm
C/O\!\to\!1$. Our results agree better with the work of
\cite{vanClepper2022} who used a more similar 1D vertical approach,
coupling a large gas-grain chemical network with pebble growth, gas
and small dust diffusion, and vertical settling, while the pebbles
forming near the midplane are assumed to permanently remove the ice
from further gas exchange.  These authors also found $\rm C/O\to 1$,
stating that going beyond unity is a challenge for their model as it
would seemingly require to dissociate the CO. It is interesting to
note that all models with complex gas phase chemistry, which
kinetically treat the chemical conversion between atomic oxygen and
the included vapour species, seem to find this behaviour.

Apart from these chemical results, our models also show a significant
response of the particle size distribution to the process of ice
formation and hence a different dynamical behaviour of the icy dust
particles. Without ice formation, we have shown that the DiffuDrift
models for passive particles approach our analytic solution for dust
settling very well, where particle diffusion is in perfect balance
with gravitational settling. The ice mostly deposits on the small
particles, because these particles provide the majority of the
surface, adding an ice layer of approximately $0.4-0.8\,\mu$m
thickness to all particles, and hence the sub-micron particles are
missing entirely after ice formation. Moreover, since the total ice
mass and the refractory dust mass are about equal, but the ice volume
exceeds the refractory dust volume by a factor of about 4, the
particle size distribution after ice formation has a pronounced peak
at about 1\,$\mu$m, and equally large particles are much lighter after
ice formation. Those particles can easily be lofted by turbulence
which substantially increases the local (dust$\,+\,$ice)/gas ratio at
intermediate heights.  {This is in some contrast to the
  \cite{Bergin2010} and \cite{Du2017} models which assume permanent
  removal by grain growth, settling and radial drift.}  The complex
interplay between the fast ice formation chemistry, and the relatively
slow coagulation and fragmentation processes, together with diffusion,
settling and radial drift requires further attention in the dynamic
modelling of volatile transport in disc evolution models.

\begin{acknowledgements}
   {This article has used AI assistance by Claude Sonnet 5.5 for
     code generation, literature research, and text improvement. In
     particular, the new pre-conditioning routine (see
     App.\,\ref{app:num_solution}) and the box-test
     (App.\,\ref{app:box_test}) have been developed with AI
     assistance, and 2/78 source code routines of DiffuDrift are now
     mostly written by the AI}. I want to thank a number of colleagues
   for fruitful discussions of the paper content, in particular
   Christiane Helling, Andrei H{\"u}nemeyer, Paul Rimmer, Manuel
   Scherf, Kundan Kadam, Sierk van Terwisga, Helmut Lammer and Martin
   Ferus.  I acknowledge the funding received from the Austrian
   Science Fund (FWF) for the support of the VeReDo research project,
   grant I6857-N.
\end{acknowledgements}

\bibliographystyle{aa}
\bibliography{references}

\cleardoublepage

\appendix

\section{Full set of equations}
\label{app:fulleqs}

The full set of equations solved in this paper with all transport
terms and full chemistry (without the background gas $\epsilon_k$),
but without nucleation and coagulation, is given by
\begin{eqnarray}
  \frac{d}{dt}\big(\rhoL_j\big)
  &=& \frac{j}{3} \sum\limits_{p=1}^{2} n_p V_{\!p}^{j/3-1}\,
      \frac{\partial V_{\!p}}{\partial t} \nonumber\\
  &+& \frac{\partial}{\partial z}\,
      \bigg(P_{\rm c}(j\!+\!1)\sum\limits_{p=1}^{2} n_p
      V_{\!p}^{j/3}\,\vdreq(V_{\!p})\bigg) \nonumber\\ 
  &+& \frac{\partial}{\partial z}\,
      \bigg(D\,\rho\,\frac{\partial L_j}{\partial z}\bigg) \\
  \label{fulleq:Lj}    
  \frac{d}{dt}\big(\rhoL_3^s\big)    
  &=& \sum\limits_{p=1}^{2} n_p \frac{\partial V_{\!p}^s}{\partial t} \nonumber\\
  &+& \frac{\partial}{\partial z}\,
      \bigg(P_{\rm c}(4)\sum\limits_{p=1}^{2} n_p
      V_{\!p}^s\,\vdreq(V_{\!p})\bigg) \nonumber\\
  &+& \frac{\partial}{\partial z}\,
      \bigg(D\,\rho\,\frac{\partial L_3^s}{\partial z}\bigg) \\
  \label{fulleq:L3s}    
  \frac{d}{dt}\big(n_i\big)
  &=& \Gamma_i^{\rm gas} - \Lambda_i^{\rm gas} 
  ~+~ \Gamma_i^\# - \Lambda_i^\#  \nonumber\\
  &+& \frac{\partial}{\partial z}\,
      \bigg(D\,\rho\,\frac{\partial(n_i/\rho)}{\partial z}\bigg)
  \label{fulleq:ni}    
\end{eqnarray}
with the two-particle representation $\rhoL_j\!=\!\sum_{p=1}^2 n_p
V_{\!p}^{j/3}$, the net volume growth rates of the different materials
$\partial V_{\!p}^s/\partial t\,(n_i,T)$, the settling velocity
$\vdreq(V)$, the chemical formation $\Gamma_i(n_j,T)$ and
destruction rates $\Lambda_i(n_j,T)$, and the relations between the
ice adsorbtion/desorption rates and the net volume growth rates
$V_0^s\,\big(\Gamma_s^\# - \Lambda_s^\#\big) = \sum_{p=1}^2
n_p\,\partial V_{\!p}^s/\partial t$ explained in
Sect.\,\ref{sec:chemeq}.  The power-correction factors $P_{\rm c}$ are
explained in the next section. All other symbols have been introduced
in the main part of the paper.

\section{The power correction scheme}
\label{app:power_correct}

Using the two-particle representation of the size moments
Eq.\,(\ref{eq:double-delta}), we can approximately express
any size-integrated property of all particles by
\begin{equation}
  \langle p_j\rangle ~= \int_{\Vl}^\infty f(V)\,V^{j/3} p(V)\,dV
  \approx \sum_{p=1}^2 n_p V_{\!p}^{j/3} p(V_{\!p}) \ ,
  \label{eq:prop_approx}
\end{equation}
where $p(V)$ is a property of the particles that depends size $V$.  In
order to improve this approximation, we write
\begin{equation}
  \langle p_j\rangle ~=~ P_{\rm c} \sum_{p=1}^2 n_p V_{\!p}^{j/3} p(V_{\!p}) \ ,
  \label{eq:prop_corr}
\end{equation}
where $P_{\rm c}$ is the cower-correction factor. For $p(V)\!=\!1$ we
get the definition of the size moments $p_j\!=\!\rho L_j$, and since
the $\{n_1, n_2, V_1, V_2\}$ are derived from $\{L_0, L_1, L_2, L_3\}$
via Eq.\,(\ref{eq:double-delta}), Eq.\,(\ref{eq:prop_corr}) is already
exact without correction $P_{\rm c}\!=\!1$ for $j\!=\!0,1,2,3$ in this
case.

However, if we consider a powerlaw property of the particles,
$p(V)\!=\!{\rm const}\,V^{\,k/3}$, the result is
\begin{equation}
  \frac{\langle p_j\rangle}{\rm const}
  = \int_{\Vl}^\infty f(V)\,V^{\,(j+k)/3}\,dV
  = P_{\rm c}(j+k) \sum_{p=1}^2 n_p V_{\!p}^{\,(j+k)/3} ,
\end{equation}
from which we find the definition of the power-correction factor
\begin{equation}
  P_{\rm c}(j) ~=~ \frac{\int_{\Vl}^\infty f(V)\,V^{j/3}\,dV}
               {\sum_{p=1}^2 n_p V_{\!p}^{j/3}}
  \label{eq:powcorr}
\end{equation}
As already stated, $P_{\rm c}(j)\!=\!1$ by definition for
$j\!=\!0,1,2,3$.  But for any $j\!<\!0$ or $j\!>\!3$ the power
correction factor will in general deviate from unity.

This matters for the settling and growth terms in the moment
equation system.  In the Epstein regime, $\vdreq\propto a\propto V^{1/3}$
and $\partial V_{\!p}/\partial t\propto a^2 \propto
V^{2/3}$. Therefore, in Eqs.\,(\ref{fulleq:Lj}) and
(\ref{fulleq:L3s}), we get correction factors $P_{\rm c}(j+1)$ and
$P_{\rm c}(4)$ for the drift terms, and correction
factors $P_{\rm c}(j-1)$ and $P_{\rm c}(2)$ for the growth terms, respectively.
However, since there is a factor $j$ in the growth term in Eq.\,(\ref{fulleq:Lj})
and $P_{\rm c}(j)\!=\!1$ by definition for $j\!=\!0,1,2,3$, the growth terms
do not need to be power-corrected. The only remaining power correction
factor is $P_{\rm c}(4)$ for the settling terms that appear in the
equations for $\rho L_3$ and $\rho L_3^s$.

In the Stokes regime, $\vdreq\propto a^2\propto V^{2/3}$
and $\partial V_{\!p}/\partial t\propto a \propto
V^{1/3}$, so we get correction factors $P_{\rm c}(j+2)$ and
$P_{\rm c}(5)$ for the drift terms, and correction
factors $P_{\rm c}(j-2)$ and $P_{\rm c}(1)$ for the growth terms
in Eqs.\,(\ref{fulleq:Lj}) and (\ref{fulleq:L3s}), respectively.
In that case, the settling terms for $\rho L_2$, $\rho L_3$ and $\rho
L_3^s$ need to be corrected, as well as the growth term for $\rho
L_1$, the latter with $P_{\rm c}(-1)$.

Analysis of Eq.\,(\ref{eq:powcorr}) shows that the power correction
factors do not depend on mean particle size or amount of particles,
but on the relative width of the size distribution function. If $f(V)$ is
narrow, all $P_{\rm c}(j)\to 1$ for all $j$. A useful measure of the
unitless relative width of the size distribution function is
\begin{equation}
  w ~=~ \frac{L_1 L_3}{L_2^2} \ , 
  \label{eq:width}
\end{equation}
for which $w\!\geq\!1$ holds, $w\!=\!1$ means a $\delta$-function.
In order to put this new concept into practise, we make an eductive
guess for the shape of $f(V)$ and define a family of such $f(V)$ with
increasing widths.  Next, for each of these distibution functions, we
compute all moments $L_j$ and calculate the $P_{\rm c}(j)$ according to
Eq.\,(\ref{eq:powcorr}) for $j=\{-1,4,5\}$ and save them in 
arrays as function of width $w$.  Later, in the
main program, we compute the width from Eq.\,(\ref{eq:width})
and interpolate all requested correction factors (only $P_{\rm c}(4)$
in this paper) in $w$ between the pre-tabulated values.

\begin{figure}
  \centering
  \includegraphics[width=75mm,trim= 10 15 10 10,clip]
                  {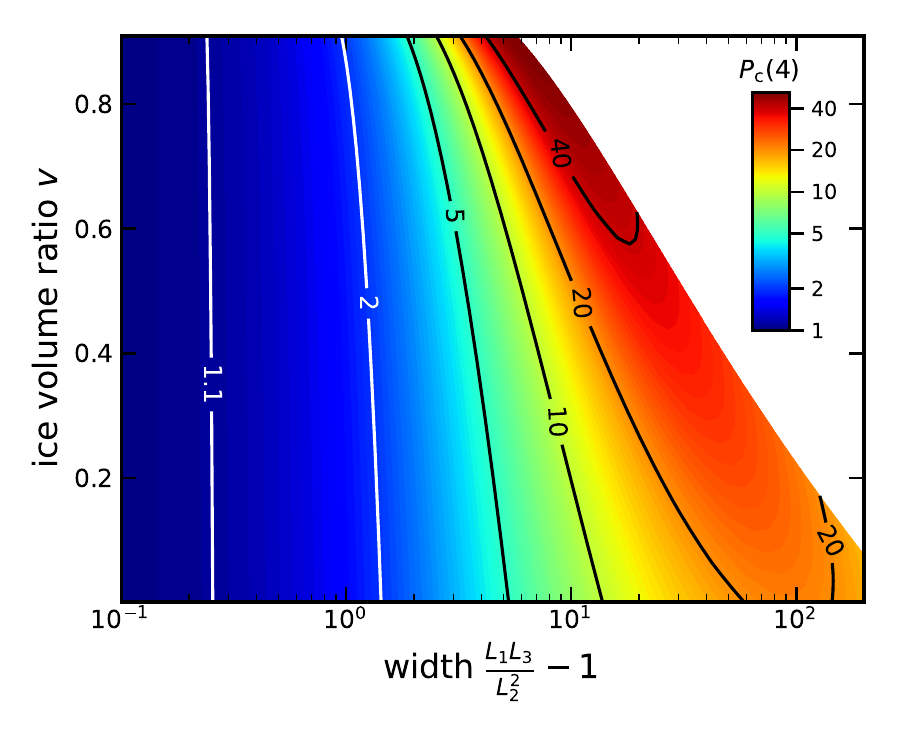}\\*[-3mm]
  \caption{The power correction factor $P_{\rm c}(4)$ for the settling in
    the large Knudsen number limiting case at $r\!=\!10$\,au as function of the
    width of the size distribution function $w$ and the ice volume
    ratio $v$.}
  \label{fig:P4}
\end{figure}

In order guess the shape of $f(V)$, we use the analytical solution
of a powerlaw size distribution function after settling, see
Sect.\,\ref{sec:settled}), where the height grid points, in
decreasing order, provide a data sequence with monotonously increasing $w$.
For the special case of icy particles considered in this paper, we
also take into account the effect of ice condensation on the particle
size distribution. We consider the ice volume ratio as second variable
\begin{equation}
  v ~=~ \frac{\sum_{s\,\in\,\rm ices} L_3^s}{L_3} \ .
  \label{eq:iceratio}
\end{equation}
At every height point, we use the bare settled $f(V)$, from the
analytic soluion, and add an ice layer of constant thickness to all
particles to match $v$, where $v$ is varied between 0 and about 1, this
provides 2D tables of $P_{\rm c}(4)$ as function of $(w,v)$. In the
main program we then 2D-interpolate the power correction factor
$P_{\rm c}(4)$ for $(w,v)$ as computed from the actual
$L_j$ and $L_3^s$. {Fig.\,\ref{fig:P4} illustrates the
  power-correction table as generated for the 10\,au model that can be
  considered as out closure condition for this model.}

\section{The numerical solution method}
\label{app:num_solution}
\def\Ff{\vec{F}_{\rm fro}}
\def\yf{\vec{y}_{\rm fro}}
\def\tstop{t_{\rm stop}}

In every grid cell DiffuDrift advances the moments of the dust size
distribution, the volumes of the condensed materials and the gas chemical
composition (solution vector $\vec{y}(t)$, see Eq.\,\ref{eq:solvec})
by means of one implicit integration step
\begin{equation}
  \vec{y}(t+\Delta t) ~=~ \vec{y}(t)
  + \Delta t\,\vec{F}\big(\vec{y}(t+\Delta t)\big) \ .
  \label{eq:backEuler}
\end{equation}
Using an implicit method is important for stability, because we have a
potentially highly stiff system with short chemical and/or
condensation relaxation timescales. And doing it in one step is the
key for good performance. The length of $\Delta t$ is limited by the
general operator splitting method, and here we require that the
molecules that enter the computation of the nucleation and growth rates
must not change their concentrations by more than a relative threshold,
here chosen to be 25\%, within $\Delta t$.

While this method itself can cope with most chemical situations well,
there are a few cases where it fails or gives inaccurate results, for
example when the particle properties change a lot within $\Delta t$
and the growth rate $\propto\!a^2$ is evaluated at the final
size, or when a particle completely sublimates and has no end state at
all. Particularly relevant cases are:
\begin{enumerate}
\item fast relaxation of one or several major condensates $s$ towards
  saturation $\partial V_s/\partial t\to 0$,
\item sublimation of one or several major volatile materials while other
  trace materials are stable or grow,
\item complete sublimation of particles.  
\end{enumerate}
In order to guide the implicit solver toward the physically correct
solution, we have developed a pre-conditioner that integrates the
properties of the two representative particles along their own
growth/sublimation paths for a \emph{frozen gas} with r.h.s. vector
$\Ff(\vec{y})$, resulting in an approximation $\yf$ for
$\vec{y}(t\!+\!\Delta t)$. Equation~(\ref{eq:backEuler}) can then
be written as
\begin{align}
  \vec{F} &= \Ff + (\vec{F} - \Ff) \\
  \vec{y}(t\!+\!\Delta t) &= \vec{y}(t)\,+\!\int_t^{t+\tstop}\!\!\Ff\,dt'
          \,+\!\int_t^{t+\Delta t}\!\!\vec{F}\,dt'
          \,-\!\int_t^{t+\tstop}\!\!\Ff\,dt' \nonumber\\
          &= \yf +\, \Delta t\,\vec{F}\big(\vec{y}(t\!+\!\Delta t)\big)
               \,-\, \tstop\,\Ff\big(\vec{y}(t\!+\!\Delta t)\big)\!\!
\end{align}
If no modifications are applied, we have $\yf\!=\vec{y}(t)$,
$\Ff\!=\!0$ and $\tstop\!=\!\Delta t$, reverting to
Eq.\,(\ref{eq:backEuler}).  With this machinery in place, we can apply
certain physical approximations to generate a $\yf$ close to
$\vec{y}(t\!+\!\Delta t)$, such that the errors of the single implicit
integration step become small. An important realisation is that
chemistry and particle growth forms a multi-dimensional relaxing
system in which every material is on its way to reach either saturation
or complete sublimation, but every material does that on its own
timescale, and the gas phase chemistry possibly introduces additional
long characteristic timescales.

The generic case of a frozen time step is that all gas concentrations
stay about constant within $\Delta t=\tstop$. We then integrate the
material volumes $V_s$ of the two representative particles along their
growth and sublimation paths with the gas concentrations frozen at
$n_i(t)$, which predicts $L_3^s$ and $L_j$ at $t\!+\!\Delta t$. The
chemical concentrations in $\yf$ are set to $n_i(t)$ minus the
molecules consumed by growth, or plus those liberated by sublimation,
up to $t\!+\!\Delta t$. $\Ff(\vec{y})$ contains the growth and
sublimation terms of $\vec{F}(\vec{y})$ for the particles described by
$\vec{y}$, but evaluated with the frozen gas concentrations
$n_i(t)$. Gas-phase chemistry and nucleation are not part of $\Ff$, so
$\vec{F}-\Ff$ comprises the chemistry, the nucleation, and the
feedback of the changing gas concentrations on the growth rates.
The more complex cases are treated as follows:

\smallskip\noindent{\sf Case 1 (relaxation towards saturation):} With
frozen gas concentrations, condensates approaching saturation would
overshoot, so the frozen path is stopped when this happens at
$\tstop\!<\!\Delta t$.  $\yf$ holds $n_i(t)$ minus the molecules
consumed up to $\tstop$ and $\Ff$ holds the respective condensation
sinks as before. With these vectors passed, the implicit integration
solves for the rest, including the relaxation through gas feedback.

\smallskip\noindent{\sf Case 2 (sublimation of volatile materials):}
In the frozen integration, every material $s$ in each representative
particle follows its own growth or sublimation rate. Thus, if passive
or growing materials are present, they form a core that sets a minimum
volume.  Particles cannot shrink below that volume while other
volatile materials with $\partial V_s/\partial t\!<\!0$ sublimate at
rates proportional to $b_s^{\rm mix}$. Particles with such a core will
therefore never be removed at $V_\ell$.  A volatile material that is
completely used up within $\Delta t$ is removed from the particle at
that time, its content is returned to the gas phase, and its rate is
switched off in both $\vec{F}$ and $\Ff$ for the subsequent implicit
step. If a material is used up in both representative particles, it is
absent from $\yf$, i.e.\ $L_3^s\!=\!0$.

\begin{figure}
  \centering
  \includegraphics[width=75mm,trim= 0 0 0 0,clip]
                  {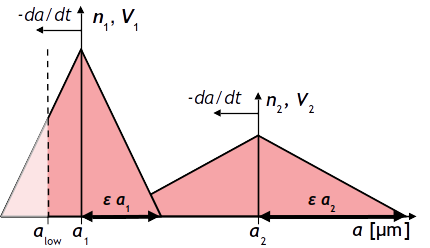} \\[-1mm]
  \caption{Smooth crossing of particles through the lower boundary
    $a_{\rm low}$ in size space. The dust moments are represented by two
    particles $\{n_1,V_1,n_2,V_2\}$ with hat functions attached. The
    volume that crosses $a_{\rm low}$ in $\Delta t$ is removed.}
  \label{fig:hats_sketch}
  \vspace*{-2mm}
\end{figure}

\smallskip\noindent{\sf Case 3 (complete sublimation).} When all
materials are sublimating and particles are expected to shrink below
the lower boundary $a_{\rm low}=(3V_\ell/4\pi)^{1/3}$ within
$\Delta t$, we switch to a representation of the moments by two
triangular distribution functions (``hats'') centred at the radii
$\{a_1,a_2\}$, with half-widths $\varepsilon\,a_i$ (see
Fig.\,\ref{fig:hats_sketch})
\begin{equation}
  f(a) = \sum_{i=1}^{2} \frac{n_i}{\varepsilon\,a_i}\,
         \max\!\left(0,\,1-\frac{|a-a_i|}{\varepsilon\,a_i}\right) \ .
  \label{eq:hats}
\end{equation}
For any $\varepsilon$ up to a maximum value $\varepsilon_{\max}$ set
by the moments, there is a unique set $\{n_1,V_1,n_2,V_2\}$ that
represents $\{L_0,L_1,L_2,L_3\}$ exactly; we use
$\varepsilon=\min(0.9,\,0.9\,\varepsilon_{\max})$. For
$\varepsilon\to 0$ the hats become the double-$\delta$ representation
of Eq.\,(\ref{eq:double-delta}). Along the frozen growth law, the
particle sizes within the hats can be followed exactly, which yields
the fraction of the size distribution that crosses $a_{\rm low}$
within $\Delta t$. This fraction is removed from $\yf$ and its content
is returned to the gas phase. In $\vec{F}$ and $\Ff$, the removal
appears as a continuous negative flux in size space,
$J(V_\ell)=f(a_{\rm low})\,\dot a(a_{\rm low})<0$, which enters the
moment equations (\ref{eq:mom}) like a negative nucleation rate. The
hats thereby prevent the removal of particles from being sudden and
discontinuous. If both representative particles disappear within
$\Delta t$, the dust components of $\vec{y}$ are set to zero and all
condensed material is returned to the gas phase.

All modifications made by the pre-conditioner credit the gas with
exactly the amounts of the elements taken from or returned by the
particles. Therefore, $\vec{F}$ and $\Ff$ are both element-conserving,
and hence the final implicit intergation step is also
element-conserving.

\section{The box test}
\label{app:box_test}

\begin{figure}
  \centering
  \includegraphics[width=55mm,trim= 0 0 0 0,clip]
                  {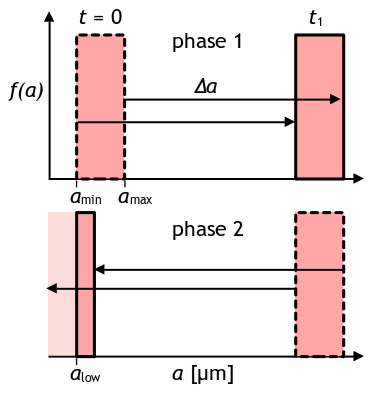} \\[-1mm]
  \caption{Setup of the box test. At $t\!=\!0$ we assume a box-like
    size distribution function $f(a)$ between $a_{\rm min}$ and
    $a_{\rm max}$. Considering a supersaturated gas in the
    Epstein regime, this box will move by $\Delta a$ in size space,
    until the partial pressure of the vapour approaches the saturation
    vapour pressure at $t_1$ (phase 1). We then suddenly increase the
    temperature, forcing the particles to shrink again (phase 2),
    until they cross the lower boundary radius $a_{\rm low}$ and
    eventually sublimate completely.}
  \label{fig:box_test_setup}
  \vspace*{-2mm}
\end{figure}

\begin{figure}
  \hspace*{-2mm}
  \includegraphics[width=90mm,trim= 0 0 0 0,clip]
                  {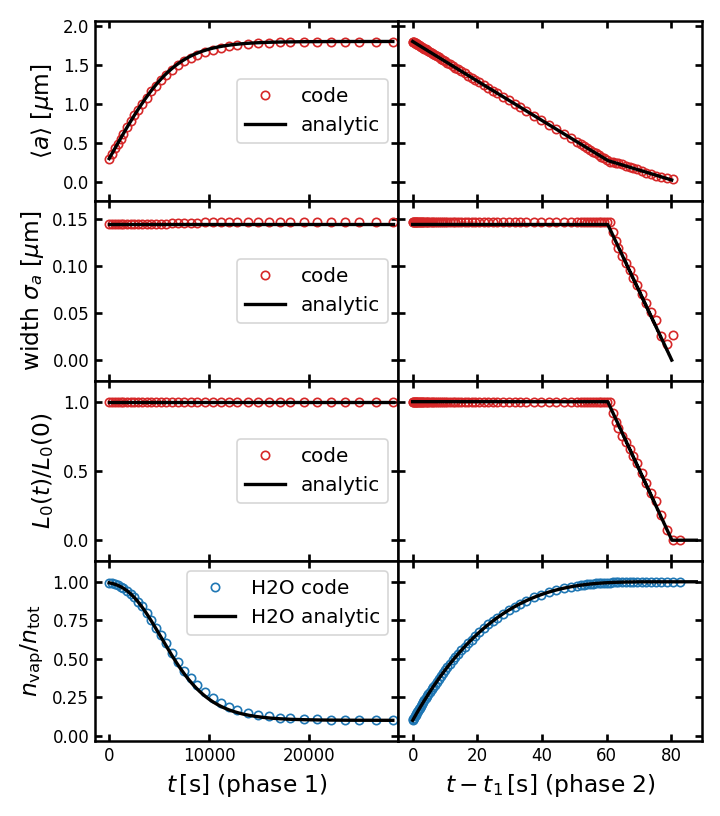} \\[-1mm]
  \caption{The upper panel shows the mean particle size $\langle
    a\rangle$, the second panel the width of the size distribution
    function $\sigma_a$, the third panel the number of particles
    divided by the initial number of particles, and the fourth panel
    shows the density of \ce{H2O} molecules in the gas phase divided by
    the total.  The left side shows the growth phase up to $t\!=\!t_1$
    (phase 1) and the right side shows the sublimation phase (phase
    2). The black lines are analytic solutions of the problem.}
  \label{fig:box_test}
\end{figure}

In order to check the validity of our moment method, including the new
hats to smoothly remove particles when they sublimate completely, we
have designed a test problem that has an analytic solution called the
``box test''.  We start with a box-like size distribution between
$a_{\rm min}(0)=\!0.05\,\mu$m and $a_{\rm max}(0)\!=\!0.55\,\mu$m, see
Fig.\,\ref{fig:box_test_setup}, with a mean radius of $\langle
a\rangle\!=\!(a_{\rm min}\!+\!a_{\rm max})/2\!=\!0.3\,\mu$m, a
width of $\sigma_a\!=\!\sqrt{\langle a^2\rangle -\langle a\rangle^2} =
\big(3/(4\pi)\big)^{1/3}\!\!\sqrt{L_2/L_0-(L_1/L_0)^2}\!=\!(a_{\rm
  max}\!-\!a_{\rm min})/\sqrt{12}\!=\!0.1443\,\mu$m. The
two-particle representation gives $a_1\!=\!\langle
a\rangle-\sigma_a\!=\!0.1557\,\mu$m and $a_2\!=\!\langle
a\rangle+\sigma_a\!=\!0.4443\,\mu$m.  During the growth phase,
we expose these particles to a supersaturated \ce{H2O} vapour until
the vapour has relaxed to saturation.  In the Epstein regime this
causes the built up of an additional constant layer of water ice on
all particles, resulting in a common shift of $f(a)$ in size space
that conserves its shape, here in particular $\sigma_a\!=\rm\!const$.
Then suddenly, at the onset of phase 2, we increase the saturation
vapour pressure (equivalent to a sudden increase of the temperature)
such that the particles are now undersaturated and start to shrink,
still with the same constant $\sigma_a$, until they start crossing
$a_{\rm low}$ and finally evaporate completely.

Figure~\ref{fig:box_test} shows the results of the box test, computed
with precision $=\!0.25$ (36 time steps in phase~1 and 80 in
phase~2). The mean particle radius $\langle a\rangle$ agrees with the
analytic solution to within 1.4\% during phase~1 and 0.3\% during
phase~2. The width of the size distribution, which should stay
constant, increases by about 3\% during phase~1. This slight overprediction
of $\sigma_a$ is expected, because the implicit remainder $\Delta
t\,(\vec{F}\!-\!\Ff)$ updates the moments linearly, with rates evaluated at
the end-of-step sizes, which adds a small positive variance of order
$(\delta a)^2$ per time step, where $\delta a$ is the size correction
carried by the remainder.  The vapour concentration $n_{\rm
  vap}/n_{\rm tot}$ deviates by at most 2.5\% in the steepest part of
the growth phase. The removal of the particles through $a_{\rm low}$
is smooth; it starts 1.4\,s later and is complete 2.3\,s later than in
the analytic solution, i.e.\ within 2--3\% of the duration of phase~2.

We ran many more tests like this one, also in the viscous regime,
where $\sigma_a$ decreases with growth, and for mixtures of two
materials.  Each time, the introduction of the frozen solution and the
hats resulted in significant improvements of the test results.

\section{Vapour pressures from desorption energies}
\label{app:pvap}

To derive the coefficients for the vapour pressure $[\rm dyn/cm^2]$
in the form
\begin{equation}
    \ln(\Pvap) = A - \frac{B}{T} + C \ln(T) 
    \label{eq:target}
\end{equation}
we consider an equilibrium between the rate of adsorption
of gas-phase molecules onto a surface and the rate of thermal
desorption from that surface
\begin{equation}
    R_\mathrm{ads} = R_\mathrm{des} \ .
\end{equation}
Assuming perfect sticking, the rate of
adsorption in the equilibrated state is given by
\begin{equation}
    R_\mathrm{ads} = \frac{\Pvap}{\sqrt{2\pi\,\msp\,kT}}
\end{equation}
The rate of first-order thermal desorption is given by
\begin{equation}
    R_\mathrm{des} = \nsurf\,\freq\,\exp\left(-\frac{\Edes}{kT}\right)
\end{equation}
where $\freq = \sqrt{2\,\nsurf \Edes/\msp} / \pi$
is the characteristic vibrational frequency of the adsorbate, $\nsurf$
is the number of binding sites per surface area and $\Edes$ is the
desorption energy [erg].  Equating the two rates and solving for
$\Pvap$ leads to
\begin{equation}
  \Pvap = \nsurf \freq \sqrt{2\pi\,\msp\,kT}\,
          \exp\left(-\frac{\Edes}{kT}\right)
\end{equation}
Taking the natural logarithm of both sides results in
\begin{equation}
  \ln(\Pvap) = \ln\left( \nsurf \freq \sqrt{2\pi\,\msp\,k} \right)
  + 0.5 \ln(T) - \frac{\Edes}{kT}
\end{equation}
Comparison to the coefficients in Eq.\,(\ref{eq:target}) results in
\begin{align}
    B &= \Edes/k \\
    C &= 0.5 \\
    A &= \ln\left( \frac{\nsurf}{\pi} \sqrt{4\pi\,\nsurf \Edes\,k} \right)
\end{align}
Table~\ref{tab:icedata} lists these coefficients for \ce{H2O}\#,
\ce{CO2}\#, \ce{CO}\# and \ce{CH4}\# when using the same desorption
energies as in ProDiMo and assuming $\nsurf\!=\!1.5\times10^{15}\rm\,cm^{-2}$.

\section{Validation of simplified network}
\label{app:networks}

\begin{table}
  \caption{Chemical networks \label{tab:networks}}
  \vspace*{-2mm}
  \begin{tabular}{ll}
    \hline
    \multicolumn{2}{c}{tiny network} \\
    \hline
    ices          & \ce{H2O}\#, \ce{CO}\#, \ce{CO2}\#, \ce{CH4}\#\\
    \multicolumn{2}{p{85mm}}{12 species$^{(1)}$: \small
                  \ce{H},\,\ce{H2},\,\ce{OH},\,\ce{H2O},\,\ce{O},\,\ce{CO},\,\ce{CO2},\,\ce{C},\,\ce{CH},\,\ce{CH2},\,\ce{CH3},\,\ce{CH4} }\\
    elements$^{(1)}$ & H, C, O \\
    const.\,rates & gain: 8,  loss: 3\\
    linear rates  & gain: 53, loss: 12\\
    square rates  & gain: 105, loss: 90\\[2mm]
    \hline
    \multicolumn{2}{c}{medium network} \\
    \hline
    ices          & \ce{H2O}\#, \ce{CO}\#, \ce{CO2}\#, \ce{CH4}\#\\
    \multicolumn{2}{p{85mm}}{24 species: \small
                  \ce{H+}\!, \ce{H}, \ce{H2}, \ce{OH}, \ce{H2O},
                  \ce{O}, \ce{CO}, \ce{CO2},
                  \ce{C+}, \ce{C}, \ce{CH}, \ce{CH2}, \ce{CH3}, \ce{CH4},
                  \ce{N}, \ce{N2}, \ce{NH}, \ce{NH2}, \ce{NH3}, \ce{CN},
                  \ce{HCN}, \ce{HNC}, \ce{HCO}, \ce{HNO} } \\
    elements      & H, C, N, O \\
    const.\,rates & gain: 20,  loss: 6\\
    linear rates  & gain: 222, loss: 24\\
    square rates  & gain: 312, loss: 304\\[2mm]
    \hline
    \multicolumn{2}{c}{full network} \\
    \hline
    ices$^{(2)}$   & \ce{H2O}\#, \ce{CO}\#, \ce{CO2}\#, \ce{CH4}\#\\
    \multicolumn{2}{p{85mm}}{119 species$^{(3)}$: \small
    \ce{H}, \ce{H+},     \ce{H-},     \ce{H2},
    \ce{H2+},    \ce{H3+}, \ce{H2}$^{\!\star}$, \ce{He},     \ce{He+},
    \ce{HeH+},   \ce{C},      \ce{C+},     \ce{C++},    \ce{CH},
    \ce{CH+},    \ce{CH2},    \ce{CH2+},   \ce{CH3},    \ce{CH3+},
    \ce{CH4},    \ce{CH4+},   \ce{CH5+},   \ce{C2},     \ce{C2+},
    \ce{C2H},    \ce{C2H+},   \ce{C2H2},   \ce{C2H2+},  \ce{C2H3},
    \ce{C2H3+},  \ce{C2H4},   \ce{C2H4+},  \ce{C2H5},   \ce{C2H5+},
    \ce{CN},     \ce{CN+},    \ce{HCN},    \ce{HCN+},   \ce{HCNH+},
    \ce{HNC},    \ce{H2CN},   \ce{OCN},    \ce{OCN+},   \ce{CO},
    \ce{CO+},    \ce{HCO},    \ce{HCO+},   \ce{CO2},    \ce{CO2+},
    \ce{HCO2+},  \ce{H2CO},   \ce{H2CO+},  \ce{CH3O},   \ce{H3CO+},
    \ce{CH2OH},  \ce{CH3OH},  \ce{CH3OH+}, \ce{CH3OH2}, \ce{CS},
    \ce{CS+},    \ce{HCS},    \ce{HCS+},   \ce{H2CS},   \ce{H2CS+},
    \ce{H3CS+},  \ce{OCS},    \ce{OCS+},   \ce{HOCS+},  \ce{N},
    \ce{N+},     \ce{N++},    \ce{NH},     \ce{NH+},    \ce{NH2},
    \ce{NH2+},   \ce{NH3},    \ce{NH3+},   \ce{NH4+},   \ce{N2},
    \ce{N2+},    \ce{HN2+},   \ce{NO},     \ce{NO+},    \ce{NO2},
    \ce{NO2+},   \ce{HNO},    \ce{HNO+},   \ce{H2NO+},  \ce{NS},
    \ce{NS+},    \ce{HNS+},   \ce{O},      \ce{O+},     \ce{O++},
    \ce{OH},     \ce{OH+},    \ce{H2O},    \ce{H2O+},   \ce{H3O+},
    \ce{SO},     \ce{SO+},    \ce{SO2},    \ce{SO2+},   \ce{HSO2+},
    \ce{S},      \ce{S+},     \ce{S++},    \ce{HS},     \ce{HS+},
    \ce{H2S},    \ce{H2S+},   \ce{H3S+},   \ce{Mg},     \ce{Mg+},
    \ce{Mg++},   \ce{Na},     \ce{Na+},    \ce{Na++},   \ce{e-} } \\
    elements      & H, He, C, N, O, S, Na, Mg \\
    const.\,rates & gain: 0,  loss: 0\\
    linear rates  & gain: 331, loss: 112\\
    square rates  & gain: 3333, loss: 2456\\
    \hline
  \end{tabular}\\[1mm]
  \footnotesize
  $^{(1)}$: All elements and species of the full network are still present as
  background gas in the tiny and medium networks, see
  Sect.\,\ref{sec:chemistry}\\
  $^{(2)}$: The original ProDiMo network also includes \ce{O}\#,
  \ce{OH}\#, \ce{C}\#, \ce{CH}\#, \ce{CH2}\# and \ce{CH3}\# to
  simulate the ice hydrogenation chemical pathways.\\
  $^{(3)}$: \ce{H2}$^{\!\star}$ means vibrationally excited molecular hydrogen.
\end{table}

\begin{figure*}
  \hspace*{-4mm}
  \begin{tabular}{lcc}
    {\bf tiny network}\hspace{15mm}
    {\sf 1\,au} & {\sf 10\,au} & {\sf 100\,au} \\[-1mm]    
    \hspace*{-1mm}
    \centerimage[page=1,height=54mm,trim=5 0 0 0,clip]
                {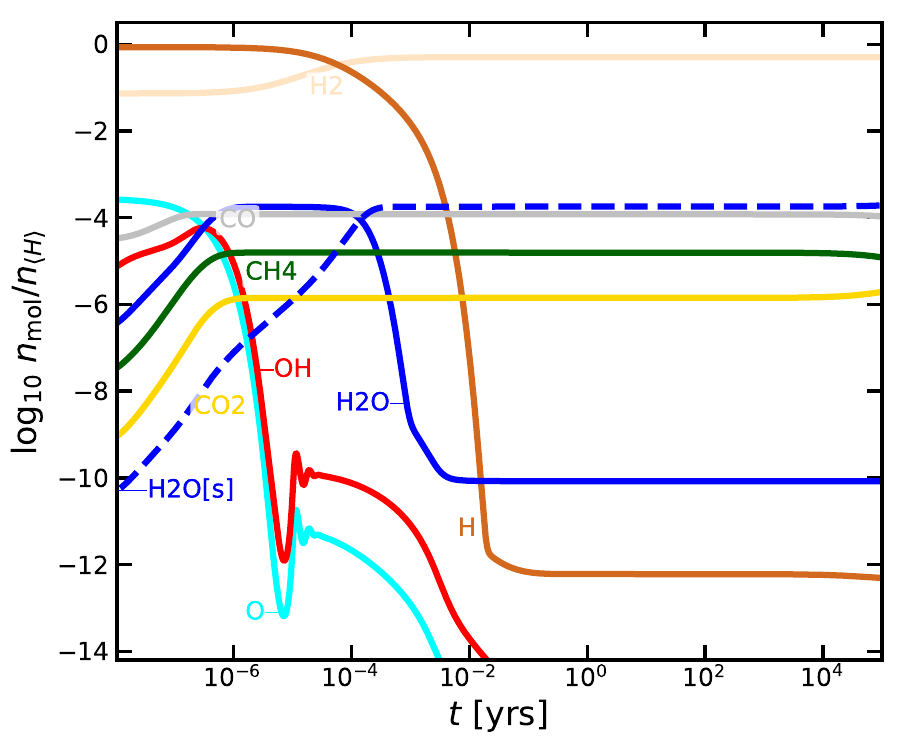} &
    \hspace*{-5mm}
    \centerimage[page=1,height=54mm,trim=28 0 0 0,clip] 
                {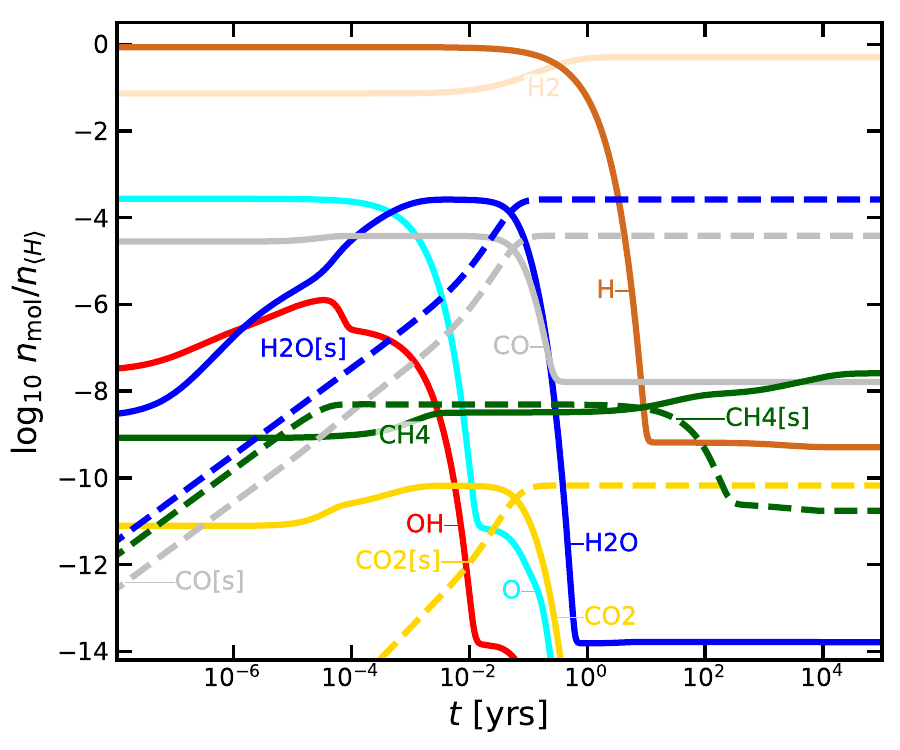} &
    \hspace*{-5mm}
    \centerimage[page=1,height=54mm,trim=28 0 0 0,clip] 
                {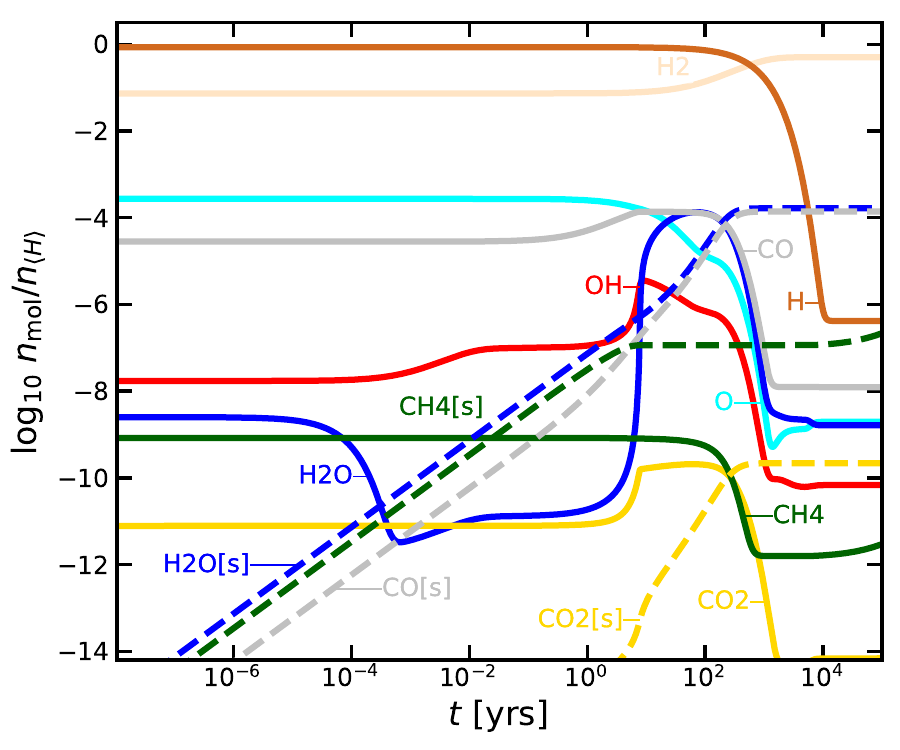} \\
                
    {\bf med.\,network} &&\\*[-1mm]
    \hspace*{-1mm}
    \centerimage[page=1,height=54mm,trim=5 0 0 0,clip]
                {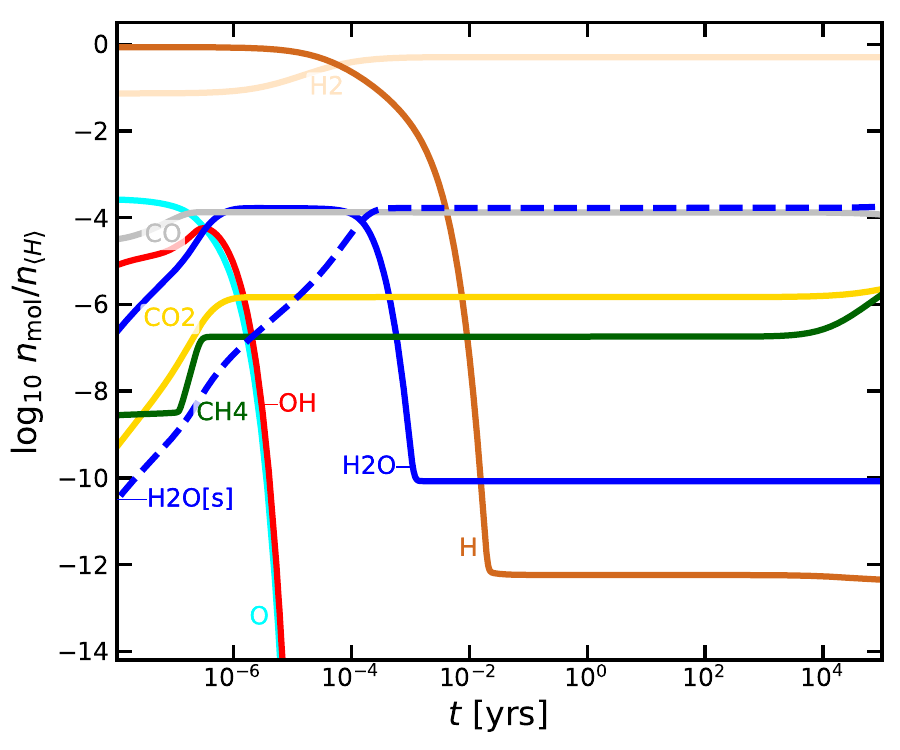} &
    \hspace*{-5mm}
    \centerimage[page=1,height=54mm,trim=28 0 0 0,clip] 
                {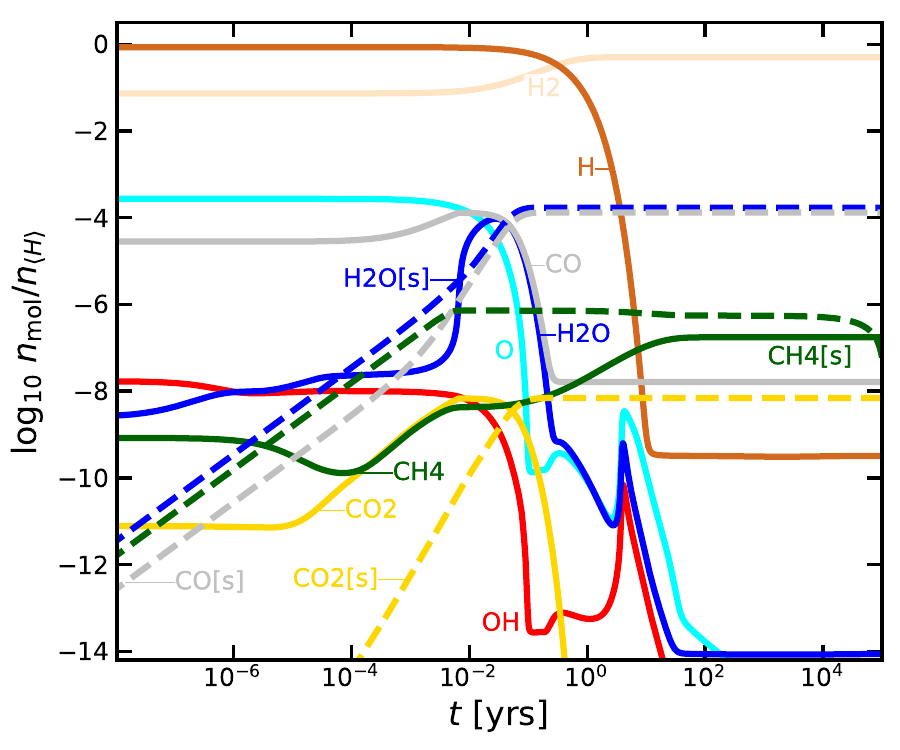} &
    \hspace*{-5mm}
    \centerimage[page=1,height=54mm,trim=28 0 0 0,clip] 
                {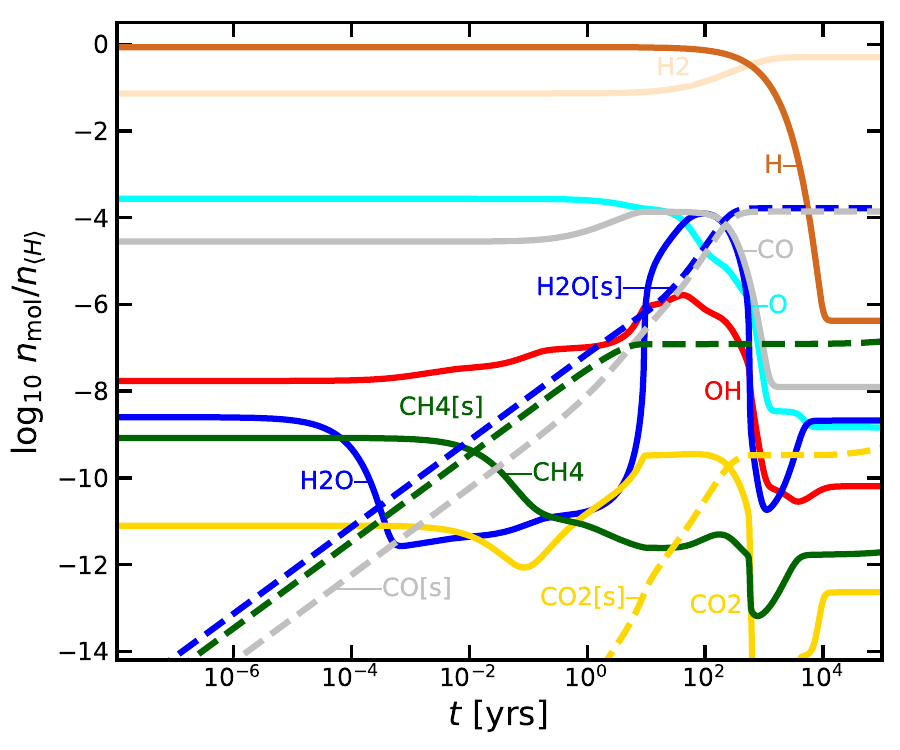} \\
                
    {\bf full network} &&\\*[-1mm]
    \hspace*{-1mm}
    \centerimage[page=1,height=54mm,trim=5 0 0 0,clip]
                {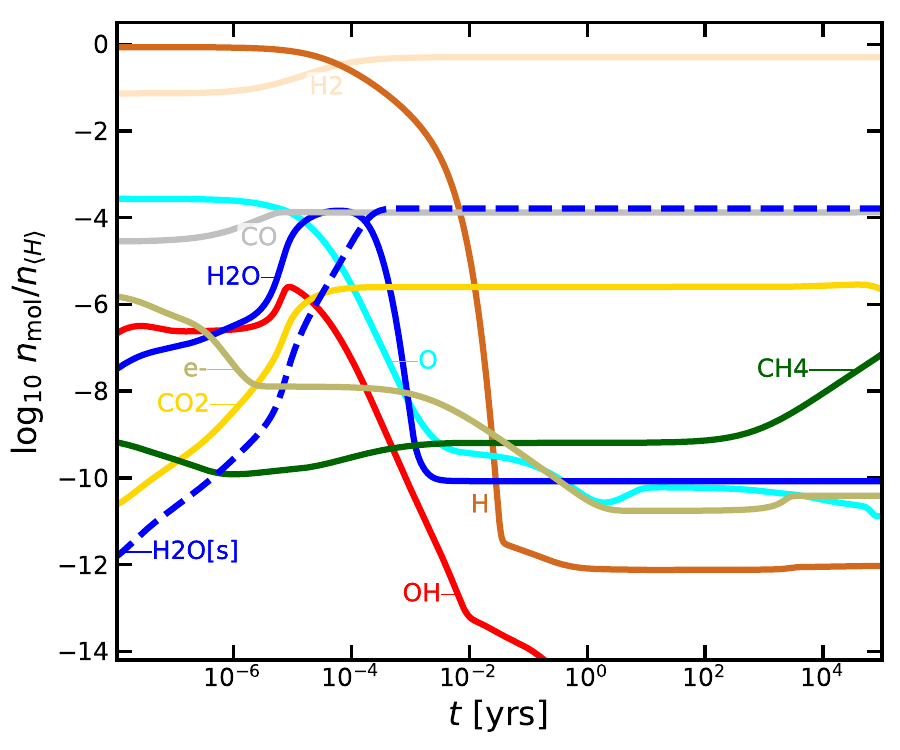} &
    \hspace*{-5mm}
    \centerimage[page=1,height=54mm,trim=28 0 0 0,clip] 
                {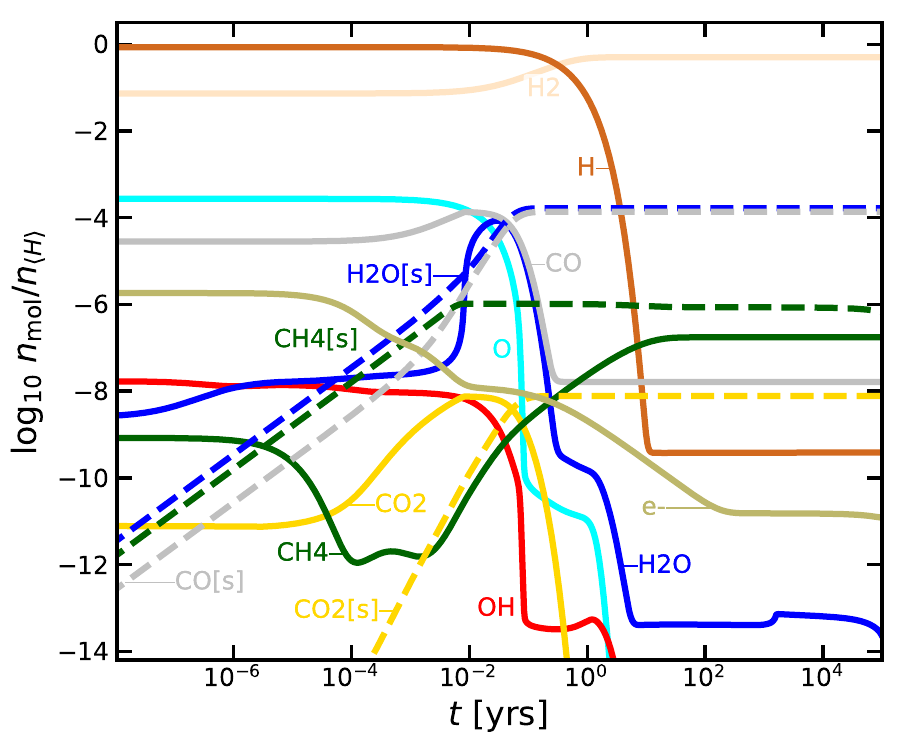} &
    \hspace*{-5mm}
    \centerimage[page=1,height=54mm,trim=28 0 0 0,clip] 
                {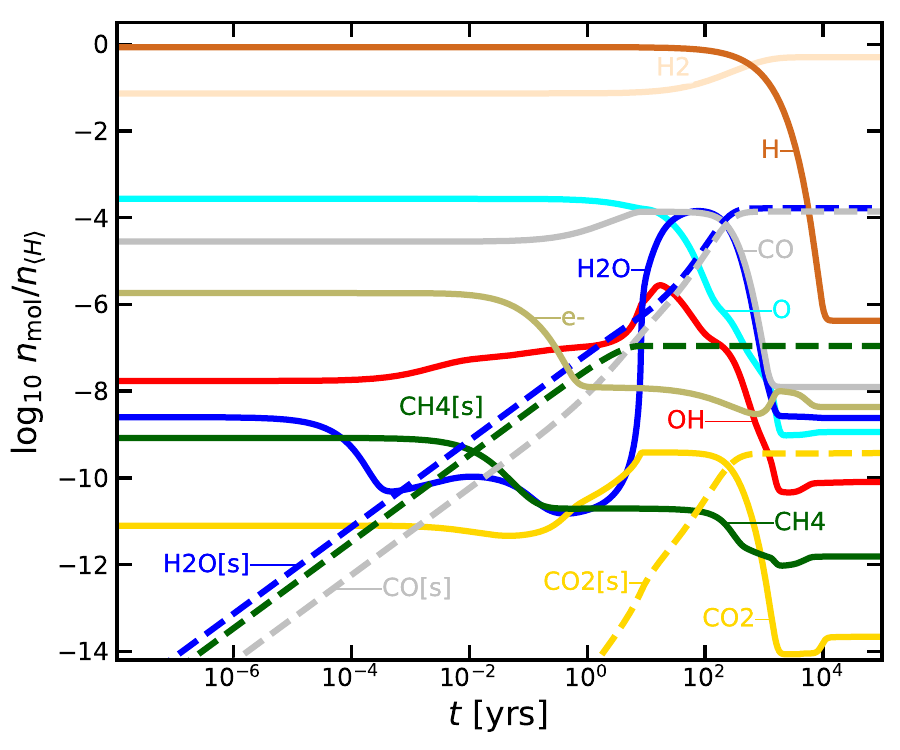} \\
                
    {\bf ProDiMo} &&\\*[-1mm]
    \hspace*{-1mm}
    \centerimage[page=1,height=54mm,trim=5 0 0 0,clip]
                {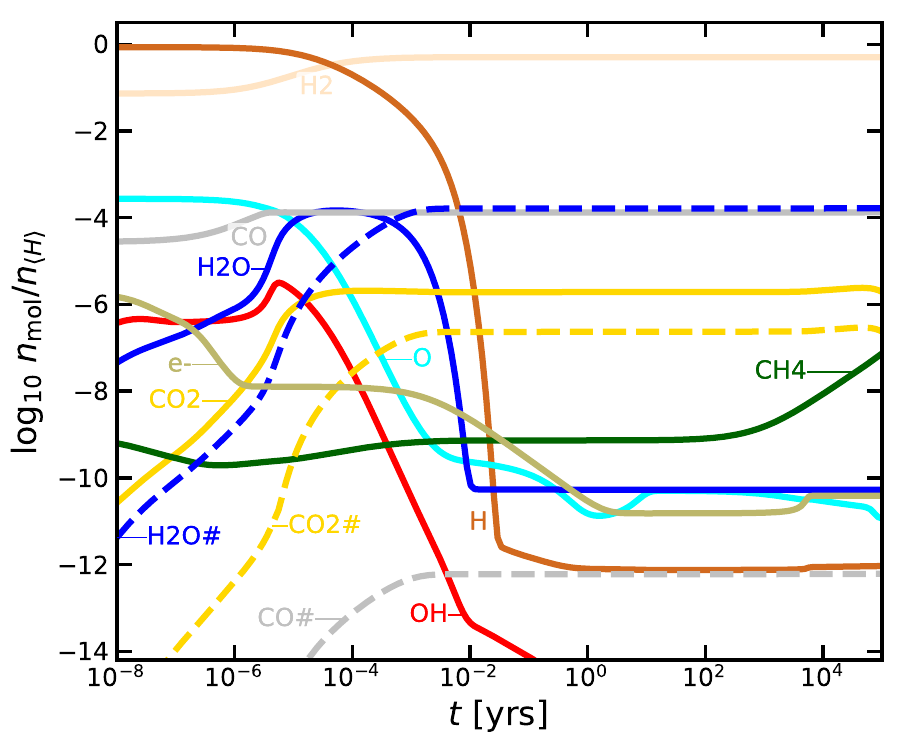} &
    \hspace*{-5mm}
    \centerimage[page=1,height=54mm,trim=28 0 0 0,clip] 
                {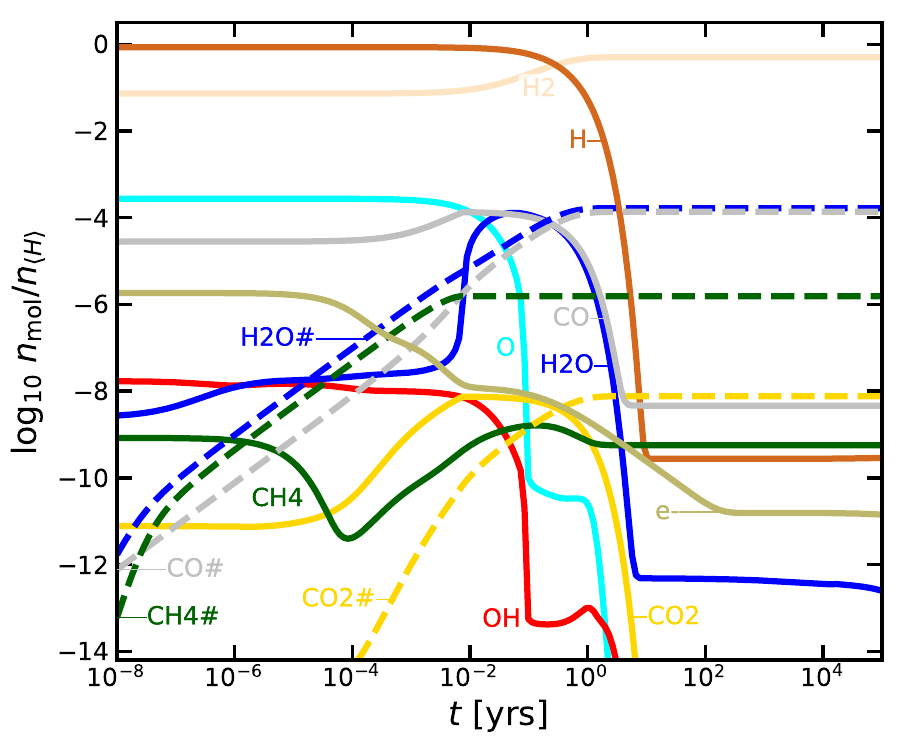} &
    \hspace*{-5mm}
    \centerimage[page=1,height=54mm,trim=28 0 0 0,clip] 
                {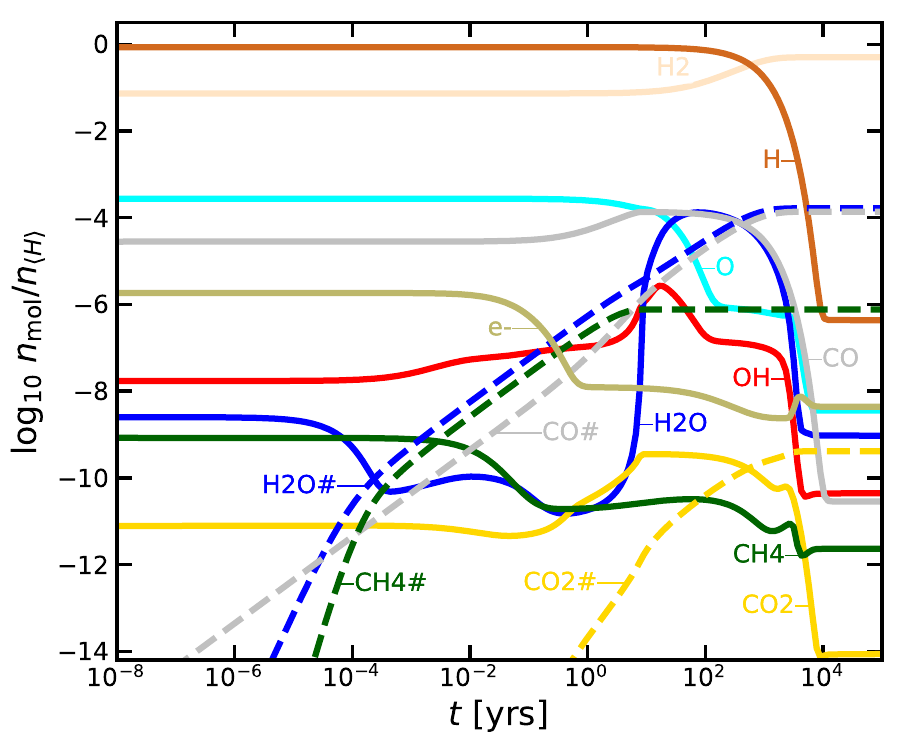} \\*[-1mm]
  \end{tabular}
  \caption{Testing the chemistry and ice formation during the first
    100\,000\,yrs at 3 selected places in the midplane, modelled with
    different chemical networks.  Ice species are plotted with dashed
    lines.  The first three rows show the results calculated with
    DiffuDrift using the tiny, the medium, and the full chemical
    network, see Table~\ref{tab:networks}. The last row shows the
    results of the original ProDiMo model, where the solid surface is
    fixed and the ice formation is treated in different ways, see
    text.  Only the chemistry is advanced here -- the density,
    temperature and radiation field at the selected places are fixed,
    and there is no diffusion nor settling of icy dust particles in
    these models.}
  \label{fig:ChemModel}
\end{figure*}

Figure~\ref{fig:ChemModel} shows the results of a number of test
models with different chemical networks of increasing size.  We select
three points in the midplane of the disc for these tests: a point at
1\,au, a point at 10\,au and a point at 100\,au, some properties of
these points are described in Table~\ref{tab:columns}.  We only
advance the chemistry at these points, {without diffusion and
  settling}, from initial concentrations listed in
Table~\ref{tab:initconc} and the initial powerlaw size distribution of
passive particles explained in Sect.~\ref{sec:settled}, for 100\,000
yrs, by simply calling the chemistry and ice formation module in
DiffuDrift again and again at this point, without using the modules
for diffusion and particle settling.  This is a relatively hard test
for the chemical networks, because the gas phase composition is
initially very different from the final relaxed chemical state, upon
which the simplified networks were built by ProDiMo, in particular
concerning the composition of the background gas.

The last row of plots in Fig.~\ref{fig:ChemModel} shows the same
simulations made with ProDiMo itself.  Here, however, the gas-ice
reactions are treated in a different way and the dust surface is
assumed to be constant, because ProDiMo does not consider the physical
growth of the dust particles by ice formation. Therefore, all
DiffuDrift models finally show an accelerated phase of ice formation
once the surface area has increased already by ice formation.

The properties of the three networks ``tiny'', ``medium'' and ``full''
are explained in Table~\ref{tab:networks}.  Already the tiny network
can reproduce the timescale for water ice formation and the final
concentrations of \ce{H2O} and \ce{CO2} in the gas phase quite
accurately in all three points. However, there are some
short-comings, in particular \ce{CH4} is over-predicted with respect to
the more sophisticated networks.

Noteworthy, all DiffuDrift models for the first column at 1\,au only
show water ice formation, but no \ce{CO2} ice and no CO ice, whereas
the original ProDiMo model predicts those in trace concentrations.
This is because of the different kinetic treatment of the ice
formation with net (deposition $-$ sublimation) rates regulated by
the supersaturation ratio in DiffuDrift, whereas ProDiMo uses absolute
adsorption and thermal desorption rates.  In DiffuDrift net growth
$\gamma_s>0$ cannot occur when $S<1$ (see Eq.\,\ref{eq:desorb}), so
$L_3^s$ stays exactly zero, whereas in ProDiMo, the molecules adsorb
anyway, but just come off quickly again, resulting in tiny yet
non-zero concentrations of thermodynamically unstable ices.

Figure~\ref{fig:ChemModel} also shows slightly different ice growth
rates at very early phases of \ce{H2O}\# and \ce{CH4}\#.  This is
because the sub-networks used in DiffuDrift to compute the ice
hydrogenation rates (Sect.\,\ref{sec:hydrogenation}) is not yet in a
stationary state.

Concerning the medium network, the ice formation timescales and the
final ice and vapour concentrations for all considered ice species
match the results of the full network well, at all considered 3
points, but some temporal deviations still occur, which are likely
artefacts due to the network reduction from 119 to 24 species.  We
have therefore decided to simply use the full network in the
DiffuDrift models for the main results presented in 
Sect.\,\ref{sec:mainresults}. The reduced tiny and medium networks can
still be used to provide quick pre-view results, but for a thorough
discussion of the time-dependent ice formation in protoplanetary
discs, we prefer to use the slow but complete full network.

\begin{table}
  \caption{Evolution of size moments in the DiffuDrift model at radius
    1\,au with full chemistry, and analytical expectations.}
  \label{tab:Lmom}
  \vspace*{-2mm}
  \begin{tabular}{c|c|cc}
    \hline
    &&&\\[-2.2ex]
    $j$
    & $L_j\,(t\!=\!0)\rm\,[cm^j/g]$
    & $L_j\,\rm(0.1\,Myr)\rm\,[cm^j/g]$
    & $L_j^{\rm ana}\rm\,[cm^j/g]$\\
    \hline
    &&&\\[-2.2ex]
    0 & $2.751\times10^{+9}$ & $2.751\times10^{+9}$ & $2.751\times10^{+9}$ \\
    1 & $3.709\times10^{+4}$ & $2.543\times10^{+5}$ & $2.544\times10^{+5}$ \\
    2 & $8.955\times10^{-1}$ & $2.391\times10^{+1}$ & $2.392\times10^{+1}$ \\
    3 & $1.771\times10^{-3}$ & $4.032\times10^{-3}$ & $4.033\times10^{-3}$ \\
    4 & $2.868\times10^{-4\;,(1)}$ & $2.883\times10^{-4\;,(2)}$ & $2.875\times10^{-4}$ \\
    \hline
  \end{tabular}\\
  \footnotesize
  $^{(1)}$: The two particle representation gives
  $L_4\!=\!7.82\times10^{-6}$ and the power-correction factor is
  $P_{\rm c}(4)\!=\!36.8$. The reported value is $P_{\rm c}(4)\times L_4$\\
  $^{(2)}$: The two particle representation gives
  $L_4\!=\!8.59\times10^{-6}$ and the power-correction factor is
  $P_{\rm c}(4)\!=\!33.7$.
\end{table}
{Concerning the evolution of the size moments we start the
  DiffuDrift models with a powerlaw size distribution with index
  $-3.5$ between $0.05\,\mu$m and 3\,mm and a dust/gas mass ratio of
  0.004, which corresponds to the size moments $L_j\!\propto\!\int
  f(a)\,a^j\,da$ with initial values listed in the left column of
  Table~\ref{tab:Lmom}.  At the end of the simulation at 1\,au with
  full chemistry, we measure the values listed in the middle column of
  Table~\ref{tab:Lmom}.  Our analytical expectations are $L_j^{\rm
    ana}\!\propto\!\int f(a)\,(a\!+\!\Delta a)^j\,da$. Computing these
  integrals numerically by means of 100 log-equidistant points with a
  unique size increase of $\Delta a\!=\!0.49\,\mu$m results in the
  values listed in the right column. This comparison shows that the
  changes of the size distribution that we see in the models are
  consistent with our analytical expectations of all particles having
  grown by 0.49\,$\mu$m in radius.}


\section{Initial molecular concentrations}
\label{app:initconc}

Table~\ref{tab:initconc} lists the assumed initial concentrations and
element abundances used in the DiffuDrift models
$n_i(z,t\!=\!0)=n_i^0$. These have been computed by ProDiMo in a model
for a molecular cloud with parameters $\nH\!=\!10^4\rm\,cm^{-3}$,
$T\!=\!10\,$K and optical extinction $A_V\!=\!10$, where we advance
the full chemistry for 100\,000\,yrs.  The initial concentrations for
this molecular cloud model are to put all elements with abundances
listed in Table~\ref{tab:initconc} into their highest ionisation state
available in the network, which are $\{\ce{H+}, \ce{He+}, \ce{C++},
\ce{N++}, \ce{O++}, \ce{S++}, \ce{Mg++}, \ce{Na++}\}$.  The duration
of 100\,000\,yrs is chosen such that the \ce{H2} and ice-formation is
yet incomplete. The resulting concentrations listed in
Table~\ref{tab:initconc} are used for the initial molecular
concentrations in the DiffuDrift models.  However, since we want to
start the DiffuDrift models strictly without any ice, the ices listed
in Table~\ref{tab:initconc} are broken up again into atoms and added
to the respective atomic initial concentrations.  {This procedure
  reflects the chemical-reset limit as denoted by \cite{Leemker2026},
  who argue for water ice being inherited rather than reformed in the
  disc. However, the ice formation and destruction timescales in our
  models turn out to be much shorter than evolutionary timescales, see
  Fig.\,\ref{fig:ChemModel}.}

\begin{table}
 \caption{Initial concentrations of molecules assumed}
 \label{tab:initconc} 
 \vspace*{-2mm}\hspace*{-1mm} 
 \resizebox{90mm}{!}{ 
 \begin{tabular}{cc|cc|cc}
 \hline
 &&&&&\\[-2.2ex]  
 \ce{H}     & $8.54\times10^{-1}$ & \ce{Mg+}   & $1.04\times10^{-8}$ &  \ce{H3O+}  & $7.26\times10^{-11}$ \\
 \ce{He}    & $9.63\times10^{-2}$ & \ce{O}\#   & $5.29\times10^{-9}$ &  \ce{CS}    & $5.06\times10^{-11}$ \\
 \ce{H2}    & $7.25\times10^{-2}$ & \ce{OH}\#  & $5.29\times10^{-9}$ &  \ce{HCN}   & $3.06\times10^{-11}$ \\
 \ce{O}     & $2.17\times10^{-4}$ & \ce{CN}    & $4.78\times10^{-9}$ &  \ce{CH3+}  & $2.64\times10^{-11}$ \\
 \ce{C}     & $8.10\times10^{-5}$ & \ce{H2O}   & $2.51\times10^{-9}$ &  \ce{HNC}   & $2.15\times10^{-11}$ \\
 \ce{N}     & $7.80\times10^{-5}$ & \ce{CH}\#  & $2.28\times10^{-9}$ &  \ce{H3+}   & $1.59\times10^{-11}$ \\
 \ce{H2O}\# & $5.45\times10^{-5}$ & \ce{CH2}\# & $2.28\times10^{-9}$ &  \ce{CH3}   & $7.96\times10^{-12}$ \\
 \ce{CO}    & $2.79\times10^{-5}$ & \ce{C}\#   & $2.28\times10^{-9}$ &  \ce{CO2}   & $7.78\times10^{-12}$ \\
 \ce{CH4}\# & $2.62\times10^{-5}$ & \ce{CH3}\# & $2.28\times10^{-9}$ &  \ce{C2H2+} & $6.20\times10^{-12}$ \\
 \ce{CO}\#  & $2.20\times10^{-6}$ & \ce{Na+}   & $2.11\times10^{-9}$ &  \ce{SO+}   & $4.71\times10^{-12}$ \\
 \ce{e-}    & $1.83\times10^{-6}$ & \ce{NO}    & $1.33\times10^{-9}$ &  \ce{OH+}   & $4.53\times10^{-12}$ \\
 \ce{H+}    & $1.15\times10^{-6}$ & \ce{He+}   & $1.24\times10^{-9}$ &  \ce{H2O+}  & $4.51\times10^{-12}$ \\
 \ce{N2}    & $6.79\times10^{-7}$ & \ce{C2H2}  & $1.05\times10^{-9}$ &  \ce{H2CO}  & $4.08\times10^{-12}$ \\
 \ce{C+}    & $5.00\times10^{-7}$ & \ce{C2}    & $9.63\times10^{-10}$ & \ce{HCO}   & $3.28\times10^{-12}$ \\
 \ce{S+}    & $1.66\times10^{-7}$ & \ce{CH4}   & $8.36\times10^{-10}$ & \ce{HCO+}  & $2.91\times10^{-12}$ \\
 \ce{S}     & $1.93\times10^{-8}$ & \ce{Mg}    & $2.87\times10^{-10}$ & \ce{NO+}   & $2.75\times10^{-12}$ \\
 \ce{OH}    & $1.70\times10^{-8}$ & \ce{C2H}   & $2.62\times10^{-10}$ & \ce{N+}    & $1.17\times10^{-12}$ \\
 \ce{CH}    & $1.18\times10^{-8}$ & \ce{Na}    & $1.77\times10^{-10}$ \\
 \hline
 \end{tabular}}\\
 \footnotesize Sorted concentrations for all species with $n_{\rm
   mol}/\nH\!>\!10^{-12}$ from ProDiMo for a molecular cloud
 model, see text. For the initial concentrations in the DiffuDrift
 models, the ices are converted into atoms and added to those
 concentrations. The total element abundances are $\epsilon_{\rm
   H}\!=\!1$, $\epsilon_{\rm
   He}\!=\!9.64\times10^{-2}$, $\epsilon_{\rm
   C}\!=\!1.38\times10^{-4}$, $\epsilon_{\rm
   N}\!=\!7.94\times10^{-5}$, $\epsilon_{\rm
   O}\!=\!3.02\times10^{-4}$, $\epsilon_{\rm
   Na}\!=\!2.29\times10^{-9}$, $\epsilon_{\rm
   Mg}\!=\!1.07\times10^{-8}$, and $\epsilon_{\rm
   S}\!=\!1.86\times10^{-7}$.
\end{table}

\end{document}